\RequirePackage{fix-cm}
\documentclass[smallcondensed]{svjour3}     
\smartqed  
\usepackage{graphicx}

\usepackage[ruled,vlined,boxed]{algorithm2e}

\usepackage{booktabs}
\usepackage{pifont}
\usepackage[compatibility=false]{caption}
\usepackage{subcaption}
\usepackage{color}
\usepackage{comment}
\usepackage{enumitem}

\usepackage{longtable}

\usepackage{mathtools}
\usepackage{amsfonts}
\usepackage{bm}

\usepackage[toc,page]{appendix}

\usepackage{ifthen}

\usepackage{cite}
\newcommand{\ie}{\emph{i.e.}\xspace}

\usepackage{hyperref}

\begin{document}

\title{Investigating the Characteristics of One-Sided Matching Mechanisms Under Various Preferences and Risk Attitudes
}

\titlerunning{Investigating the Characteristics of One-Sided Matching Mechanisms}        

\author{Hadi Hosseini        \and
        Kate Larson \and
        Robin Cohen
}


\institute{Hadi Hosseini \at
	Department of Computer Science, Rochester Institute of Technology, Rochester, NY, USA \\
	\email{hhvcs@rit.edu}           
	\and
	Kate Larson \at
	Cheriton School of Computer Science, University of Waterloo, Waterloo, ON, Canada \\
	\email{klarson@uwaterloo.ca}           
	\and 
	Robin Cohen \at
	Cheriton School of Computer Science, University of Waterloo, Waterloo, ON, Canada \\
	\email{rcohen@uwaterloo.ca}           
}


\date{Received: date / Accepted: date}

\maketitle

\begin{abstract}

One-sided matching mechanisms are fundamental for assigning a set of indivisible objects to a set of self-interested agents when monetary transfers are not allowed. 
Two widely-studied randomized mechanisms in multiagent settings are the Random Serial Dictatorship (RSD) and the Probabilistic Serial Rule (PS).
Both mechanisms require only that agents specify ordinal preferences and have a number of desirable economic and computational properties. However, the induced outcomes of the mechanisms are often incomparable and thus there are challenges when it comes to deciding which mechanism to adopt in practice.

In this paper, we first consider the space of general ordinal preferences and provide empirical results on the (in)comparability of RSD and PS. We analyze their respective economic properties under general and lexicographic preferences.
We then instantiate utility functions with the goal of gaining insights on the manipulability, efficiency, and envyfreeness of the mechanisms under different risk-attitude models. Our results hold under various preference distribution models, which further confirm the broad use of RSD in most practical applications.


\end{abstract}

%
%
\section{Introduction}


One-sided matching mechanisms have been extensively adopted in many resource allocation settings such as assigning dormitory rooms or offices to students, students to public schools, college courses to students, organs and medical resources to patients, and members to subcommittees \cite{roth2004kidney,ashlagi2013mix,budish2012multi,manlove2013algorithmics}.
Two prominent (randomized) matching mechanisms that only elicit ordinal preferences from agents are \emph{Random Serial Dictatorship} (RSD)~\cite{abdulkadirouglu1998random} and \emph{Probabilistic Serial Rule} (PS)~\cite{bogomolnaia2001new}. Both mechanisms have important economic properties and are practical to implement. 
The RSD mechanism has strong truthful incentives but guarantees neither efficiency nor envyfreeness. PS satisfies efficiency and envyfreeness; however, it is susceptible to manipulation. 
Therefore, there are subtle points to be considered when deciding which mechanism to use. For example, given a particular preference profile, the mechanisms often produce random assignments which are simply incomparable and thus, without additional knowledge of the underlying utility models of the agents, it is difficult to determine which is the ``better'' outcome. 
Furthermore, properties like efficiency, truthfulness, and envyfreeness can depend on whether there is underlying structure in the preferences, and even in general preference models it is valuable to understand under what conditions a mechanism is likely to be efficient, truthful, or envyfree as this can guide designers choices.



%
In this paper, we study the comparability of PS and RSD when there is only one copy of each object, and analyze the space of all preference profiles for different numbers of agents and objects. 
Working in the space of general ordinal preferences, we provide empirical results on the (in)comparability of RSD and PS and analyze their respective economic properties.
%
%
We show that despite the inefficiency of RSD, the fraction of random assignments at which PS stochastically dominates RSD vanishes, especially when the number of agents is less than or equal to the available objects. We also investigate the manipulability of PS and show that PS is almost always 99\% manipulable for any combination of agents and objects, and the fraction of strongly manipulable profiles goes to one as the ratio of objects to agents increases.
We show similar trends under lexicographic preferences, and further present results on envy of RSD. Our results show that although the fraction of envious agents grows with the number of agents, there is a sudden drop in the fraction of envious agents when there are equal number of agents and objects.

In Section \ref{sec:risk}, we instantiate utility functions for agents to gain deeper insights on the manipulability, social welfare, and envyfreeness of the two mechanisms under different risk attitudes.
Our main result is that under risk aversion, the social welfare of RSD is as good as PS but RSD does create envy among the agents (though the fraction of envious profiles and the total envy are small). Moreover, when the number of agents and objects are equal, RSD assignments are less likely to be dominated by PS, and overall RSD assignments create negligible envy among agents.
We also show that PS is highly susceptible to manipulation in almost all combinations of agents and objects. The fraction of manipulable profiles and the gain from manipulation rapidly increases, particularly when agents become more risk averse. In Section \ref{sec:dist}, we consider two statistical preference distribution models, namely Mallows Models and Polya-Eggenberger Urn Models, and show that the same patterns and trends hold for various combinations of agents and objects, when varying risk parameters and utility functions.

%
%
\section{Preliminaries}
In this section, we describe the basic one-sided matching problem and introduce the two mechanisms we study in detail, Random Serial Dictatorship (RSD) \cite{abdulkadirouglu1998random} and Probabilistic Serial Rule (PS) \cite{bogomolnaia2001new}. We then introduce a number of properties and criteria used to evaluate these mechanisms.


A one-sided matching problem consists of a set of $n$ agents, $N$, and a set of $m$ indivisible objects, $M$.\footnote{This problem is sometimes called the assignment problem or house allocation problem in the literature.} Each agent $i\in N$ has a private strict preference ordering, $\succ_i$, over $M$ where $a\succ_i b$ indicates that agent $i$ prefers to receive object $a$ over object $b$. We represent the preference ordering of agent $i$ by the ordered list of objects $\succ_{i} = a \succ_{i} b \succ_{i} c$ or $\succ_{i} = (abc)$, for short.
We let $\mathcal{P}$ denote the set of all complete and strict preference orderings over $M$. A \emph{preference profile} $\succ\in\mathcal{P}^n$ specifies a preference ordering for each agent, and we use the standard notation $\succ_{-i}=(\succ_1,\ldots, \succ_{i-1},\succ_{i+1},\ldots, \succ_n)$ to denote preferences orderings of all agents except $i$ and thus $\succ=(\succ_i,\succ_{-i})$.

The goal in a one-sided matching problem is to assign the objects in $M$ to the agents in $N$ according to preference profiles, under the constraint that no object can be assigned to more than one agent. If $m=n$ then this means that each agent will receive exactly one object, however if $m<n$ then some agents will receive no object and if $m>n$ then some agents may receive multiple objects. An assignment is represented as a matrix

\begin{equation*}
A =
\begin{pmatrix}
A_{1} \\
A_{2} \\
\vdots \\
A_{n}
\end{pmatrix}
=
\begin{pmatrix}
A_{1,1} & A_{1,2} & \ldots & A_{1,m} \\
A_{2,1} & A_{2,2} & \ldots & A_{2,m} \\
\vdots & \vdots & \ddots & \vdots \\
A_{n,1} & A_{n,2} & \ldots & A_{n,m}
\end{pmatrix}
\end{equation*}

where $A_{i,j}\in[0,1]$ is the probability that agent $i$ is assigned object $j$.
We let $\mathcal{A}$ denote the set of all \emph{feasible} assignments where an assignment $A\in\mathcal{A}$ is \emph{feasible} if and only if $\forall j\in M$, $\sum_{i\in N}A_{i,j} = 1$. If $A\in\mathcal{A}$ is such that $A_{i,j}\in\{0,1\}$ then we say that $A$ is a \emph{deterministic} assignment; otherwise, $A$ is a \emph{random} assignment.
Every random assignment can be represented as a convex combination of deterministic assignments~\cite{von1953certain}, and thus we view random assignments as a probability distribution over a set of deterministic assignments.

\subsection{Matching Mechanisms}


In general, a \emph{matching mechanism}, $\mathcal{M}$, is a mapping from the set of preference profiles, $\mathcal{P}^n$ to the set of feasible assignments, $\mathcal{A}$. That is, $\mathcal{M}:\mathcal{P}^n\mapsto \mathcal{A}$. We focus our attention on two widely studied  mechanisms for one-side matching: Random Serial Dictatorship (RSD)~\cite{abdulkadirouglu1998random} and Probabilistic Serial Rule (PS)~\cite{bogomolnaia2012probabilistic}.

RSD relies on the concept of priority orderings over agents. Such an ordering is an ordered list of agents where the first agent gets to select its most preferred object from the set of objects, the second agent then selects its most preferred object from the set of remaining objects and so on until no objects remain.\footnote{When $n < m$ and agents can receive more than one object, RSD requires a careful method for the picking sequence at each priority ordering to ensure strategyproofness. This picking sequence should be based on an arbitrary serial dictatorship quota mechanism, which directly affects the efficiency and envy of the assignments \cite{hosseini2015strategyproof,bouveret2014manipulating}.
	For simplicity, we use the variant of RSD based on a \emph{quasi-dictatorial} mechanism \cite{papai2000strategyproof} where the first agent selects its most preferred $(m - n + 1)$ objects, and the rest of the agents choose one object each.}
%
Given a preference profile $\succ\in\mathcal{P}^n$, RSD returns an assignment $RSD(\succ)\in\mathcal{A}$ which is a uniform distribution over all deterministic assignments induced from all possible priority orderings over the set of agents. RSD has been widely adopted for fair and strategyproof assignments for the school choice problem, course assignment, house allocation, and room assignment \cite{abdulkadiroglu2009strategy,sonmez2010course,abdulkadirouglu1998random,abdulkadirouglu1999house}

PS treats objects as a set of divisible goods of equal size and simulates a simultaneous eating algorithm. Each agent starts ``eating'' its most preferred object, all at the same rate. Once an object is gone (eaten away) then the agent starts eating its next preferred object among the remaining objects. This process terminates when all objects have been ``eaten''.
Given a preference profile $\succ\in\mathcal{P}^n$, $PS(\succ)\in \mathcal{A}$ is a random assignment where $A_{i,j}$ is the probability (fraction) that object $j$ is assigned to (or ``eaten by'') agent $i$.

\subsection{General Properties}

In this section we define key properties for matching mechanisms. 
To evaluate the quality of a random assignment, we use first-order stochastic dominance~\cite{hadar1969rules,bogomolnaia2001new}. Given a random assignment $A_{i}$, the probability that agent $i$ is assigned an object that is at least as good as object $\ell$ is defined as follows
\begin{equation}
w(\succ_{i}, \ell, A_{i}) = \sum_{j\in M: j \succeq_{i} \ell} A_{i,j}
\end{equation}

We say an agent always prefers assignment $A_{i}$ to $B_{i}$, if for each object $\ell$ the probability of assigning an object at least as good as $\ell$ under $A_{i}$ is greater or equal that of $B_{i}$, and strictly greater for some object.

\begin{definition}[Stochastic Dominance]
	Given a preference ordering $\succ_{i}$, random assignment $A_{i}$ stochastically dominates (sd) assignment $B_{i} (\neq A_{i})$ if
	\begin{equation}
	\forall \ell \in M,\ w(\succ_{i}, \ell, A_{i}) \geq w(\succ_{i}, \ell, B_{i})
	\end{equation}
\end{definition}


A matching mechanism is \emph{sd}-efficient if at all preference profiles $\succ \in \mathcal{P}^{n}$, for all agents $i\in N$, the prescribed assignment is not stochastically dominated by any other assignment.
\begin{definition} [\emph{sd}-Efficiency]
	A random assignment is {sd}-efficient if for all agents, it is not stochastically dominated by any other random assignment.
\end{definition}

%


An important desirable property in matching mechanisms is strategyproofness, that is the mechanism is designed so that no agent has incentive to misreport its preferences.

\begin{definition} [\emph{sd}-Strategyproofness]
	Mechanism $\mathcal{M}$ is sd-strategyproof if at all preference profiles $\succ \in \mathcal{P}^{n}$, for all agents $i \in N$, and for any misreport $\succ'_{i} \in \mathcal{P}^{n}$, such that $A = \mathcal{M}(\succ)$ and $A' = \mathcal{M}(\succ'_{i},\succ_{-i})$, we have:
	\begin{equation}\label{eq:strat-proof}
	\forall \ell \in M,\ w(\succ_{i}, \ell, A_{i}) \geq w(\succ_{i}, \ell, A'_{i})
	\end{equation}
\end{definition}

\emph{Sd}-strategyproofness is a strict requirement. It implies that under any utility model consistent with the preference orderings, no agent can improve her expected utility by misreporting.
%
We say that a mechanism is \emph{weakly sd-strategyproof} if at all preference profiles there is no misreport such that for all $\ell \in M$, $w(\succ_{i}, \ell, A'_{i}) \geq w(\succ_{i}, \ell, A_{i})$ with at least one $\ell' \in M$ such that $w(\succ_{i}, \ell', A'_{i}) > w(\succ_{i}, \ell', A_{i})$.
Clearly, \emph{sd}-strategyproofness implies weak \emph{sd}-strategyproofness but the converse does not hold.



An assignment is \emph{manipulable} if it is not \emph{sd}-strategyproof. If there exists some agent who strictly benefits from the manipulation, (\ie the mechanism is not even weakly \emph{sd}-strategyproof) then we say the assignment is \emph{sd}-manipulable (or strictly manipulable).


%


%



We are also interested in whether mechanisms are fair and use the notion of envyfreeness to this end.
An assignment is  \emph{sd}-envyfree if each agent strictly prefers her random allocation to any other agent's assignment.


\begin{definition} [\emph{sd}-Envyfreeness]
	Given agent $i$'s preference $\succ_{i}$, assignment $A_{i}$ is sd-envyfree if for all agents $\forall k\neq i \in N$,
	\begin{equation}\label{eq:envyfree}
	\forall \ell \in M,\ w(\succ_{i}, \ell, A_{i}) \geq w(\succ_{i}, \ell, A_{k})
	\end{equation}
\end{definition}
We say an assignment is weakly \emph{sd}-envyfree if the inequality in Equation~\ref{eq:envyfree} is strict for some $\ell \in M$, but there exists at least one $\ell'$ for which the inequality in Equation~\ref{eq:envyfree} does not hold.
A matching mechanism satisfies \emph{sd}-envyfreeness if at all preference profiles $\succ \in \mathcal{P}^{n}$, it induces \emph{sd}-envyfree assignments for all agents.

Lastly, we are interested in investigating efficiency, manipulation, and envy of the random mechanisms when preferences are lexicographic. Under lexicographic preferences, given two allocations, an agent prefers the one in which there is a higher probability for getting the most-preferred object.


\begin{definition}[Lexicographic Dominance]
	Given a preference ordering $\succ_{i}$, random assignment $A_{i}$ lexicographically dominates (\emph{ld}) assignment $B_{i}$ if 
	\begin{align}
	&\exists\ \ell \in M: w(\succ_{i}, \ell, A_{i}) > w(\succ_{i}, \ell, B_{i})\ \wedge\  \\ 
	&\forall\ k \succ_{i} \ell:  w(\succ_{i}, \ell, A_{i}) = w(\succ_{i}, \ell, B_{i}). \nonumber
	\end{align} 
\end{definition}

We say that allocation $A$ \emph{lexicographically dominates} another allocation $B$ if there exists no agent $i\in N$ that lexicographically prefers $B_{i}$ to $A_{i}$.
Thus, an allocation mechanism is lexicographically efficient (\emph{ld}-efficient) if for all preference profiles its induced allocation is not lexicographically dominated by any other random allocation. 

\subsection{Properties of RSD and PS}
\begin{table}
	\centering
	\caption{\label{my-label}Properties of PS and RSD.}
	\begin{tabular}{@{}lllll@{}}
		\toprule\toprule
		& \multicolumn{2}{c}{$n \geq m$} & \multicolumn{2}{c}{$n < m$} \\ \cmidrule(l){2-3} \cmidrule(l){4-5}
		&      PS     &    RSD &     PS      &    RSD      \\ \midrule 
		\emph{sd}-strategyproof &      weak     &    \ding{51} &     \ding{55}      &    \ding{51}  \\
		\emph{sd}-efficiency    &     \ding{51}      & \ding{55}      &     \ding{51}      &    \ding{55}      \\
		\emph{sd}-envyfree        &     \ding{51}      & weak      &    \ding{51}       &   weak      \\ \bottomrule\bottomrule
	\end{tabular}
\end{table}

The theoretical properties of PS and RSD have been well studied in the economics literature~\cite{bogomolnaia2001new}, and we summarize the results in Table~\ref{my-label}. Both mechanisms are ex post efficient, that is, their \textit{realized outcomes} cannot be improved without making at least one agent worse off. PS has been shown to be both \emph{sd}-envyfree and \emph{sd}-efficient. However, it is not even weakly \emph{sd}-strategyproof when $n < m$ \cite{kojima2010incentives} and is only weakly \emph{sd}-strategyproof when $n \geq m$. On the other hand, RSD is always \emph{sd}-strategyproof, but it is only weakly \emph{sd}-envyfree and is not \emph{sd-efficient}.
Example~\ref{example:rsd-noSD} illustrates the \emph{sd}-inefficiency of RSD.

\begin{example}\label{example:rsd-noSD}
	Suppose there are four agents $N = \{1,2,3,4\}$ and four objects $M = \{a,b,c,d\}$. Consider the following preference profile $\succ = ((abcd), (abcd), (badc), (badc))$. Table~\ref{tab:example-inefficient} shows the outcomes for $PS(\succ)$ and $RSD(\succ)$.
	In this example, all agents strictly prefer the assignment induced by PS over the RSD assignment. Thus, RSD is inefficient at this preference profile.
	\begin{table}
		\caption{\label{tab:example-inefficient}Example showing the inefficiency of RSD}
		\begin{subtable}{0.49\linewidth}
			\centering
			\caption{Assignment under $PS(\succ)$}
			\begin{tabular}{ccccc}
				\hline
				& $a$ & $b$ & $c$ & $d$\\ \hline \hline
				$A_1$ & $1/2$ & $0$ & $1/2$ & $0$ \\
				$A_2$ & $1/2$ & $0$ & $1/2$ & $0$ \\
				$A_3$ & $0$ & $1/2$ & $0$ & $1/2$ \\
				$A_4$ & $0$ & $1/2$ & $0$ & $1/2$ \\
				\hline
			\end{tabular}
		\end{subtable}
		\begin{subtable}{0.49\linewidth}
			\centering
			\caption{Assignment under $RSD(\succ)$}
			\begin{tabular}{ccccc}
				\hline
				& $a$ & $b$ & $c$ & $d$\\ \hline \hline
				$A_1$ & $5/12$ & $1/12$  & $5/12$ & $1/12$  \\
				$A_2$ & $5/12$ & $1/12$  & $5/12$ & $1/12$  \\
				$A_3$ & $1/12$ & $5/12$ & $1/12$  & $5/12$ \\
				$A_4$ & $1/12$ & $5/12$ & $1/12$  & $5/12$ \\
				\hline
			\end{tabular}
			
		\end{subtable}
	\end{table}
	
\end{example}

\section{Incomparability of RSD and PS}

We argue that the theoretical findings on RSD and PS do not necessarily provide enough guidance to  a market designer trying to select the correct mechanism for a specific setting. For example, while we know that PS is \emph{sd}-efficient and RSD is not, this does not mean that PS assignment always stochastically dominate the assignments prescribed by RSD.

\begin{table}
	\caption{\label{tab:incomparability}Incomparability of RSD and PS}
	\begin{subtable}{0.49\linewidth}
		\centering
		\caption{Assignment under $PS(\succ)$}
		\begin{tabular}{ c c c c }
			\hline
			& $a$ & $b$ & $c$ \\ \hline \hline
			$A_{1}$ & $1/2$ & $0$ & $1/2$ \\
			$A_2$ & $1/2$ & $1/4$ & $1/4$ \\
			$A_3$ & $0$ & $3/4$ & $1/4$  \\
			\hline
		\end{tabular}
	\end{subtable}
	\begin{subtable}{0.49\linewidth}
		\centering
		\caption{Assignment under $RSD(\succ)$}
		\begin{tabular}{ c c c c }
			\hline
			& $a$ & $b$ & $c$ \\ \hline \hline
			$A_1$ & $1/2$ & $0$  & $1/2$ \\
			$A_2$ & $1/2$ & $1/6$  & $1/3$ \\
			$A_3$ & $0$ & $5/6$ & $1/6$ \\
			\hline
		\end{tabular}
		
	\end{subtable}
	
\end{table}

\begin{example}\label{example:non-comparable}
	Suppose there are three agents $N = \{1,2,3\}$ and three objects $M = \{a,b,c\}$. Consider the following preference profile $\succ = ((acb), (abc), (bac))$. Table~\ref{tab:incomparability} shows $PS(\succ)$ and $RSD(\succ)$. Neither assignment  dominates the other since agent 1 is ambivalent between the two assignments while agent 2 prefers $PS(\succ)$ and agent 3 prefers $RSD(\succ)$.
	
\end{example}

If we knew the utility functions of the agents, consistent with their ordinal preferences, then we might be able to use the notion of (utilitarian) social welfare to help determine the better assignment.\footnote{Given utility functions for the agents, where $u_i(j)$ is the utility agent $i$ derives from being assigned object $j$, the (utilitarian) social welfare of an assignment $A$ is $\sum_i \sum_j A_{i,j}u_i(j)$.} However, it is easy to construct different utility functions for the agents in Example~\ref{example:non-comparable} where both $RSD$ and $PS$ maximize social welfare.
Similarly, the envy of RSD and the manipulability of PS both depend on the structure of preference profiles, and thus, a compelling question, that justifies studying the practical implications of deploying a matching mechanism, is to analyze the space of preference profiles to find the likelihood of inefficient, manipulable, or envious assignments under these mechanisms.
In Example~\ref{example:non-comparable}, for instance, if utilities of agents 1 and 2 are 10, 9, and 1, and agent 3's utility is 10, 6, and 4 for the first, second, and third objects respectively, then PS assignment outperforms that of RSD with respect to social welfare because 
$
(\frac{1}{2} \cdot 10 + \frac{1}{2} \cdot 9 + (0) \cdot 1) + (\frac{1}{2} \cdot 10 + \frac{1}{4} \cdot 9 + \frac{1}{4} \cdot 1) +  (\frac{3}{4} \cdot 10 + (0) \cdot 6 + \frac{1}{4} \cdot 4) > 
(\frac{1}{2} \cdot 10 + \frac{1}{2} \cdot 9 + (0) \cdot 1) + (\frac{1}{2} \cdot 10 + \frac{1}{6} \cdot 9 + \frac{1}{3} \cdot 1) +  (\frac{5}{6} \cdot 10 + (0) \cdot 6 + \frac{1}{6} \cdot 4)
$.
However, if utility functions change such that all agents have the same utilities of 10, 9, and 1 for the first, second, and third objects respectively, then the social welfare under RSD outperforms that of PS because 
$
(\frac{1}{2} \cdot 10 + \frac{1}{2} \cdot 9 + (0) \cdot 1) + (\frac{1}{2} \cdot 10 + \frac{1}{6} \cdot 9 + \frac{1}{3} \cdot 1) +  (\frac{5}{6} \cdot 10 + (0) \cdot 9 + \frac{1}{6} \cdot 1) >
(\frac{1}{2} \cdot 10 + \frac{1}{2} \cdot 9 + (0) \cdot 1) + (\frac{1}{2} \cdot 10 + \frac{1}{4} \cdot 9 + \frac{1}{4} \cdot 1) +  (\frac{3}{4} \cdot 10 + (0) \cdot 9 + \frac{1}{4} \cdot 1)
$.

\section{General and Lexicographic Preferences}\label{sec:gen}

The theoretical properties of PS and RSD only provide limited insight into their practical applications. In particular, when deciding which mechanism to use in different settings, the incomparability of PS and RSD leaves us with an ambiguous choice in terms of efficiency, manipulability, and envyfreeness. Thus, we examine the properties of RSD and PS in the space of all possible preference profiles as well as under lexicographic preferences.
Lexicographic preferences are present in various applications and have been extensively studied in artificial intelligence and multiagent systems as a means of assessing allocations based on ordinal preferences~\cite{domshlak2011preferences,saban2013note,fishburn1974lexicographic}.
Under lexicographic preferences, an allocation that assigns a higher probability to the top ranked object is always preferred to any other allocation, regardless of the probabilities assigned to objects in the next positions. Only when two allocations assign equal probabilities to the top ranked object, the probability of the next preferred object is considered.
In the rest of this paper, we denote the efficiency, strategyproofness, manipulability, and envyfreeness with \emph{ld}- (lexicographically dominate) prefix.

\begin{figure*}[t]
	\centering
	\begin{subfigure}[b]{0.49\textwidth}
		\includegraphics[width=\textwidth]{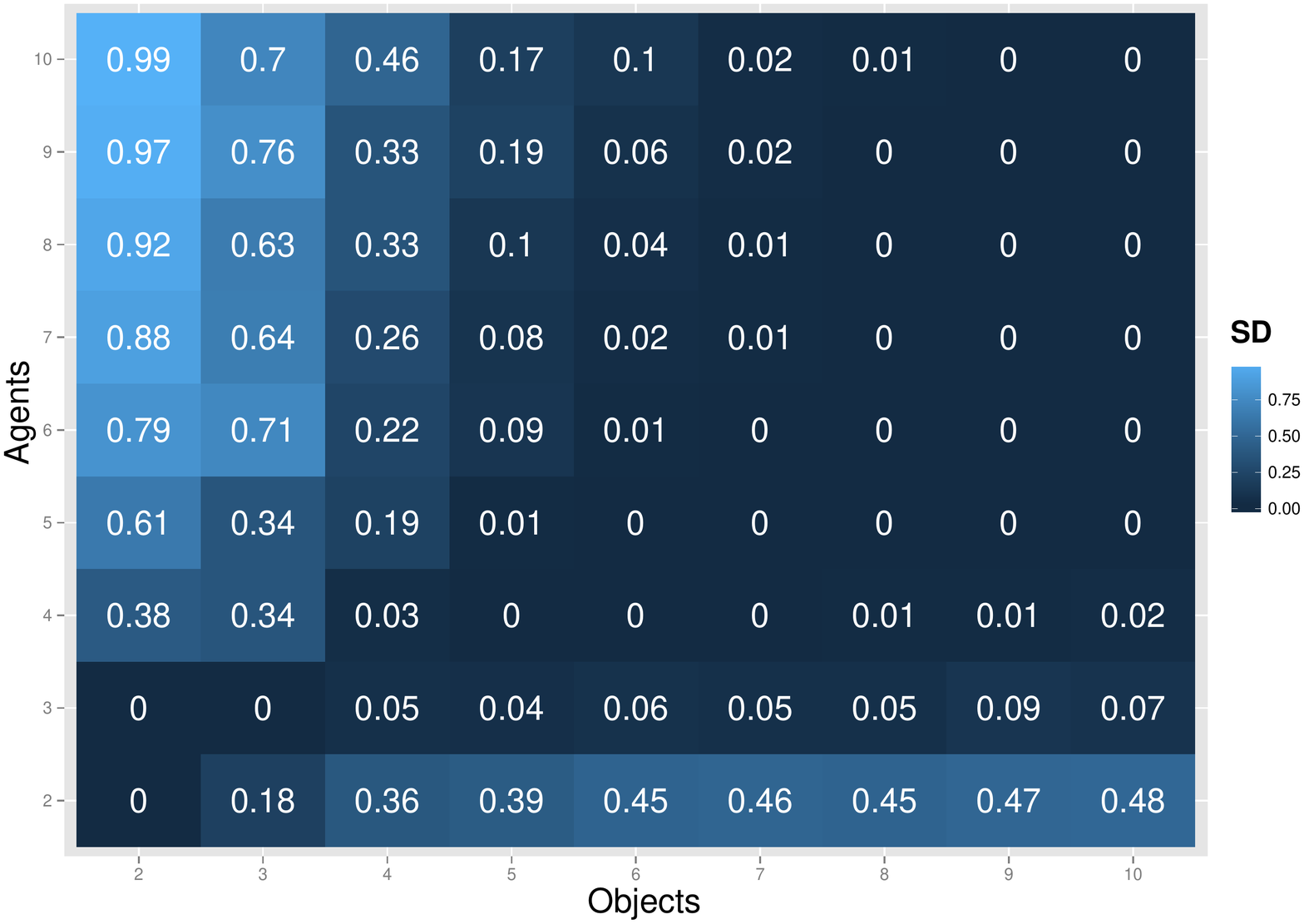}
		\caption{The fraction that PS stochastically dominates RSD.}
		\label{fig:sd-efficiency}
	\end{subfigure}
	\begin{subfigure}[b]{0.49\textwidth}
		\includegraphics[width=\textwidth]{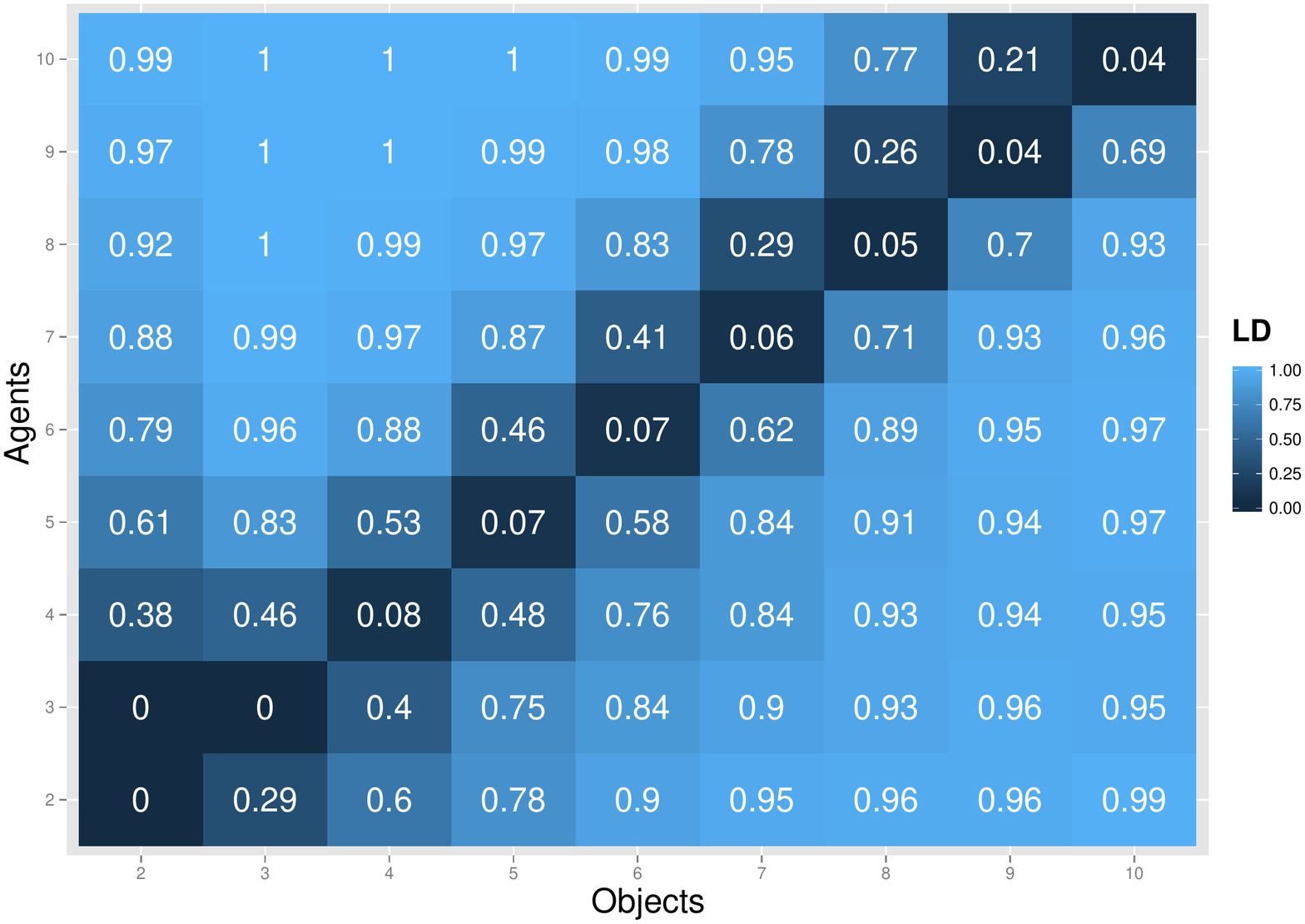}
		\caption{The fraction that PS lexicographically dominates RSD.}
		\label{fig:ld-efficiency}
	\end{subfigure}
	\caption{The fraction of preference profiles under which PS dominates RSD.
	}
	\label{Fig:efficiency}
\end{figure*}


The number of all possible preference profiles is super exponential $(m!)^{n}$. For each combination of $n$ agents and $m$ objects we performed a brute force coverage of all possible preference profiles. Thus, for all subsequent figures each data point shows the fraction of all possible preference profiles. 
For the cases of $n = 10$ and $m \in \{9, 10\}$, we randomly generated 1,000 instances by sampling from a uniform preference profile distribution. For each preference profile, we ran both PS and RSD mechanisms and compared their outcomes in terms of the stochastic dominance relation. Appendix \ref{app:numerial} illustrates our numerical results.
Note that not only is computing RSD probabilities \#P-complete (and thus intractable) \cite{aziz2013computational,saban2015complexity}, but checking the desired properties such as envyfreeness, efficiency, and manipulablity of random allocations is shown to be NP-hard for general settings \cite{aziz2015fair,aziz2015manipulating}. Thus, for larger settings even if we randomly sample preference profiles it is not easy to verify the aforementioned properties. 

\subsection{Preliminary Results}

Our experimentation disclose several intriguing observations, confirming theoretical results and providing additional insights into matching markets.
A preliminary look at our empirical results illustrates the following: when $m \leq 2, n \leq 3$, PS coincides exactly with RSD, which results in the best of the two mechanisms, \ie, both mechanisms are \emph{sd}-efficient, \emph{sd}-strategyproof, and \emph{sd}-envyfree. 
Another interesting observation is that when $m = 2$, for all $n$, PS is \emph{sd}-strategyproof (although the PS assignments are not necessarily equivalent to assignments induced by RSD), RSD is \emph{sd}-envyfree, and for most instances when $m = 2$, PS stochastically dominates RSD, particularly when $n \geq 4$.


%
%

\subsection{Efficiency}

Our first finding is that the fraction of preference profiles at which RSD and PS prescribe identical random assignments goes to $0$ when $n$ grows. 
There are two conclusions that one can draw. First, this result confirms the theoretical results of Manea on asymptotic inefficiency of RSD~\cite{manea2009asymptotic}, in that, in most instances, the assignments induced by RSD are not identical to the PS assignments.
Second, this result suggests that the incomparability of outcomes is significant, that is, the social welfare of the random outcomes is highly dependent on the underlying utility models. 




The fraction of preference profiles $\succ\in \mathcal{P}^{n}$ for which RSD is stochastically dominated by PS at $\succ$ converges to zero as $\frac{n}{m} \to 1$. 
Figure~\ref{fig:sd-efficiency} shows that when $m$ grows beyond $m > 5$, due to incomparability of RSD and PS with regard to the stochastic dominance relation, the RSD assignments are rarely stochastically dominated by \emph{sd}-efficient assignments prescribed by PS.

We also see similar results when we restrict ourselves to lexicographic preferences (Figure~\ref{fig:ld-efficiency}). The fraction of preference profiles $\succ\in \mathcal{P}^{n}$ for which RSD is lexicographically dominated by PS at $\succ$ converges to zero as $\frac{n}{m} \to 1$.

For lexicographic preferences, we also observe that the fraction of preference profiles for which PS assignments strictly dominate RSD-induced allocations goes to 1 when the number of agents and objects diverge. The fraction of preference profiles $\succ\in \mathcal{P}^{n}$ for which RSD is lexicographically dominated by PS at $\succ$ converges to 1 as $|n - m|$ grows. Intuitively, when some agents can receive more than one object ($n < m$) or when there are not sufficient objects ($n > m$) for all agents, in each realized ordering of agents by RSD, those with higher priority are treated very differently than those in lower priority. Thus, the RSD outcome tend to be unfair and undesirable for most agents.

One immediate conclusion is that although RSD does not guarantee either \emph{sd}-efficiency or \emph{ld}-efficiency, in most settings when $\frac{n}{m}\to 1$ (and also $n \leq m$ for \emph{sd}-efficiency according to Figure \ref{fig:sd-efficiency}), neither of the two mechanisms is preferred in terms of efficiency. Hence, one cannot simply rule out the RSD mechanism. 

\subsection{Manipulability of PS}

\begin{figure*}
	\centering
	\begin{subfigure}[b]{0.499\textwidth}
		\centering
		\includegraphics[width=\textwidth]{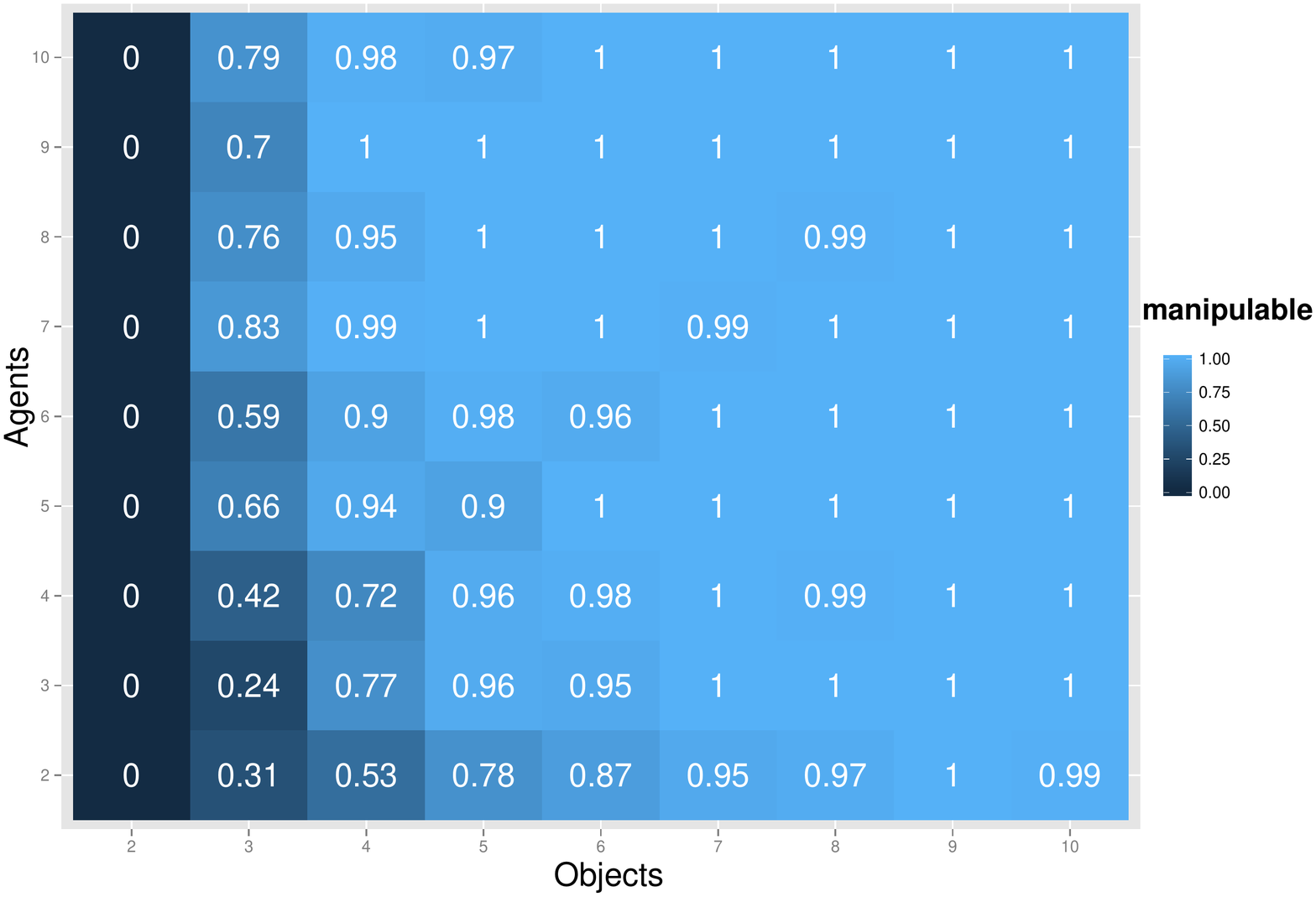}
		\caption{The fraction of manipulable preference profiles under PS.}
		\label{fig:manip}
	\end{subfigure}~
	\begin{subfigure}[b]{0.499\textwidth}
		\includegraphics[width=\textwidth]{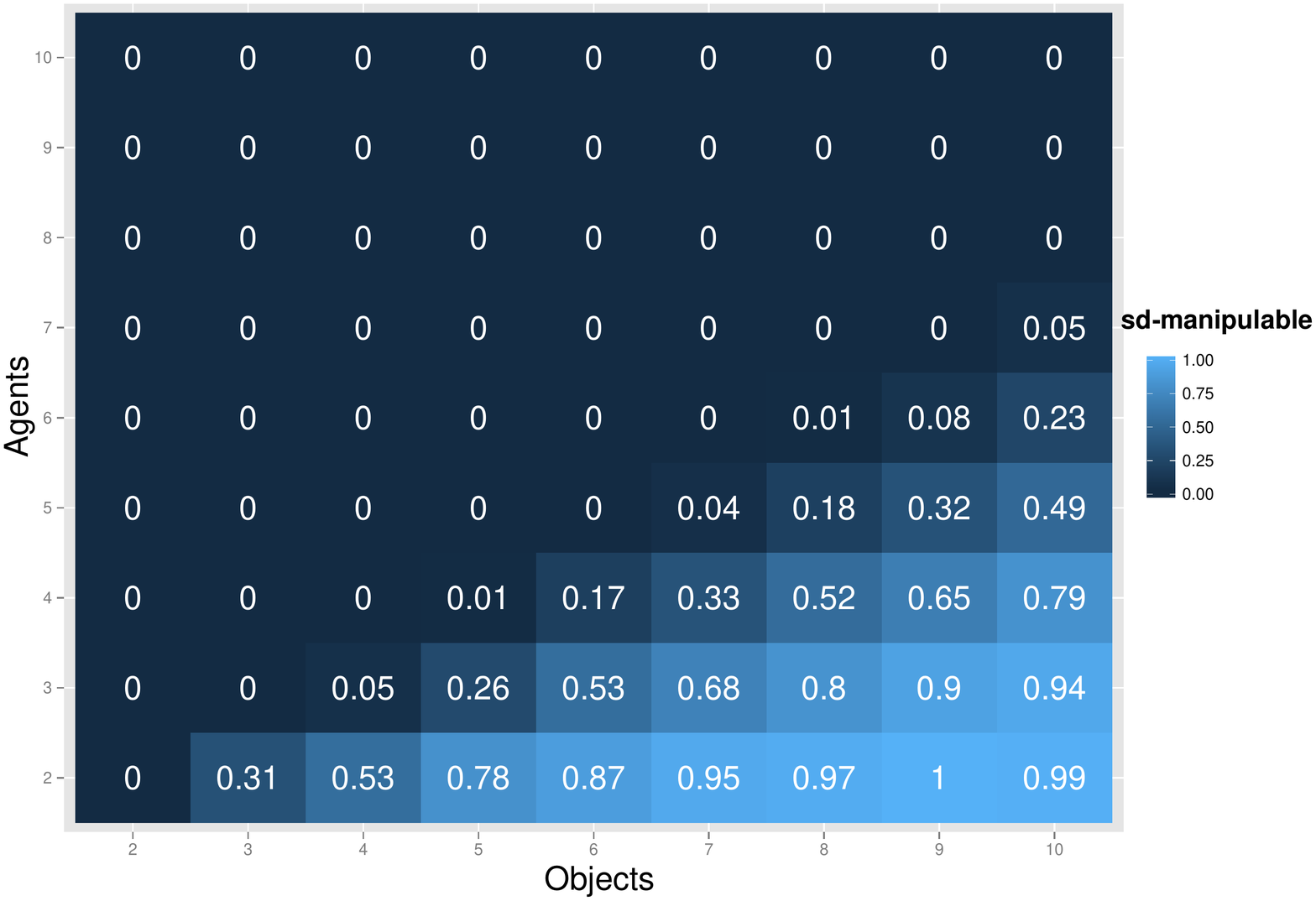}
		\caption{The fraction of \emph{sd}-manipulable profiles under PS.}
		\label{fig:sd-manipulable}
	\end{subfigure}
	\caption{Heatmaps illustrating the manipulablity of PS.} 
	\label{Fig:manipulability}
\end{figure*}

One critical issue with deploying PS is that it does not provide incentives for honest reporting of preferences.
Although for $n\geq m$ PS is weakly \emph{sd}-strategyproof \cite{bogomolnaia2001new} and \emph{ld}-strategyproof \cite{schulman2012allocation}, when $n < m$ PS no longer satisfies these two properties.\footnote{A recent experimental study on the incentive properties of PS shows that human subjects are less likely to manipulate the mechanism when misreporting is a Nash equilibrium. However, subjects' tendency for misreporting is still significant even when it does not improve their allocations~\cite{hugh2013experimental}. 
}
The real concern is that, in the absence of strategyproofness, PS allocations are only efficient (or envyfree) with respect to the reported preferences. Thus, if an agent decides to manipulate the outcome by misreporting its preferences, PS will no longer guarantee efficiency, nor envyfreeness with respect to the true underlying preferences.
Thus, we are interested in understanding the degree to which PS allocations are manipulable.

%


Figure~\ref{Fig:manipulability} shows that the fraction of manipulable profiles goes to 1 as $n$ or $m$ grow. PS is almost 99\% manipulable for $n > 5, m > 5$. 
Another interesting observation is that, for all $n < m$, the fraction of \emph{sd}-manipulable preference profiles goes to 1 as $m - n$ grows (Figure~\ref{fig:sd-manipulable}). 
These results imply that when agents are permitted to receive more than a single object, agents can strictly benefit from misreporting their preferences. 

Moreover, at those instances of problem where PS is \emph{sd}-strategyproof, the assignment prescribed by PS most often coincides with the RSD induced assignment. 
For example, when $n = m = 5$, PS is only \emph{sd}-strategyproof at $11\%$ of preference profiles, $7\%$ of which are identical to the assignments induced by RSD. This insight further confirms the vulnerability of PS to misreporting (See Table \ref{tab:numerialOrdinalResults} for detailed numerical results).

\begin{figure}[h!]
	\centering
	\includegraphics[width=.7\linewidth]{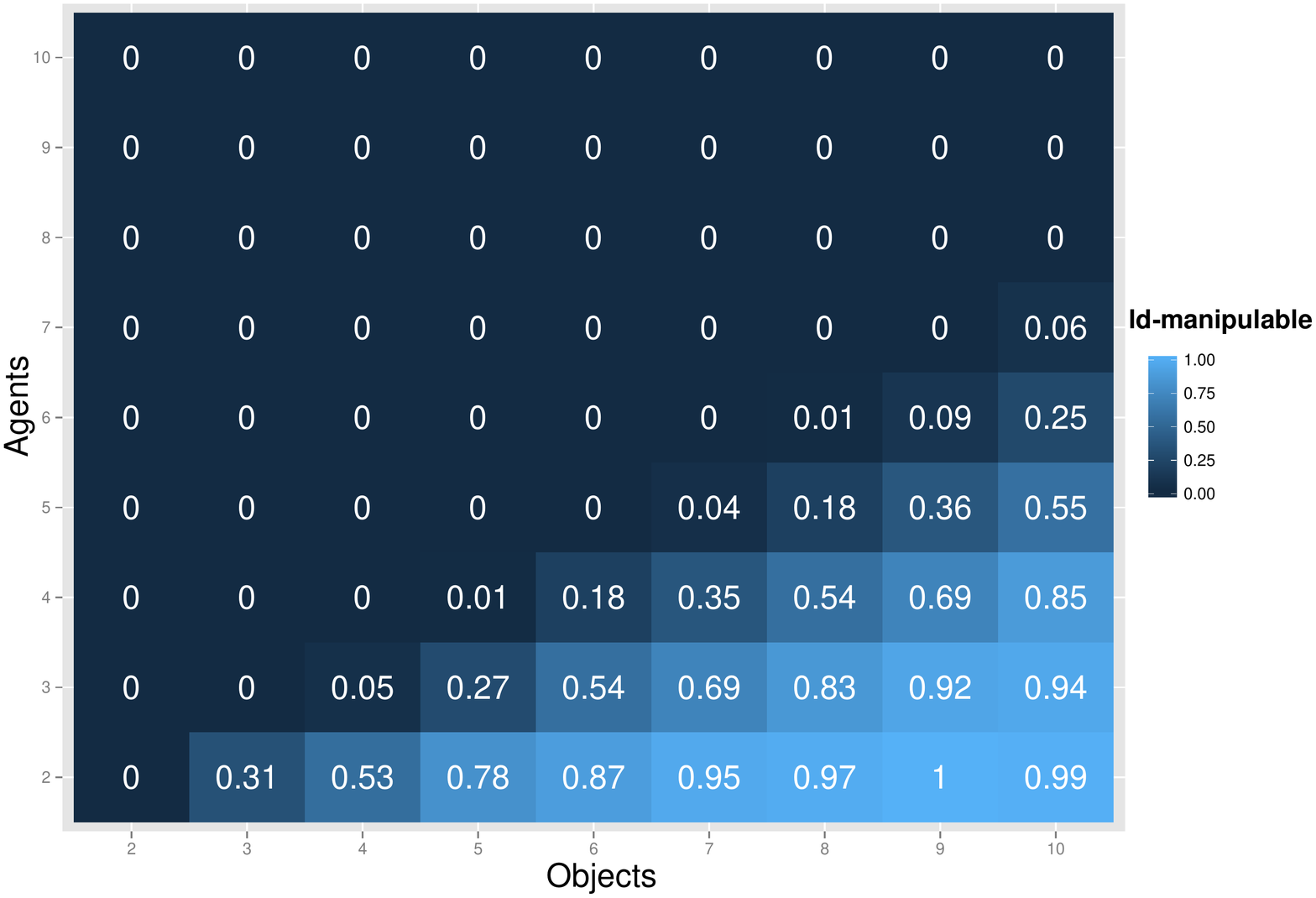}
	\caption{The fraction of \emph{ld}-manipulable profiles under PS.}
	\label{fig:ld-manipulable}
\end{figure}

As illustrated in Figure \ref{fig:ld-manipulable}, the manipulability of PS under lexicographic preferences has a similar trend when there are more objects than agents ($n < m$) and the fraction of \emph{ld}-manipulable preference profiles converges to 1 even more rapidly when $\frac{m}{n}$ grows.


\subsection{Envy in RSD}


The PS mechanism has a desirable fairness property and is guaranteed to satisfy \emph{sd}-envyfreeness, whereas RSD is not \emph{sd}-envyfree. To further investigate the envy among agents under RSD, we measured the fraction of agents that are weakly \emph{sd}-envious of at least one other agent. 



\begin{figure*}[t]
	\centering
	\begin{subfigure}[b]{0.499\textwidth}
		\includegraphics[width=\textwidth]{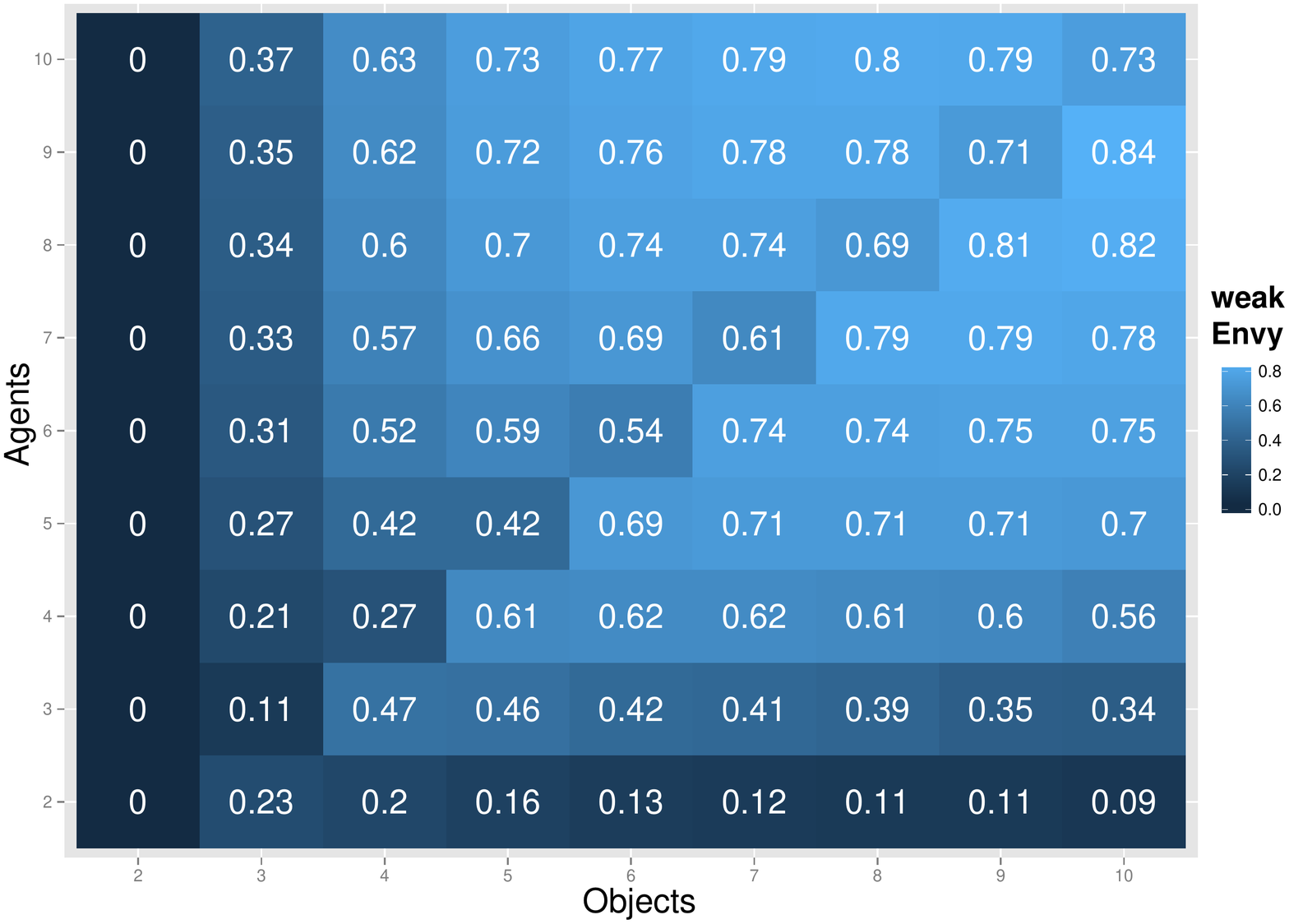}
		\caption{A heatmap showing the percentage of envious agents.}
		\label{Fig:a}
	\end{subfigure}~
	\begin{subfigure}[b]{0.499\textwidth}
		\centering
		\includegraphics[width=\linewidth]{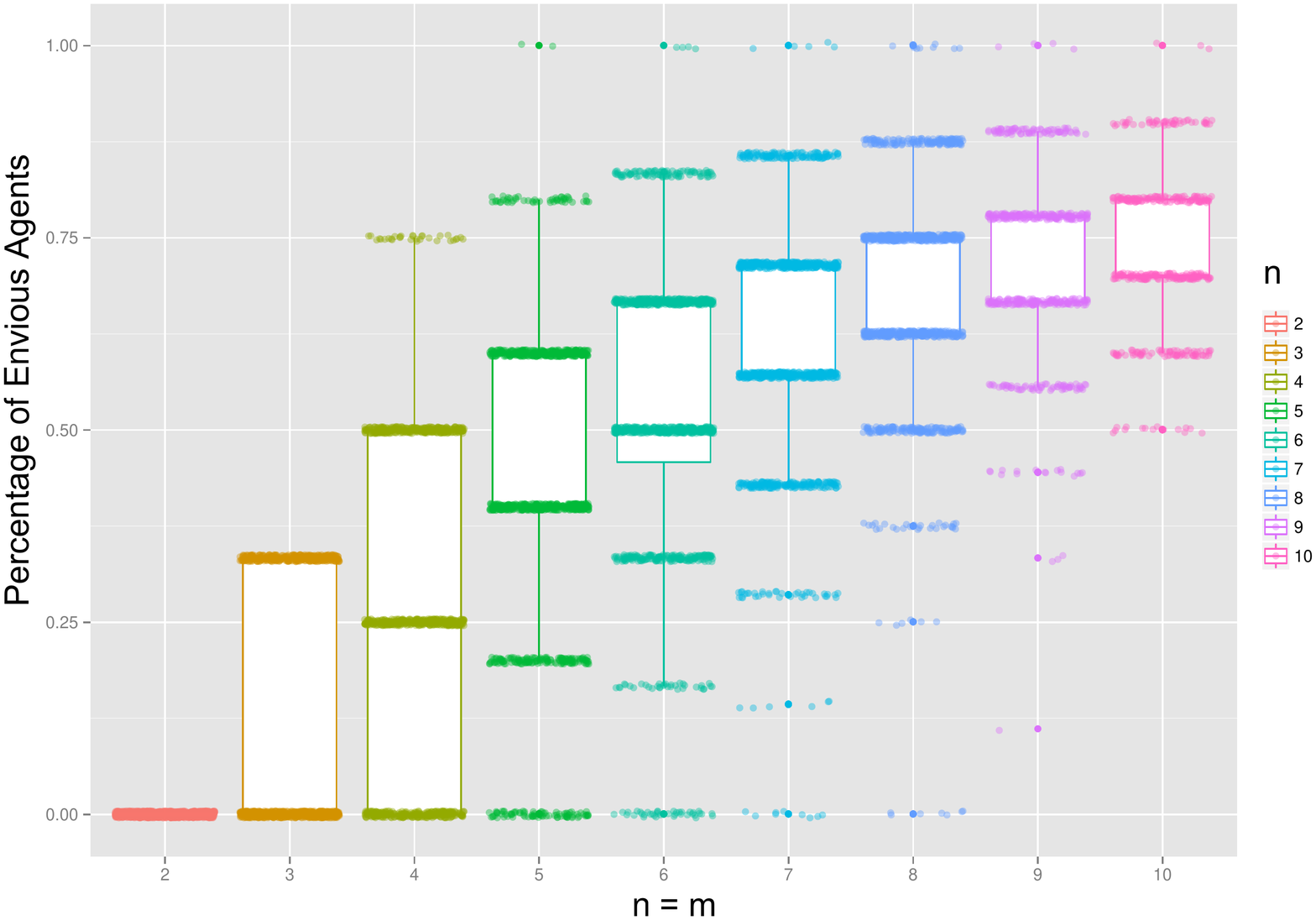}
		\caption{Boxplots showing the various envy profiles for $n = m$. The Y axis represents the percentage of envious agents.}
		\label{Fig:EnvyEqual}
	\end{subfigure}
	\caption{Plots representing the percentage of (weakly) envious agents under RSD.}
	\label{Fig:RSD-Envy}
\end{figure*}

Figure \ref{Fig:RSD-Envy} shows that for RSD, the percentage of agents that are weakly envious increases with the number of agents. 
Figure~\ref{Fig:a} reveals an interesting observation: fixing any $n>3$, the percentage of agents that are (weakly) envious grows with the number of objects, however, there is a sudden drop in the percentage of envious agents when there are equal number of agents and objects. 

For better understanding of the population of agents who feel (weakly) envious under RSD, we illustrate the various envy profiles based on the percentage of envious agents in all instances of the problem when $n = m$ (Figure \ref{Fig:EnvyEqual}). One observation is that there are few distinct envy profiles at each $n$, each representing a particular class of preference profiles, and by increasing $n$, the fraction of agents that are envious of at least one other agent increases.


\section{Utility Models}\label{sec:risk}

Given a utility model consistent with an agent's preference ordering, we can find the agent's expected utility for a random assignment.
Let $u_{i}$ denote agent $i$'s Von Neumann-Morgenstern (VNM) utility model consistent with its preference ordering $\succ_{i}$. That is, $u_{i}(a) > u_{i}(b)$ if and only if $a \succ_{i} b$.
Then, agent $i$'s expected utility for random assignment $A_{i}$ is $\mathbb{E}(u_{i} | A_{i}) = \sum_{j\in M} A_{i,j} u_{i}(j)$.

We say that agent $i$ (strictly) prefers assignment $A_{i}$ to $B_{i}$ if and only if $\mathbb{E}(u_{i} | A_{i}) > \mathbb{E}(u_{i} | B_{i})$.
A mechanism is strategyproof if there exists no agent that can improve its expected utility by misreporting its preference ordering.

\begin{definition} [Strategyproof]
	Mechanism $\mathcal{M}$ is strategyproof if for all agents $i \in N$, and for any misreport $\succ'_{i} \in \mathcal{P}^{n}$, such that $A = \mathcal{M}(\succ)$ and $A' = \mathcal{M}(\succ'_{i},\succ_{-i})$, given a utility model $u_{i}$ consistent with $\succ_{i}$, we have $\mathbb{E}(u_{i} | A_{i}) \geq \mathbb{E}(u_{i} | A'_{i})
	$.
\end{definition}


A matching mechanism is envyfree if for all preference profiles it prescribes an envyfree assignment.

\begin{definition} [Envyfreeness]
	Assignment $A$ is envyfree if for all $i,k \in N$, given utility model $u_{i}$ consistent with $\succ_{i}$, we have $\mathbb{E}(u_{i} | A_{i}) \geq \mathbb{E}(u_{i} | A_{k})$.
\end{definition}

Given utility functions for the agents, the (utilitarian) \emph{social welfare} of an assignment $A$ is $\sum_i \mathbb{E}(u_{i} | A_{i})$.
A random assignment $A$ is \emph{sd}-efficient if and only if there exists a profile of utility values consistent with $\succ$ such that $A$ maximizes the social welfare ex ante \cite{bogomolnaia2001new,mclennan2002ordinal}. This existence result does not shed light on the social welfare when comparing two random assignments, since an assignment can be \emph{sd}-efficient but may not have desirable ex ante social welfare. 
Consider the following random assignments: assignment $A$ which is \emph{sd}-efficient and assignment $B \neq A$ which is not stochastically dominated by $A$. Given a preference profile, $A$ is guaranteed to maximize the social welfare for at least one profile of consistent utilities. However, there may be other profiles of utilities consistent with preferences at which $B$ maximizes the sum of utilities (social welfare). 

\begin{example}
	Consider the problem introduced in Example \ref{example:non-comparable} with assignments illustrated in Table \ref{tab:incomparability}. Let's assume that all agents have the same utility model $u_{1} = u_{2} = u_{3}$ where the utilities are $10, 9, 0$ for the first, second, and third objects respectively.
	The sum of expected utilities under the PS assignment is
	$
	(\frac{1}{2} \cdot 10 + \frac{1}{2} \cdot 9 + 0) +
	(\frac{1}{2} \cdot 10 + \frac{1}{4} \cdot 9 + \frac{1}{4} \cdot 0) + 
	(\frac{3}{4} \cdot 10 + 0 \cdot 9 + \frac{1}{4} \cdot 0)
	$,
	while the sum of expected utilities under the RSD allocation is
	$
	(\frac{1}{2} \cdot 10 + \frac{1}{2} \cdot 9 + 0) +
	(\frac{1}{2} \cdot 10 + \frac{1}{6} \cdot 9 + \frac{1}{3} \cdot 0) +
	(\frac{5}{6} \cdot 10 + 0 \cdot 9 + \frac{1}{6} \cdot 0)
	$. It is easy to see that for this profile, the ex ante social welfare under RSD is larger than that of PS. 
\end{example}

Thus, given a profile of utilities we investigate the (ex ante) social welfare of the assignments under PS and RSD.

\subsection{Instantiating Utility Functions}

To deepen our understanding as to the performance of the two mechanisms, we investigate different utility models. In particular we look at the performance of the mechanisms when the agents are all risk neutral (\ie have linear utility functions), when agents are risk seeking and when agents are risk averse.

Our first utility model is the well-studied linear utility model. Given an agent $i$'s preference ordering $\succ_i$, we let $r(\succ_i,j)$ denote the rank of object $j$. For example, given preference ordering $a\succ_i b\succ_i c$ then $r(\succ_i, a)=1$, $r(\succ_i,b)=2$ and $r(\succ_i,c)=3$. The utility function for agent $i$, given object $j$ is $u_i(j)=m-r(\succ_i,j)$.

\begin{figure}
	\centering
	\includegraphics[width=.85\linewidth]{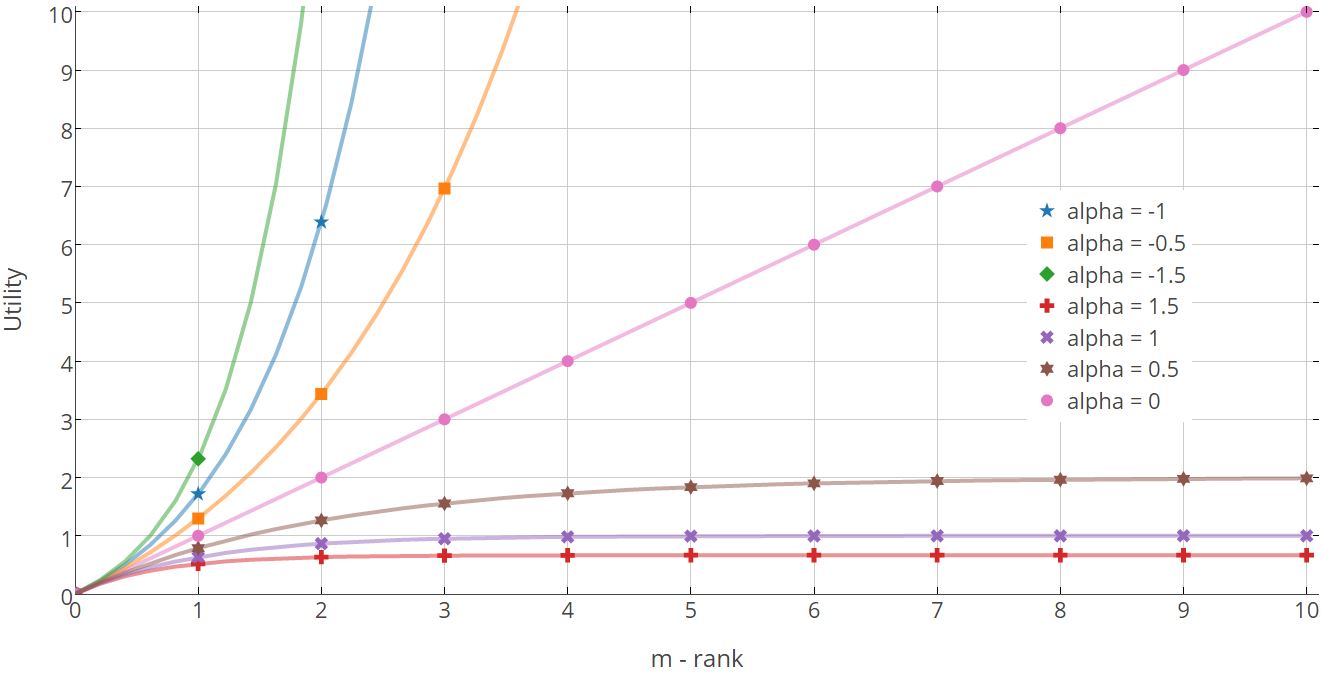}
	\caption{Utility values for various $\alpha$ under risk taking, risk neutral, and risk averse models. There are eleven objects ranked from 1 to 11, with linear utilities from 0 (the last object) to 10 (the top choice). The trendlines fit exponential trends to the discrete alpha parameters.}
	\label{fig:curvature}
\end{figure}

We use an \emph{exponential} utility model to capture risk attitudes beyond risk-neutrality. An exponential utility has been shown to provide an appropriate translation for individuals' utility models \cite{arrow1974essays}. 
In particular, we define the exponential utility as follows:
\begin{gather}
u_i(j) = 
\begin{cases}
(1-e^{-\alpha(m-r(\succ_i,j))})/\alpha, & \alpha \neq 0\\
m-r(\succ_i,j), & \alpha = 0
\end{cases}
\end{gather}
The parameter $\alpha$ represents the agent's risk attitude. If $\alpha>0$ then the agent is risk averse, while if $\alpha<0$ then the agent is risk seeking. When $\alpha = 0$ then the agent is risk neutral and we have a linear utility model. The value $|\alpha|$ represents the intensity of the attitude. That is, given two agents with $\alpha_1 > \alpha_2 > 0$, we say that agent 1 is more risk averse than agent 2. Similarly if $\alpha_1 < \alpha_2 < 0$ then agent 1 is more risk seeking than agent 2.
Figure \ref{fig:curvature} illustrates the risk curvature for various risk taking and risk averse $\alpha$ parameters.

Table \ref{tab:sample utility} shows sample utility values for various risk taking, neutral, and risk averse utility profiles. These values show how a utility for objects in various ranking positions will change according to risk attitude models. Note that we do normalize the utilities such that all utilities add up to 1.

\begin{table}[]
	\centering
	\caption{Sample utility values when there are 3 objects under different risk attitudes and risk intensities.}
	\begin{tabular}{@{}c|lllll@{}}
		\toprule
		rank /\ $\alpha$ & $\alpha = -2$       & $\alpha=-1$       & $\alpha = 0$ & $\alpha = 1$        & $\alpha = 2$        \\ \hline
		1 & $26.799$ & $6.389$ & $2$ & $0.865$ & $0.491$ \\
		2 & $3.195$  & $1.718$ & $1$ & $0.632$ & $0.432$ \\
		3 & $0$      & $0$     & $0$ & $0    $ & $0    $ \\ \bottomrule
	\end{tabular}
	
	\label{tab:sample utility}
\end{table}

\section{Results}

For our experiments, we vary three parameters: the number of agents $n$, the number of objects $m$, and the risk attitude factor $\alpha$. Each data point in the graphs shows the average over all possible preference profiles.
We study the same settings as in Section \ref{sec:gen} when $n \geq m$ and $n < m$. For each utility function, we look at homogeneous populations of agents where agents have the same risk attitudes but may have difference ordinal preferences.



To compare the social welfare, we investigate the percentage change (or improvement) in social welfare of PS compared to RSD under various utility models. That is, 

\begin{equation*}
\frac{\sum_{i} \mathbb{E}(u_{i} | PS(\succ)) - \sum_{i} \mathbb{E}(u_{i} | RSD(\succ))} {\sum_{i} \mathbb{E}(u_{i} | RSD(\succ))}.
\end{equation*}

To measure the manipulability of PS, we are interested in answering two key questions: \textit{i}) In what fraction of profiles is PS manipulable by at least one agent? and \textit{ii}) If manipulation is possible, what is the average percentage of maximum gain? That is,
\begin{equation*}
\max_{i} \{ \frac{\mathbb{E}(u_{i} | PS(\succ'_{i}, \succ_{-i})) - \mathbb{E}(u_{i} | PS(\succ))} {\mathbb{E}(u_{i} | PS(\succ))} \}.
\end{equation*}

To study the envy under the RSD mechanism, we consider two measures: \textit{i}) the fraction of envious agents, and \textit{ii}) the total envy felt by all agents.

\subsection{Linear Utility Model}

\begin{figure*}
	\centering
	\begin{subfigure}[t]{0.55\textwidth}
		\centering
		\includegraphics[width=\textwidth]{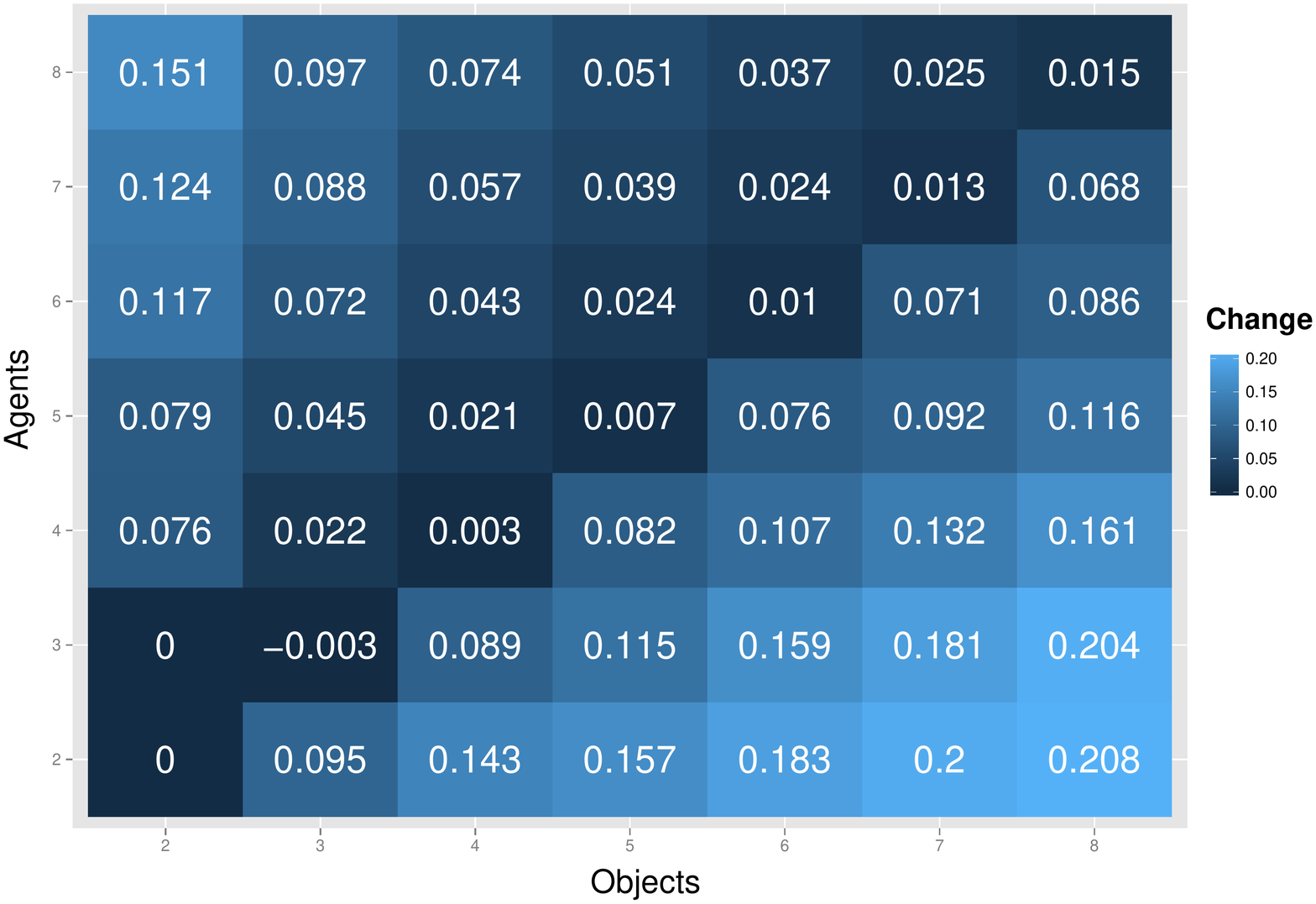}
		\caption{Welfare change, Linear}
		\label{}
	\end{subfigure}~
	
	\begin{subfigure}[t]{0.47\textwidth}
		\centering
		\includegraphics[width=\textwidth]{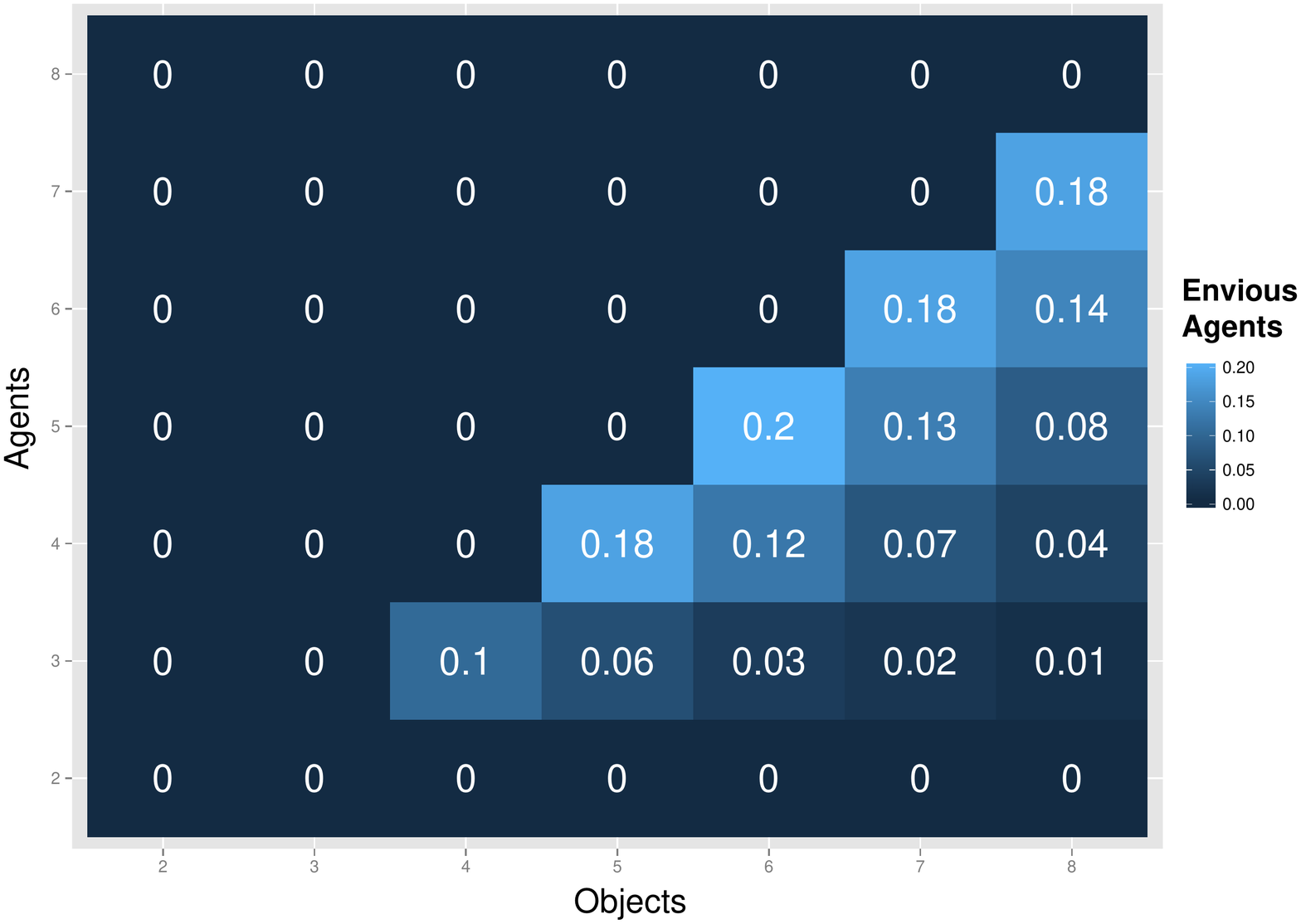}
		\caption{Envious Agents, Linear}
		\label{}
	\end{subfigure}~
	\begin{subfigure}[t]{0.47\textwidth}
		\centering
		\includegraphics[width=\textwidth]{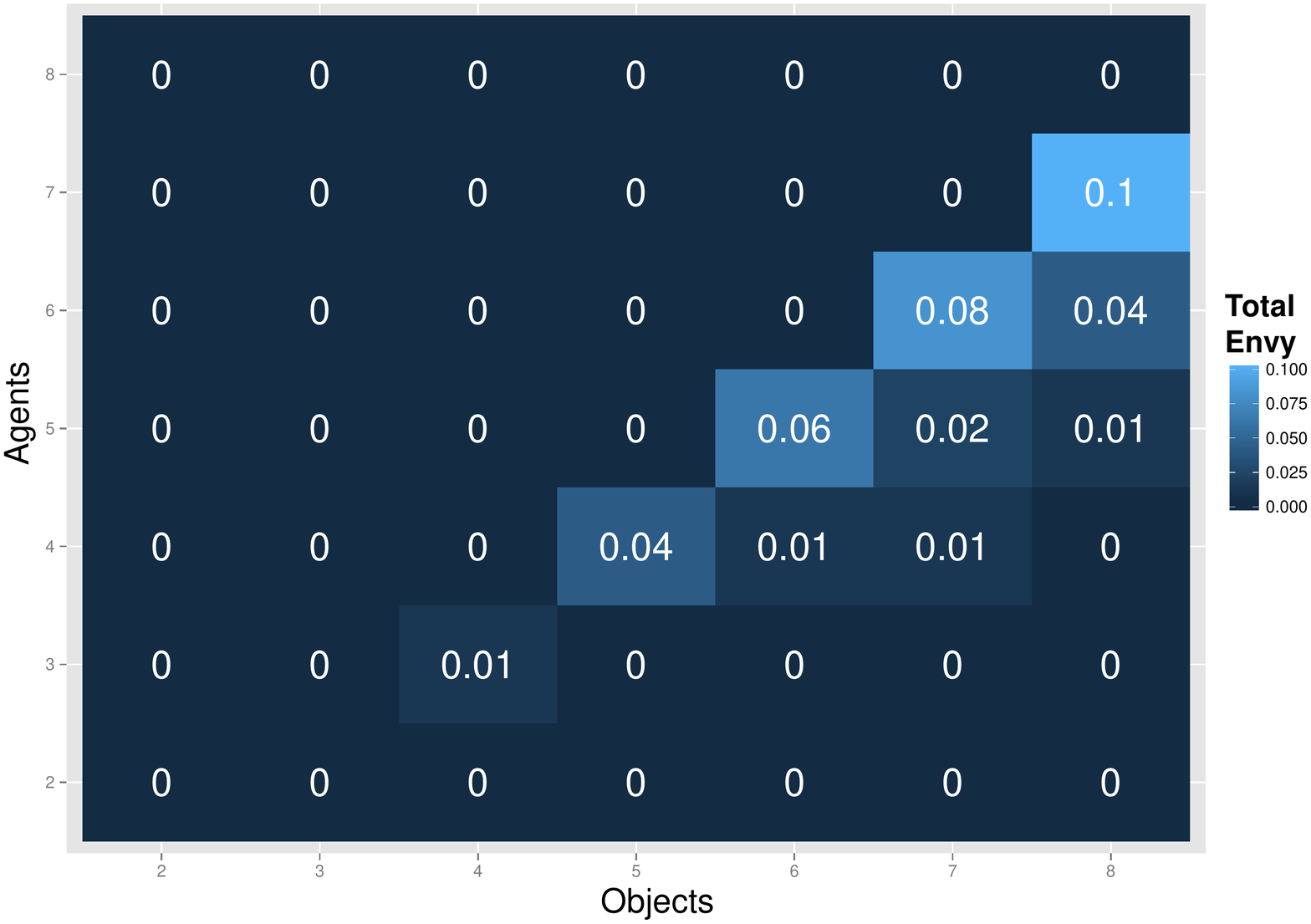}
		\caption{Total Envy, Linear}
		\label{}
	\end{subfigure}

	\begin{subfigure}[t]{0.47\textwidth}
		\centering
		\includegraphics[width=\textwidth]{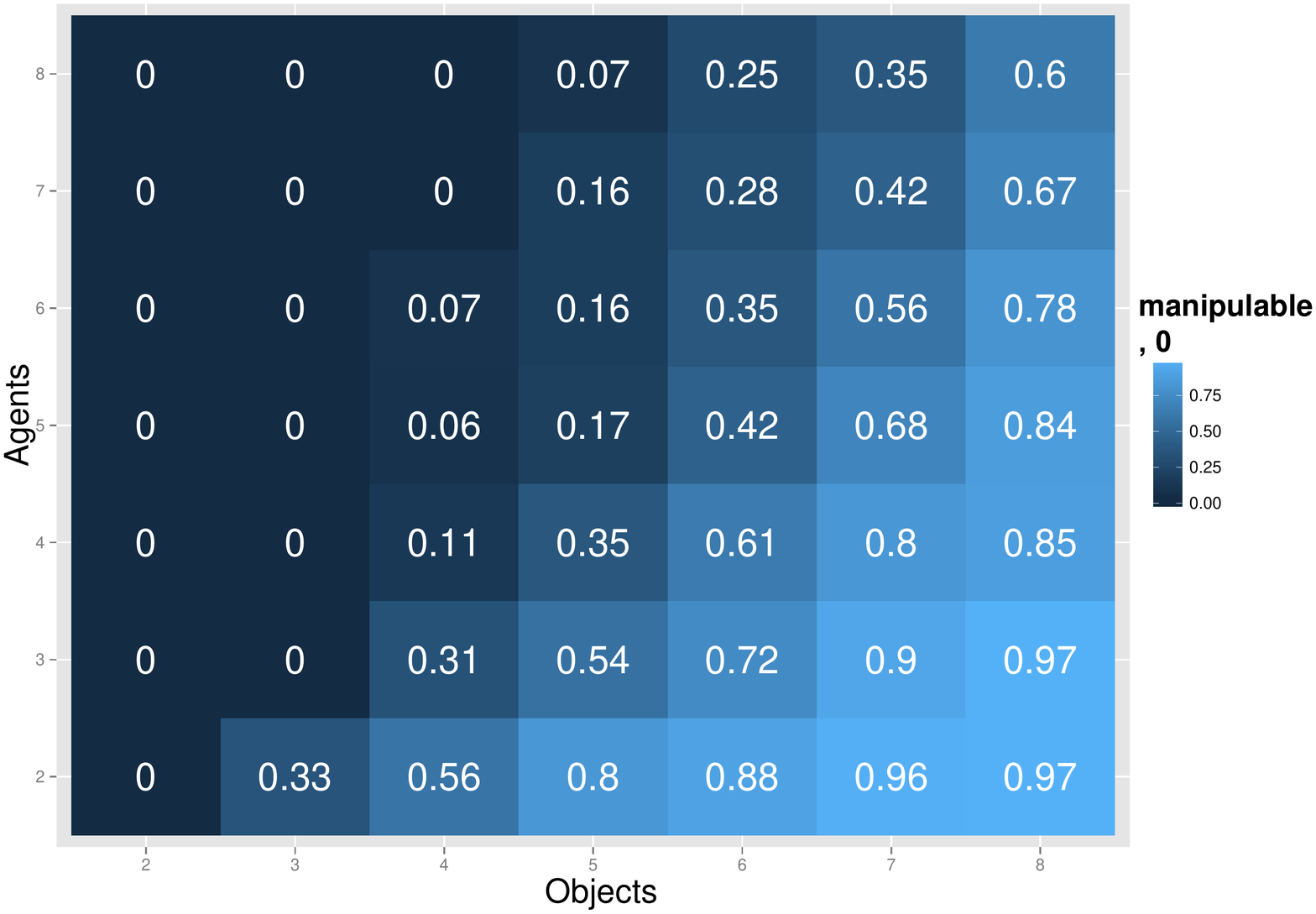}
		\caption{Manipulation, Linear}
		\label{}
	\end{subfigure}~
	\begin{subfigure}[t]{0.47\textwidth}
		\includegraphics[width=\textwidth]{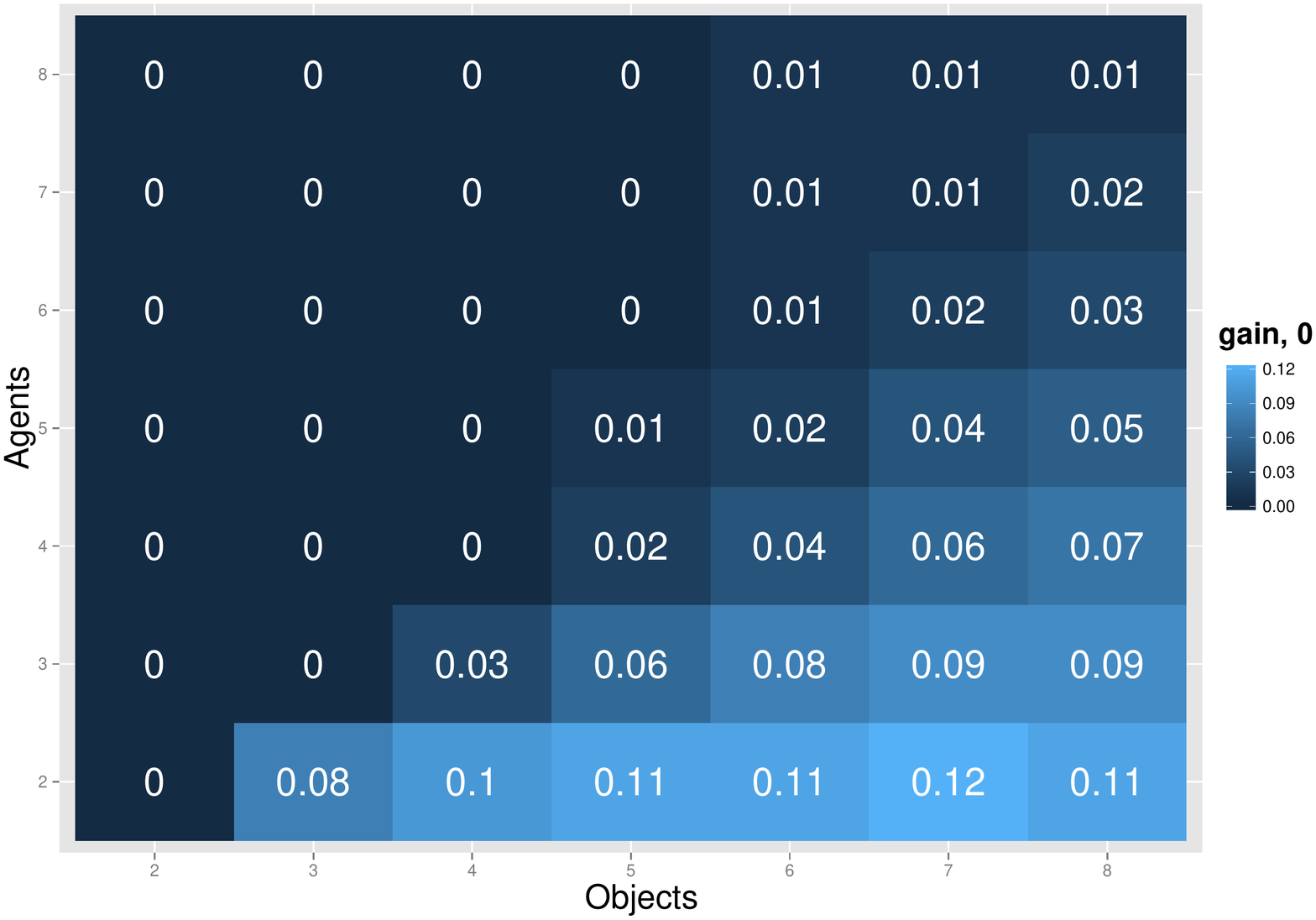}
		\caption{Manipulation gain, Linear}
		\label{}
	\end{subfigure}
	\caption{Linear Utility: Welfare Change, Envy, and Manipulation.} 
	\label{fig:linear util}
\end{figure*}

We first looked at how RSD and PS perform under the assumption that the utility models are linear (Figure \ref{fig:linear util}). In most cases, the social welfare under PS increases compared to RSD; however, the social welfare of PS is very close to that of RSD when $n= m$ (less than $0.015$ overall  improvement in all cases). Interestingly, under RSD the fraction of envious agents gets close to $0$ when $n \geq m$.
With regards to strategyproofness, PS is manipulable in most combinations of $n$ and $m$ and the fraction of manipulable profiles and the utility gain from manipulation increases as the number of objects compared to agents increases.

\subsection{Risk Seeking}
\begin{figure*}
	\centering
	\begin{subfigure}[t]{0.49\textwidth}
		\centering
		\includegraphics[width=\textwidth]{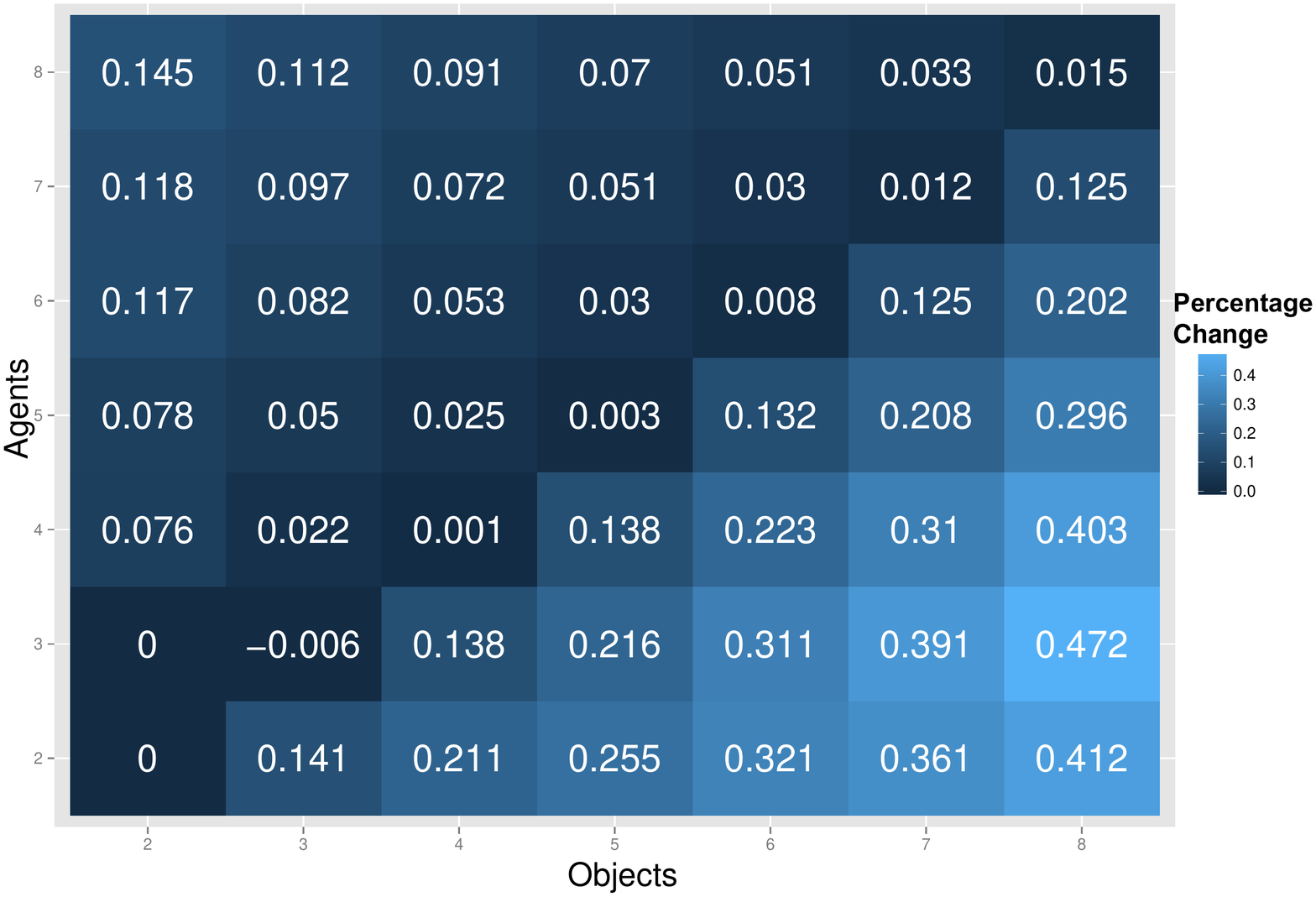}
		\caption{Risk seeking, $\alpha = -0.5$.}
		\label{half}
	\end{subfigure}~
	\begin{subfigure}[t]{0.49\textwidth}
		\centering
		\includegraphics[width=\textwidth]{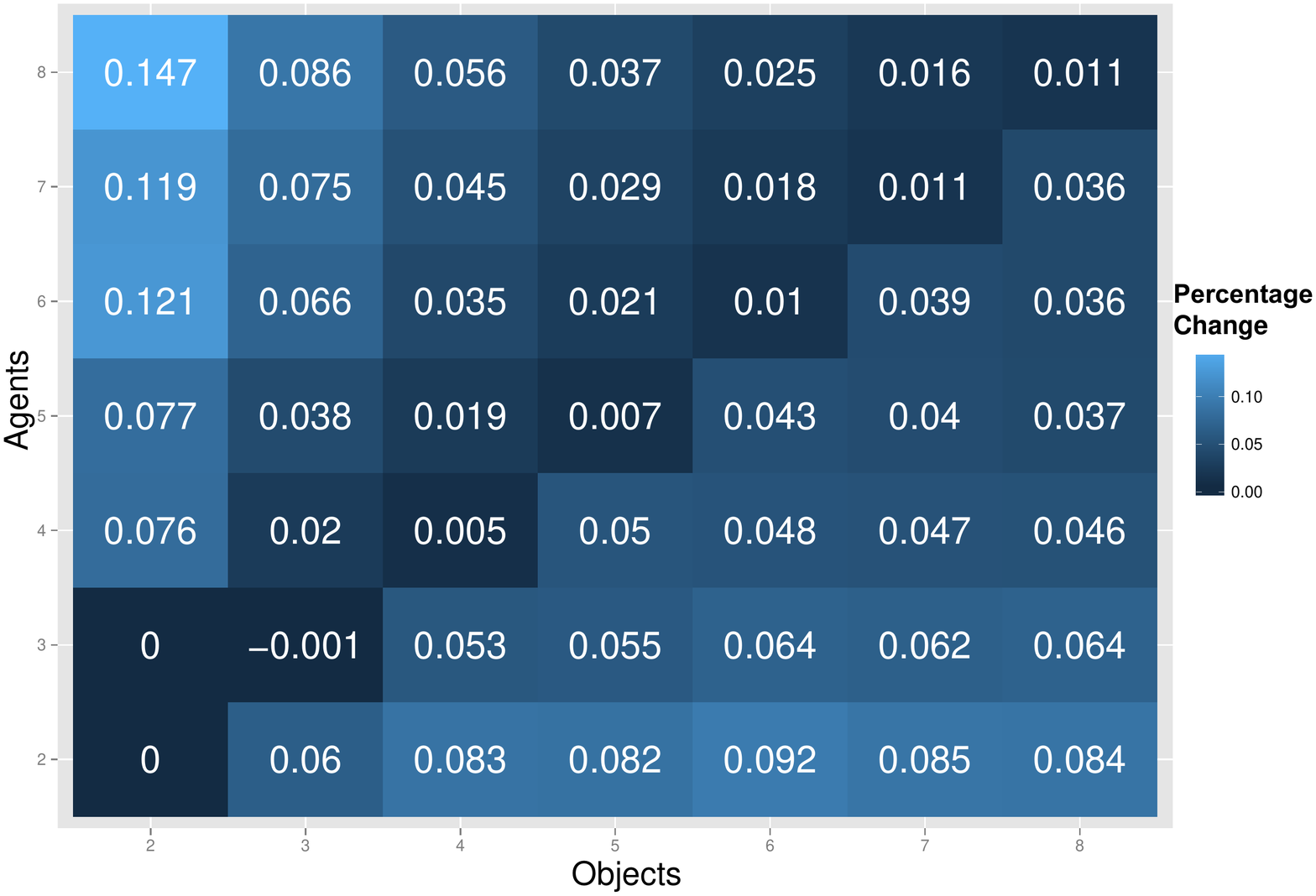}
		\caption{Risk averse, $\alpha = 0.5$.}
		\label{halfaverse}
	\end{subfigure}
	\begin{subfigure}[t]{0.49\textwidth}
		\centering
		\includegraphics[width=\textwidth]{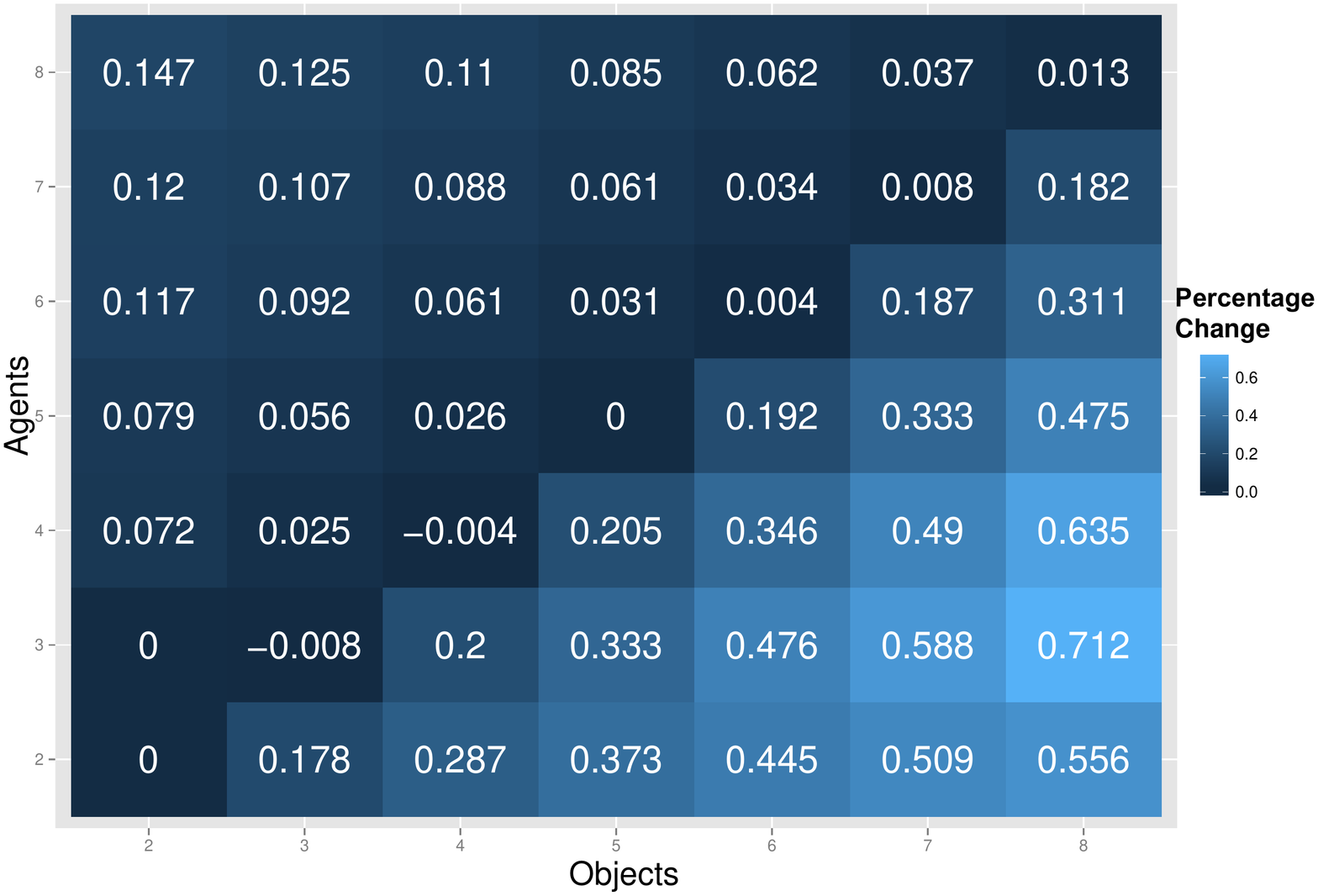}
		\caption{Risk seeking, $\alpha = -1$.}
		\label{one}
	\end{subfigure}~
	\begin{subfigure}[t]{0.49\textwidth}
		\centering
		\includegraphics[width=\textwidth]{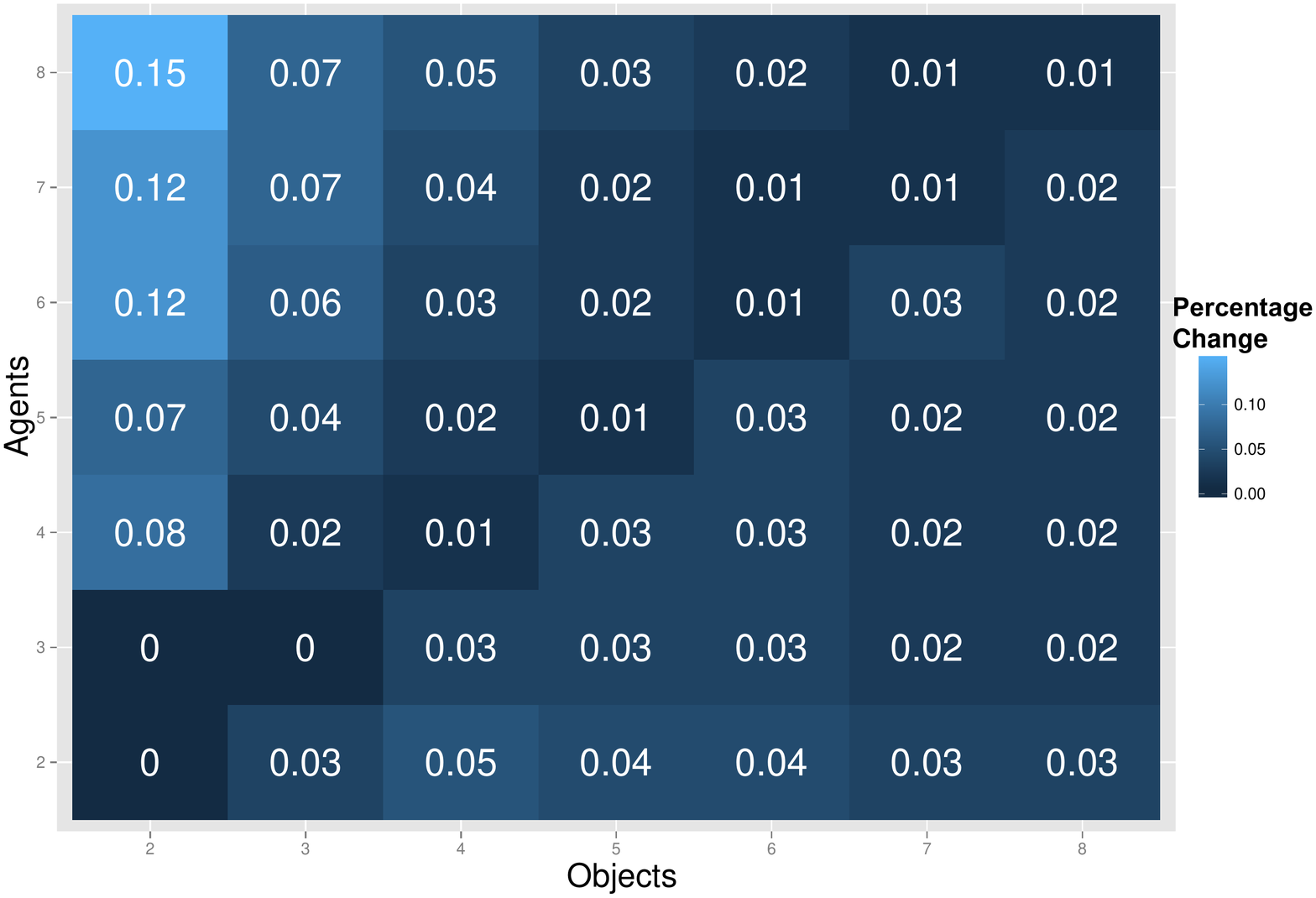}
		\caption{Risk averse, $\alpha = 1$.}
		\label{oneaverse}
	\end{subfigure}
	\begin{subfigure}[t]{0.49\textwidth}
		\includegraphics[width=\textwidth]{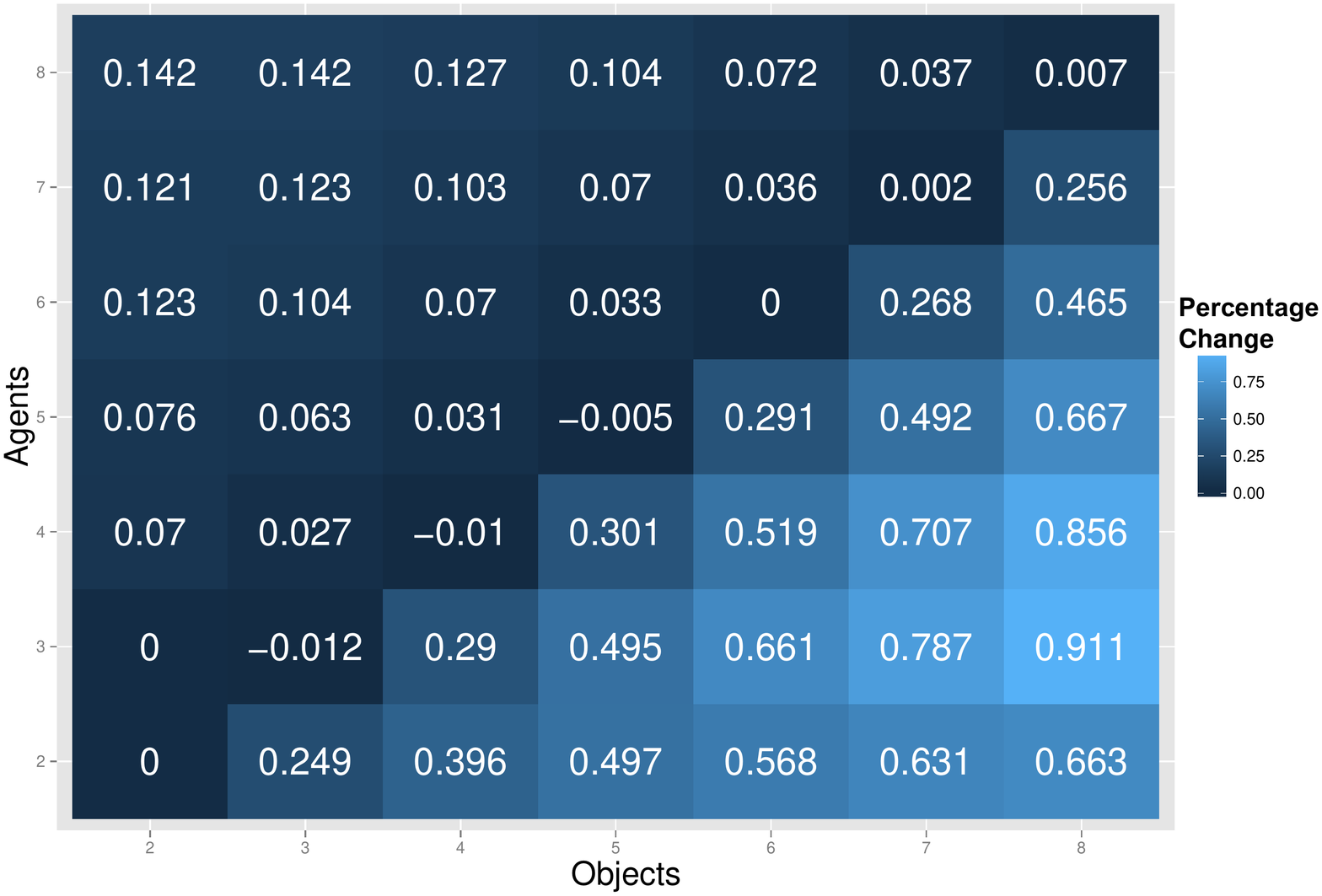}
		\caption{Risk seeking, $\alpha = -2$.}
		\label{two}
	\end{subfigure}~
	\begin{subfigure}[t]{0.49\textwidth}
		\includegraphics[width=\textwidth]{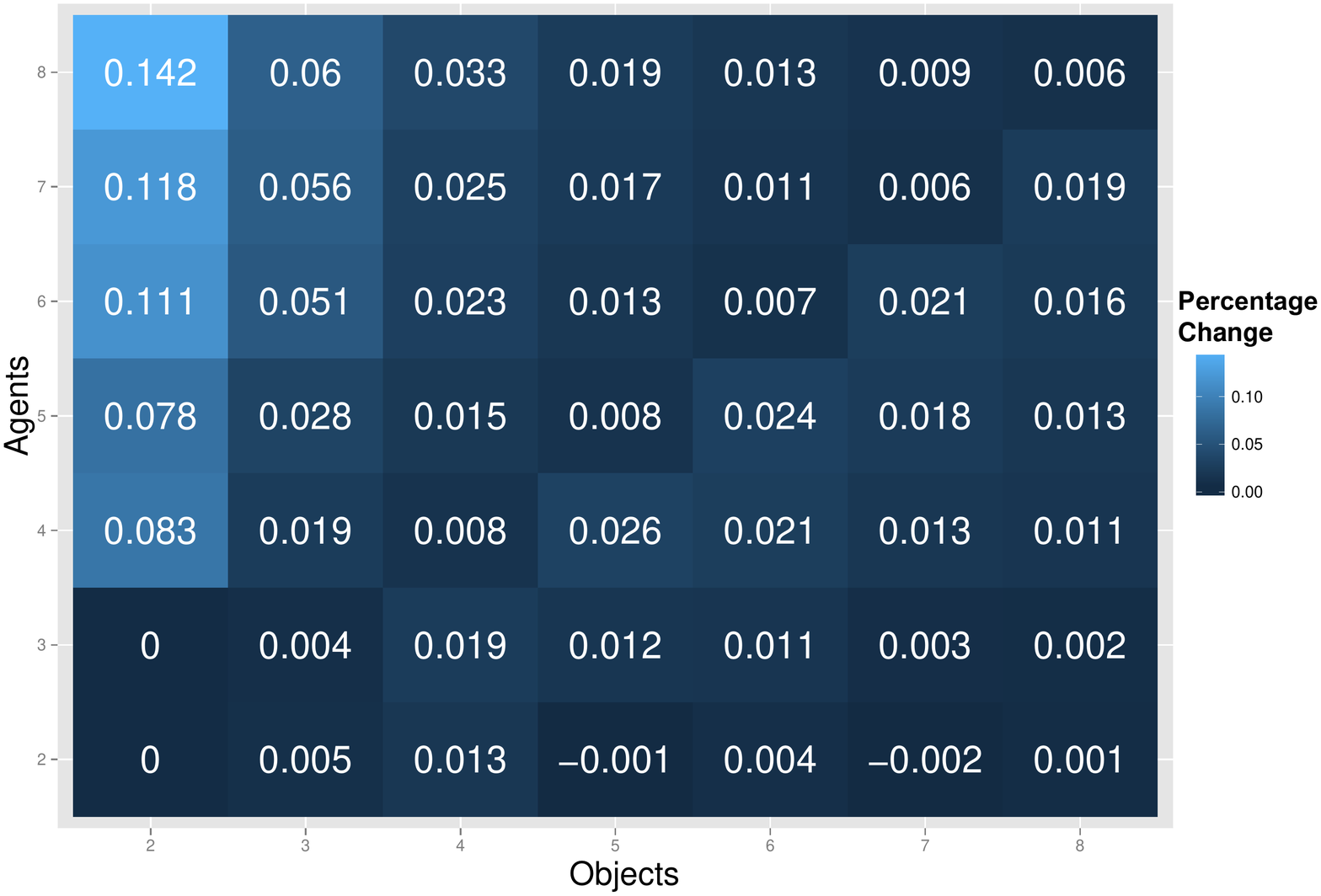}
		\caption{Risk averse, $\alpha = 2$.}
		\label{twoaverse}
	\end{subfigure}
	\caption{The percentage change in social welfare. The negative values show that RSD outperforms PS.} 
	\label{fig:percentageChange}
\end{figure*}

Figure \ref{fig:percentageChange} presents our results in terms of percentage change in social welfare between PS and RSD. Positive numbers show the percentage of improvement in social welfare from PS to RSD. Negative values represent those cases where RSD has increased social welfare compared to PS.

\textbf{Social welfare}:
Fixing $\alpha < 0$, for $n\geq m$ when $\frac{n}{m}$ grows PS improves the social welfare compared to RSD in all instances of the problem and the percentage of improvement also increases. A similar trend holds when varying risk intensity $\alpha$ for fixed $n$ and $m$ where $n\neq m$.
For $n < m$, when $\frac{m}{n}$ grows the fraction of profiles at which PS has higher social welfare compared to RSD rapidly increases and the percentage change is also noticeably larger, quickly getting close to 90\% improvement (Figures \ref{half}, \ref{one}, and \ref{two}). This social welfare gap between PS and RSD grows as the risk intensity $|\alpha|$ increases.
Surprisingly, this trend changes for equal number of agents and objects $n = m$: the more risk-seeking agents are (larger $|\alpha|$), RSD becomes more desirable than PS, and in fact, RSD improves the social welfare in more instances.


\begin{figure}
	\begin{subfigure}[t]{0.49\textwidth}
		\centering
		\includegraphics[width=\textwidth]{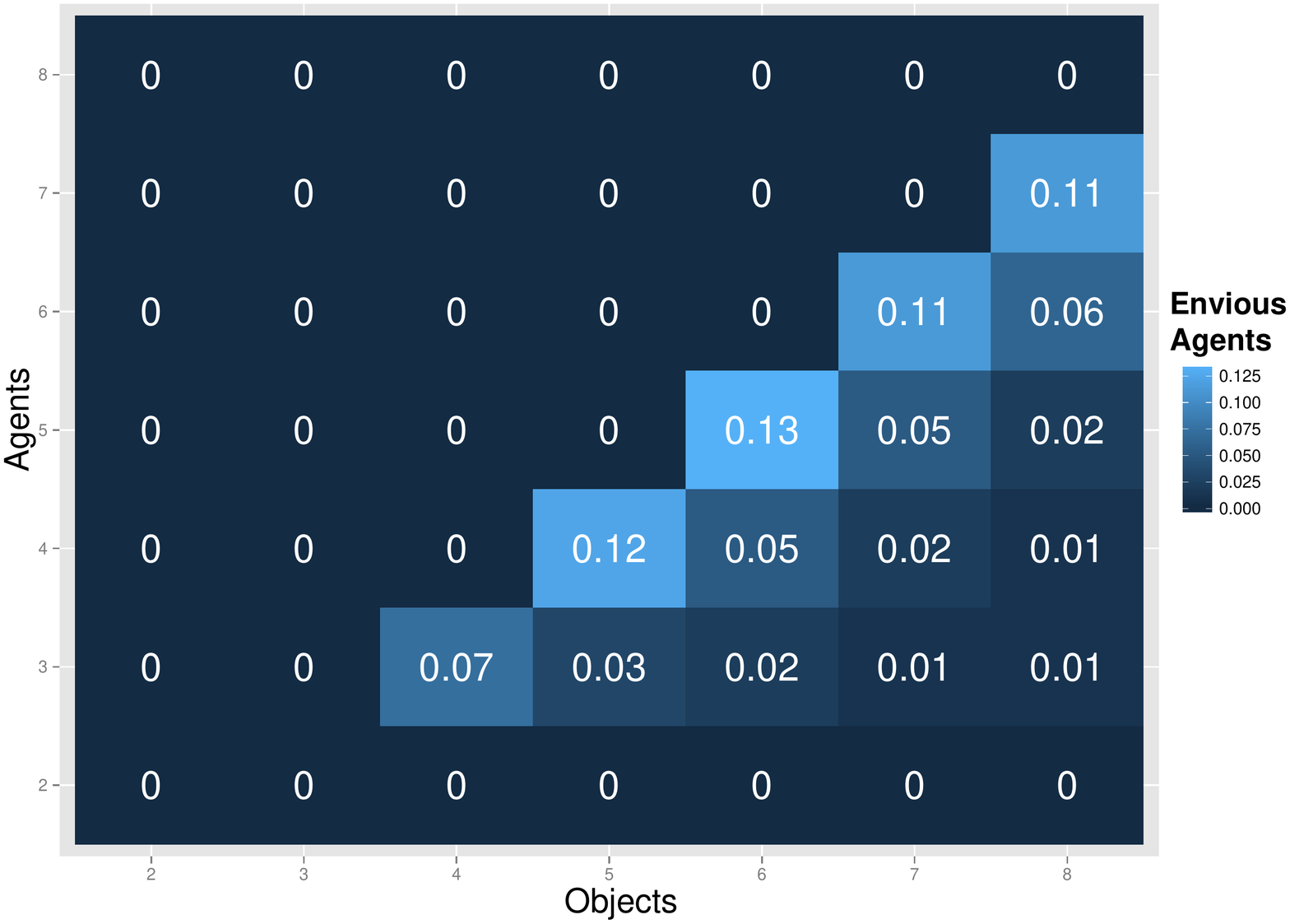}
		\caption{Fraction of envious agents, $\alpha = -0.5$.}
		\label{}
	\end{subfigure}~
	\begin{subfigure}[t]{0.49\textwidth}
		\centering
		\includegraphics[width=\textwidth]{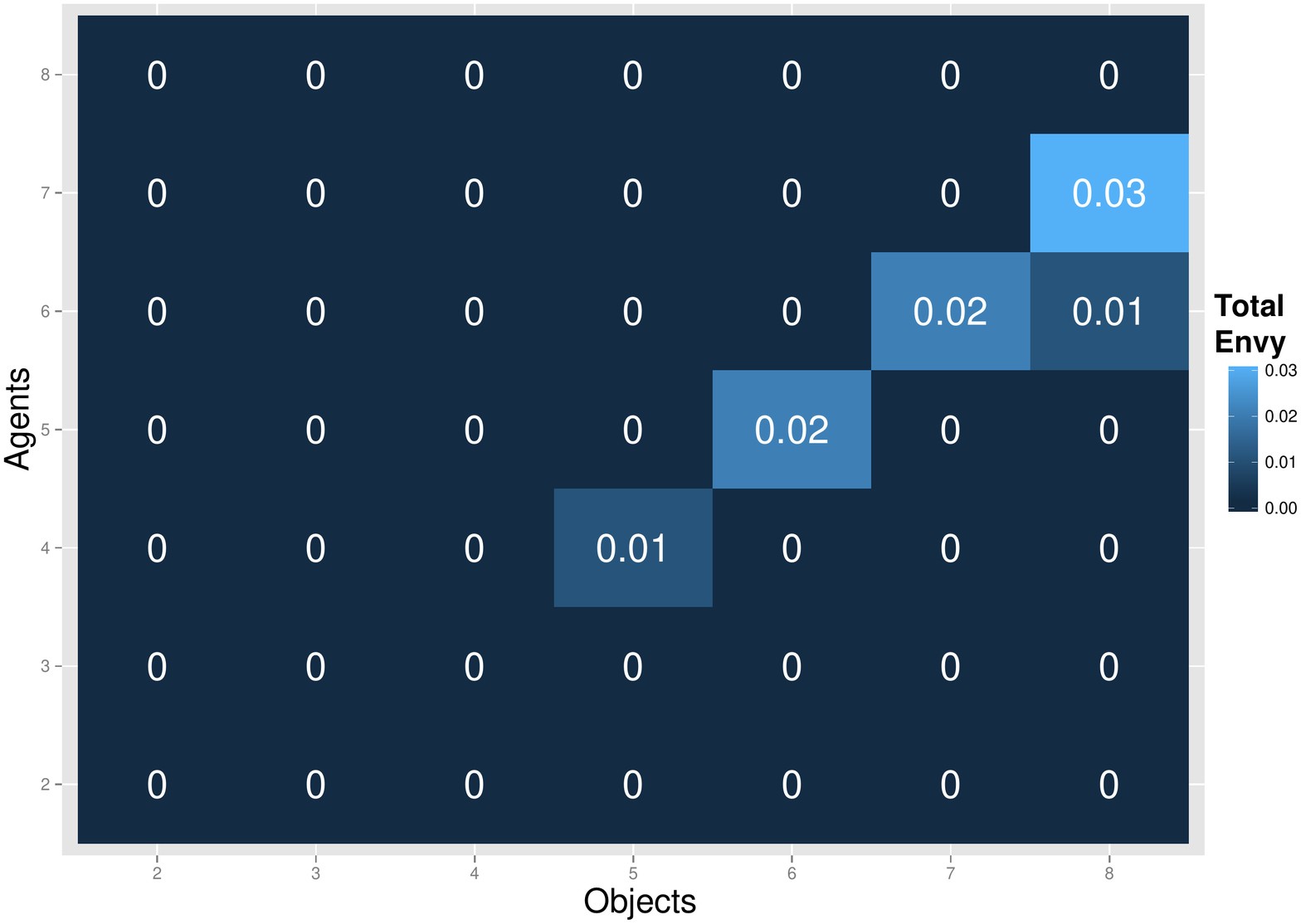}
		\caption{Total envy, $\alpha = -0.5$.}
		\label{}
	\end{subfigure}
	\begin{subfigure}[t]{0.49\textwidth}
		\centering
		\includegraphics[width=\textwidth]{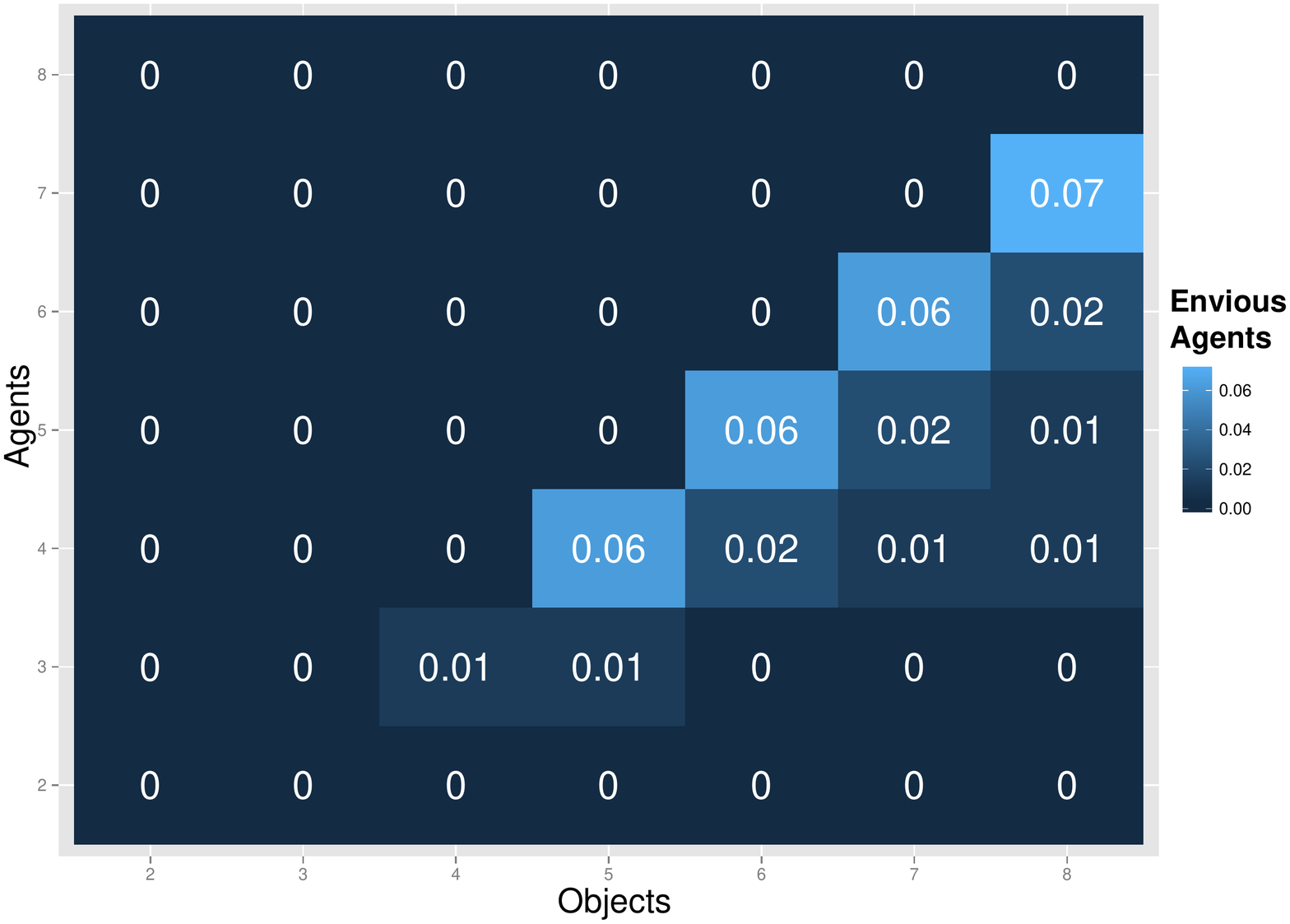}
		\caption{Fraction of envious agents, $\alpha = - 1$.}
		\label{}
	\end{subfigure}~
	\begin{subfigure}[t]{0.49\textwidth}
		\centering
		\includegraphics[width=\textwidth]{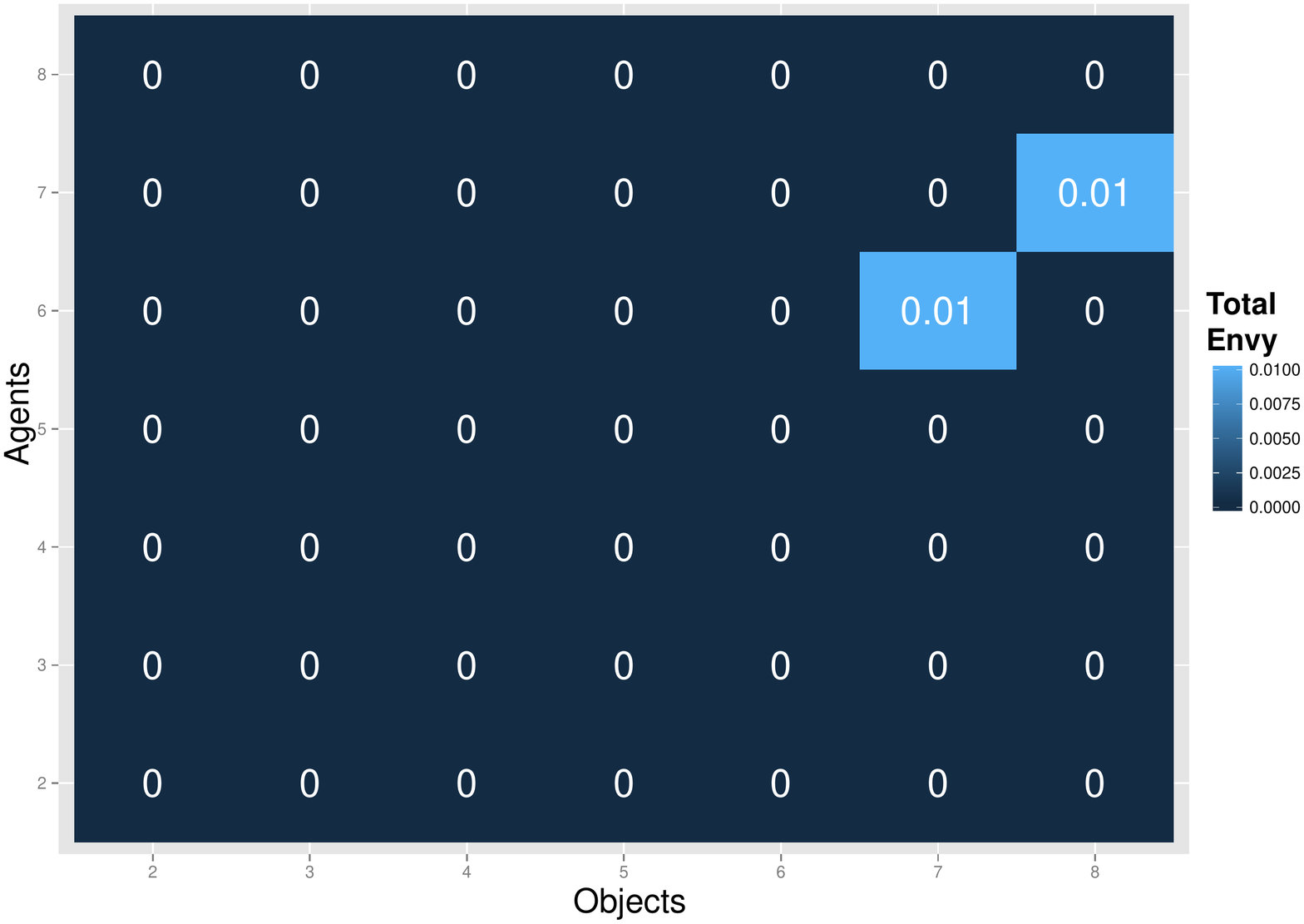}
		\caption{Total envy, $\alpha = - 1$.}
		\label{}
	\end{subfigure}
	\begin{subfigure}[t]{0.49\textwidth}
		\centering
		\includegraphics[width=\textwidth]{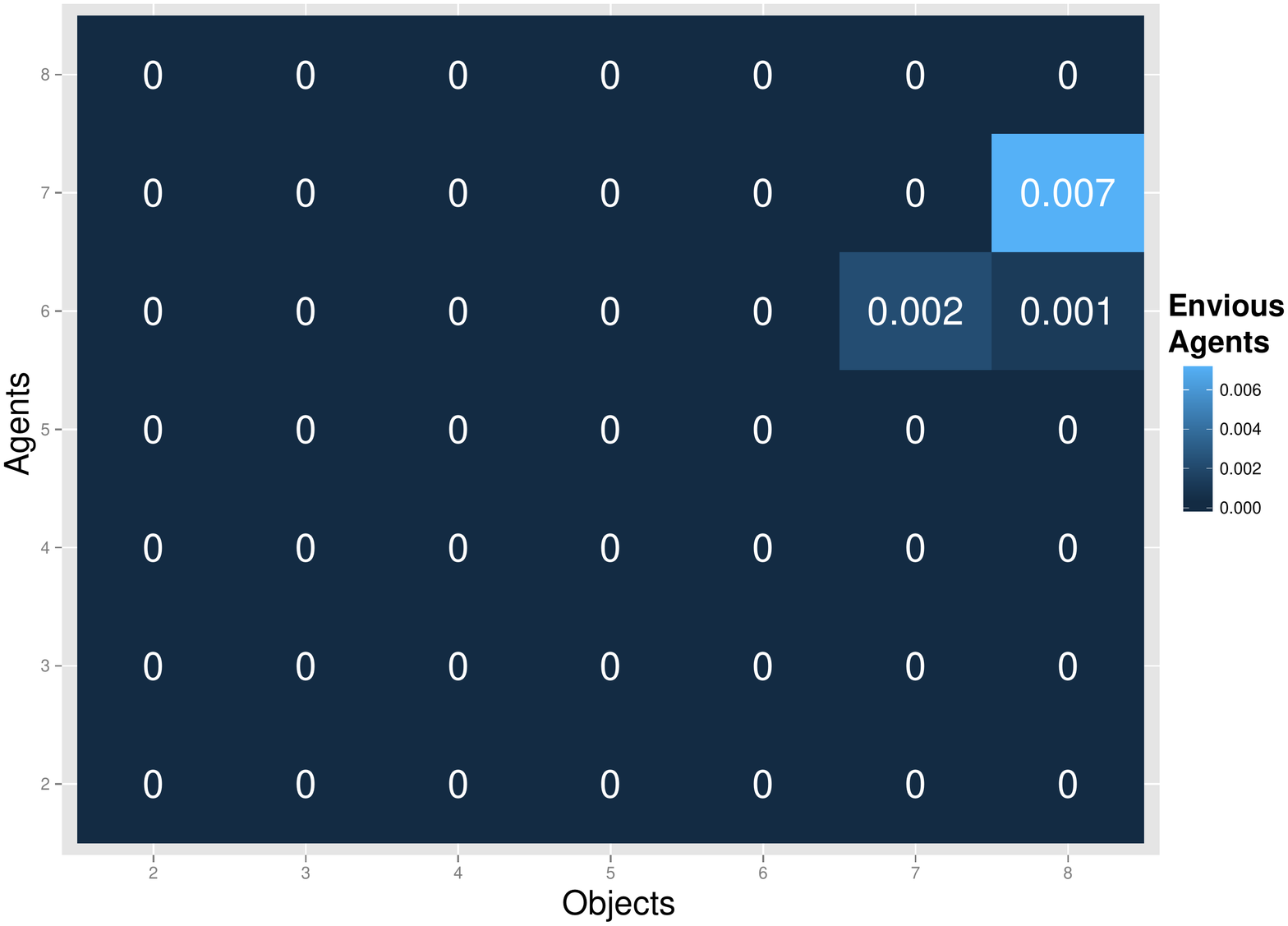}
		\caption{Fraction of envious agents, $\alpha = - 2$.}
		\label{}
	\end{subfigure}~
	\begin{subfigure}[t]{0.49\textwidth}
		\centering
		\includegraphics[width=\textwidth]{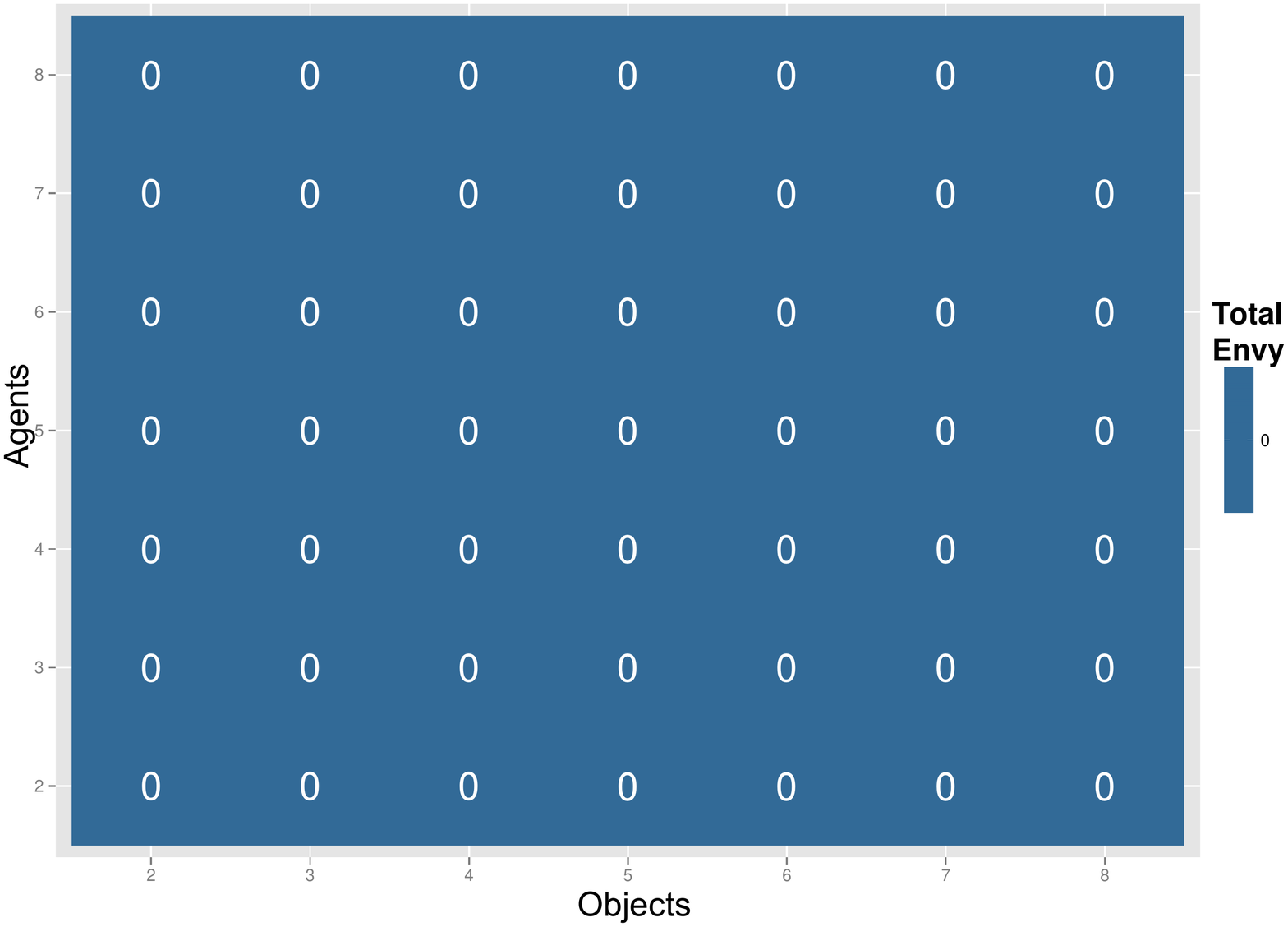}
		\caption{Total envy, $\alpha = - 2$.}
		\label{}
	\end{subfigure}
	\caption{The fraction of envious agents and total envy perceived by agents under risk-seeking utilities.} 
	\label{fig:envyRiskTaking}
\end{figure}

\textbf{Envy}:
Figure \ref{fig:envyRiskTaking} shows that for $n \geq m$, the fraction of envious agents under all profiles vanishes and RSD becomes envyfree. This is more evident when agents are more risk-seeking. Intuitively, these observations confirm the theoretical findings about the envyfreeness of RSD under lexicographic preferences \cite{hosseini2015strategyproof}. This is because one can consider lexicographic preferences as risk-seeking preferences where an object in a higher ranking is infinitely preferred to all objects that are ranked less preferably \cite{hosseini2015strategyproof}.
When $n  < m$, our quasi-dictatorial extension of RSD creates some envy among the agents, because the agent with the highest priority receives $m - n + 1$ objects, while all other agents receive at most one object. An interesting result is the envy created by RSD starts to fade out when the risk intensity $|\alpha|$ increases.

\begin{figure*}
	\begin{subfigure}[t]{0.49\textwidth}
		\centering
		\includegraphics[width=\textwidth]{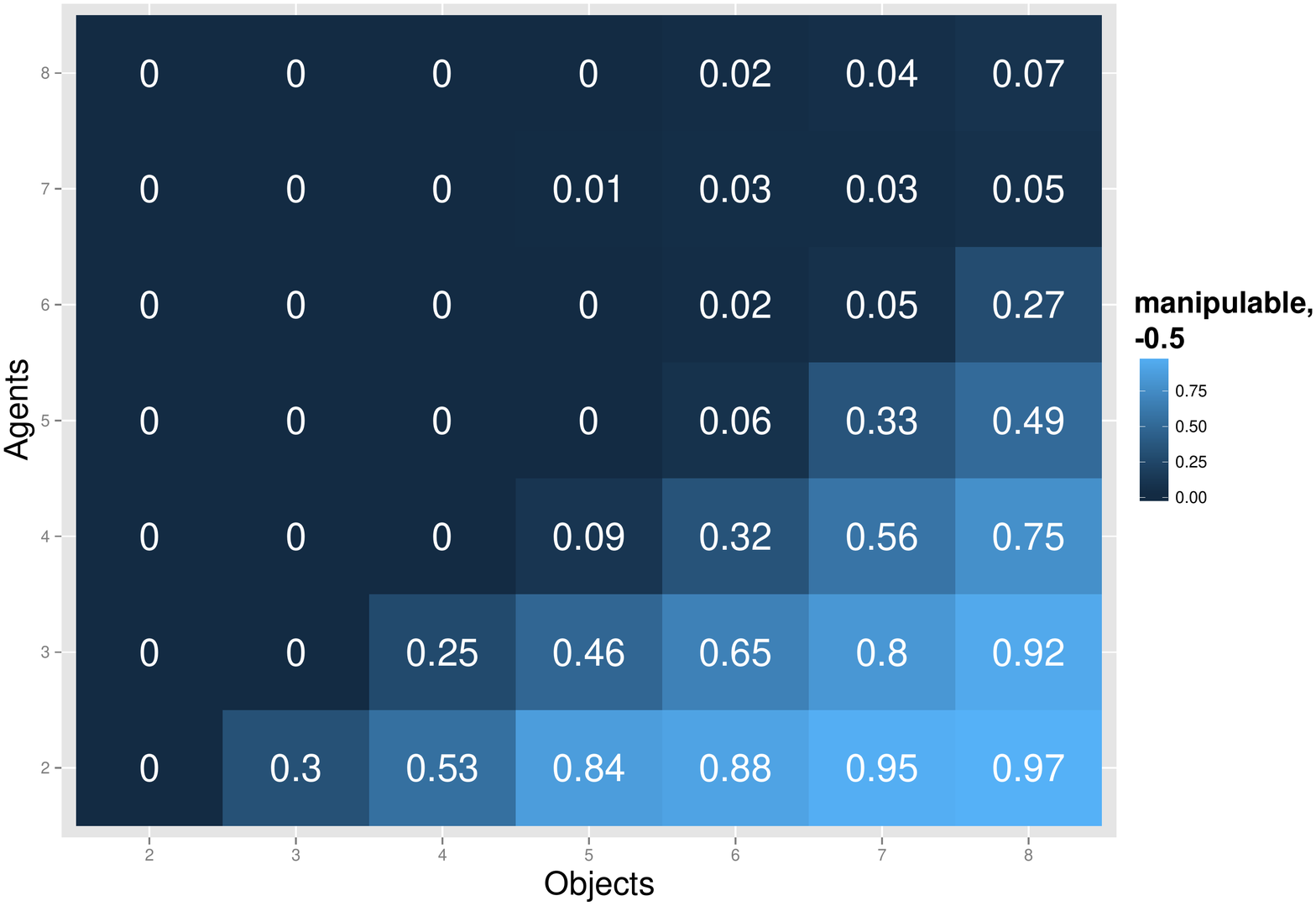}
		\caption{Manipulation, $\alpha = -0.5$.}
		\label{}
	\end{subfigure}~
	\begin{subfigure}[t]{0.49\textwidth}
		\centering
		\includegraphics[width=\textwidth]{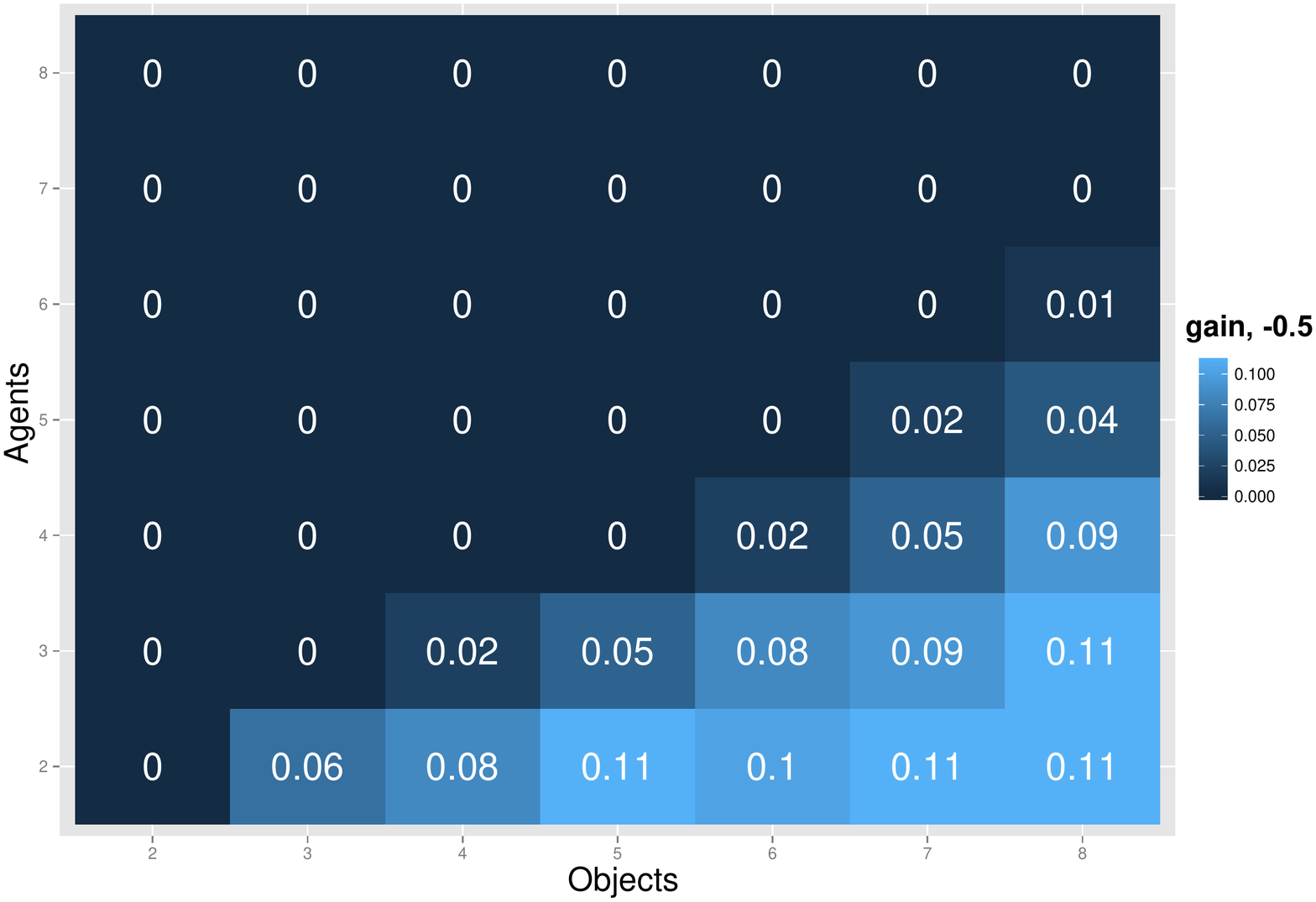}
		\caption{Gain, $\alpha = -0.5$.}
		\label{}
	\end{subfigure}
	\begin{subfigure}[t]{0.49\textwidth}
		\centering
		\includegraphics[width=\textwidth]{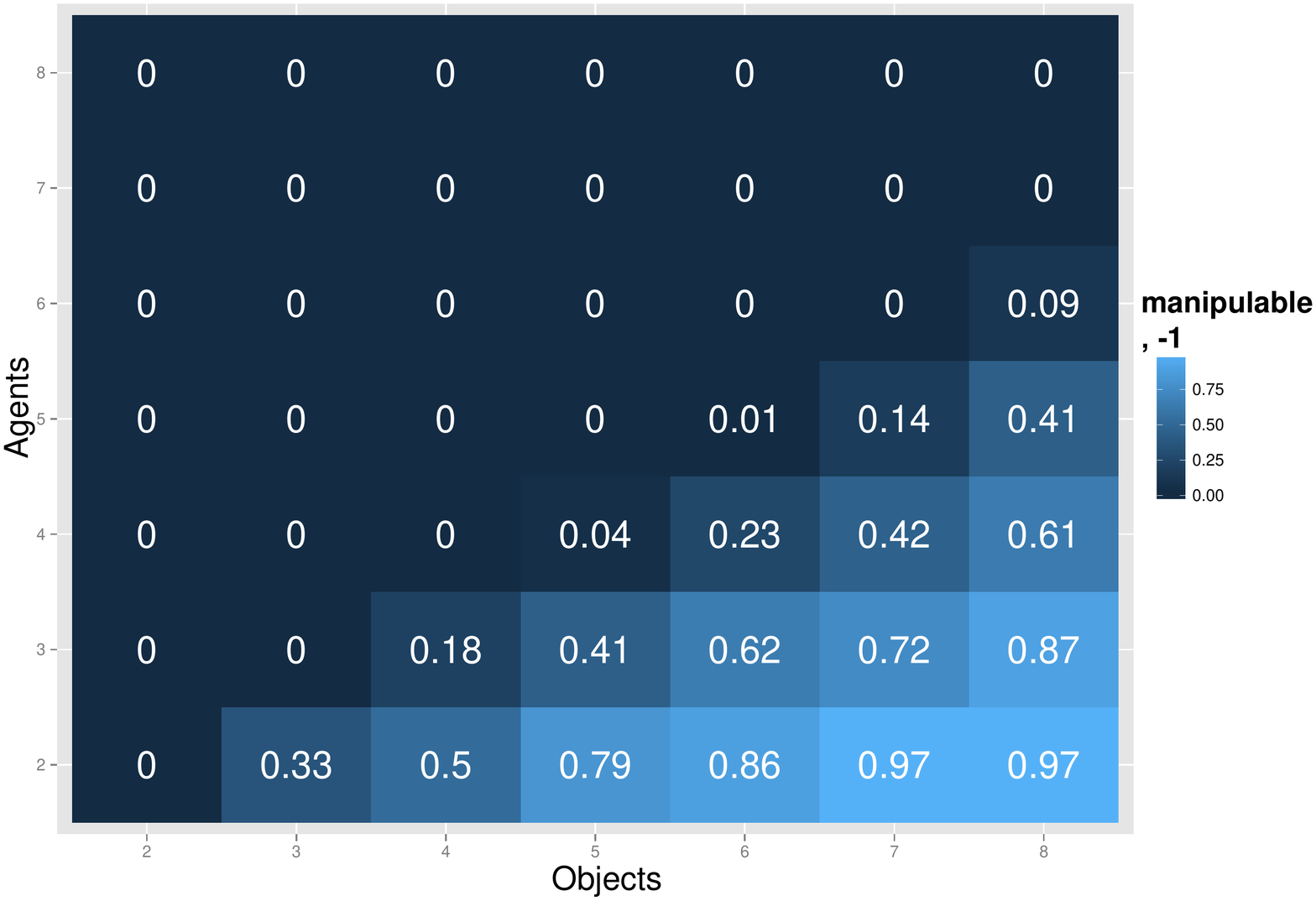}
		\caption{Manipulation, $\alpha = -1$.}
		\label{}
	\end{subfigure}~
	\begin{subfigure}[t]{0.49\textwidth}
		\centering
		\includegraphics[width=\textwidth]{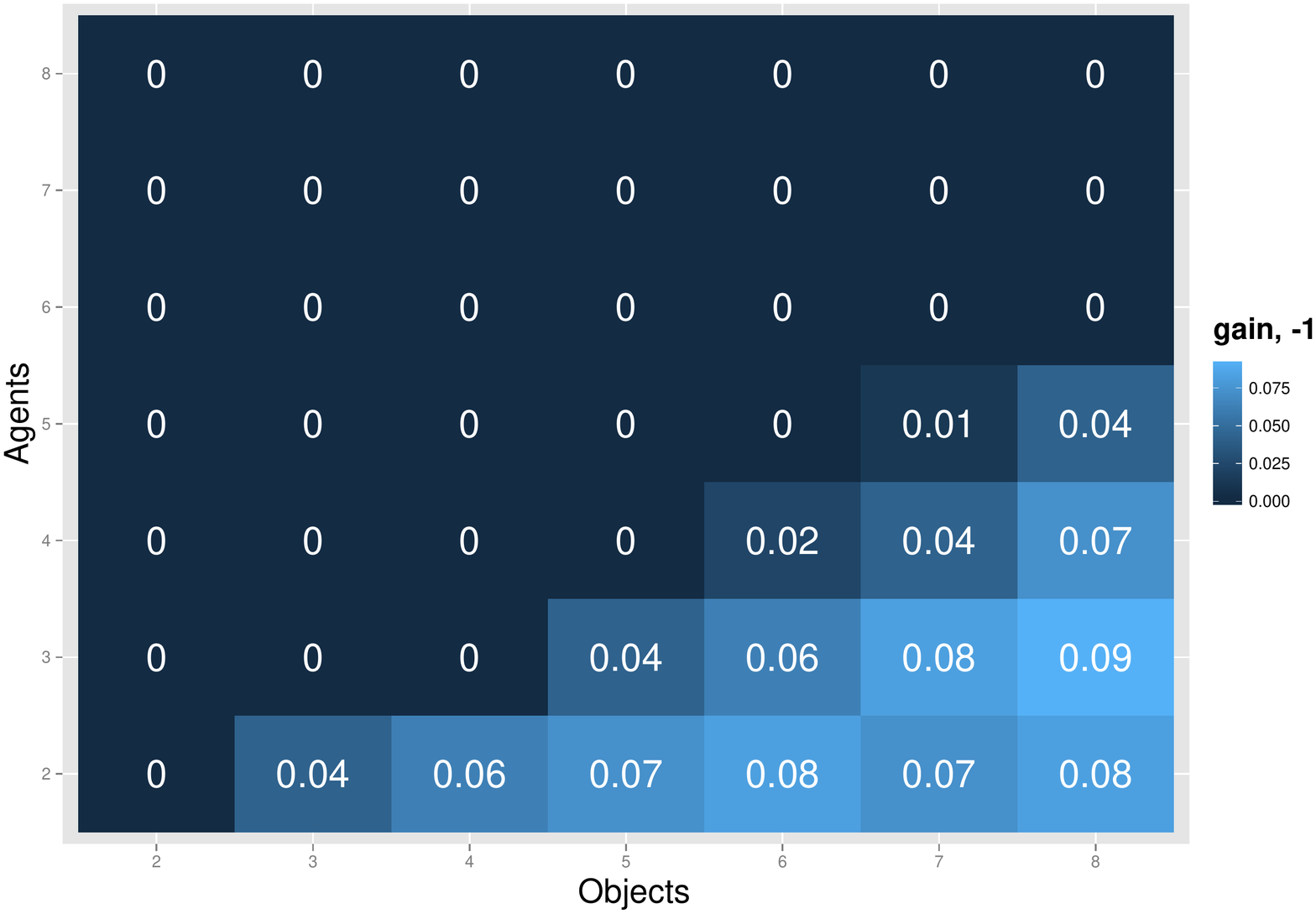}
		\caption{Gain, $\alpha = -1$.}
		\label{}
	\end{subfigure}
	\begin{subfigure}[t]{0.49\textwidth}
		\includegraphics[width=\textwidth]{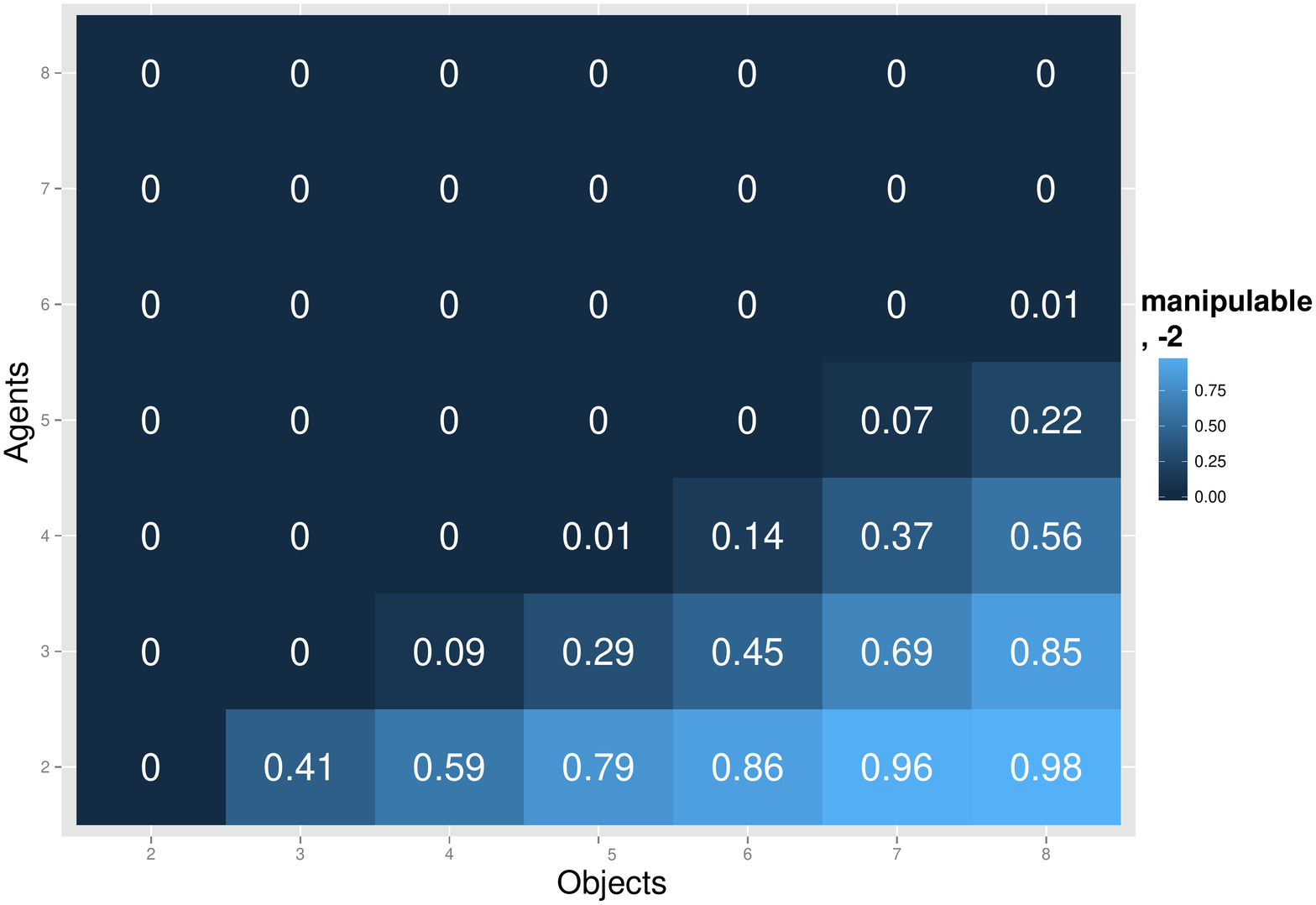}
		\caption{Manipulation, $\alpha = -2$.}
		\label{}
	\end{subfigure}~
	\begin{subfigure}[t]{0.49\textwidth}
		\includegraphics[width=\textwidth]{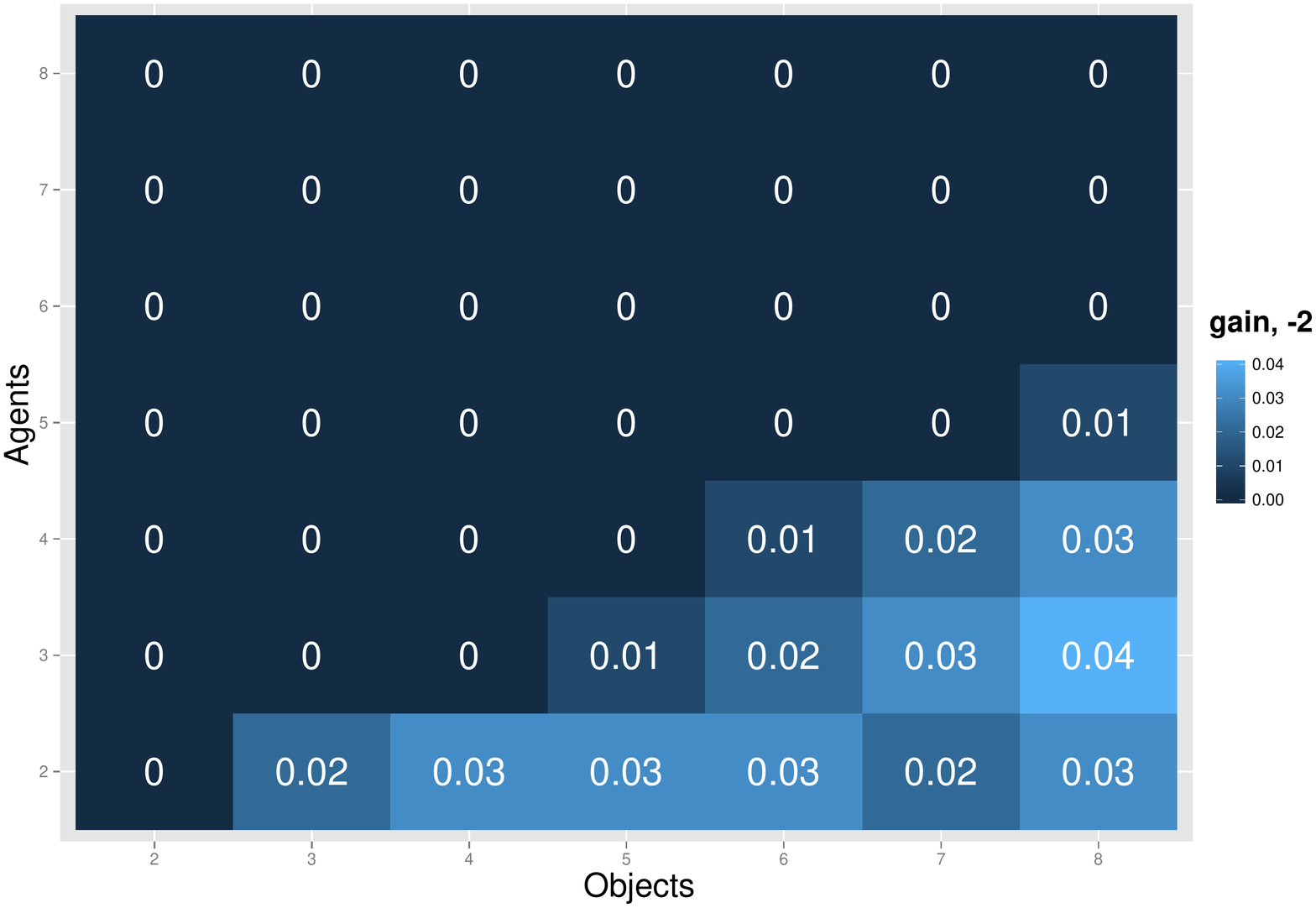}
		\caption{Gain, $\alpha = -2$.}
		\label{}
	\end{subfigure}
	\caption{The fraction of manipulable instances and manipulation gain of PS under risk-seeking preferences.} 
	\label{fig:manipRisk}
\end{figure*}

\textbf{Manipulability}: 
Figure \ref{fig:manipRisk} shows the manipulability of the PS assignments when agents are risk seeking. 
We see that the possibility of manipulation (and any gain) decreases as the risk intensity increases. When $n \geq m$ the fraction of manipulable profiles goes to $0$ the more risk seeking agents become. However, when $n < m$ even though the fraction of manipulable profiles (and manipulation gain) decreases, the fraction of manipulable profiles goes to 1 as $\frac{m}{n}$ grows.

\subsection{Risk Aversion}

\textbf{Social welfare}:
Figures \ref{halfaverse}, \ref{oneaverse}, and \ref{twoaverse} show that, fixing risk factor $\alpha > 0$, when $\frac{n}{m}$ grows, PS assignments are superior to that of RSD in terms of social welfare in more instances, and the percentage change in social welfare increases.
Fixing risk factor $\alpha > 0$ and when $\frac{m}{n}$ grows, RSD is more likely to have the same social welfare as PS, and in fact in some  instances the social welfare under RSD is better than the social welfare under PS.
Fixing $m$ and $n$, when the risk intensity $\alpha$ increases RSD is more likely to have the same social welfare as PS, that is, the welfare gap between PS and RSD closes when agents are more risk averse ($\alpha$ increases). This result is insightful and states that under risk aversion the random allocations prescribed by RSD are either as good as PS or in some cases even are superior to the allocations prescribed by PS due to the underlying shape of the utility models.
Figure \ref{fig:changeAll} illustrates the percentage change in social welfare based on the difference between available objects and agents ($m-n$) for risk seeking, linear, and risk averse utilities with different risk intensities.

\begin{figure*}
	\begin{subfigure}[t]{0.48\textwidth}
		\centering
		\includegraphics[width=\textwidth]{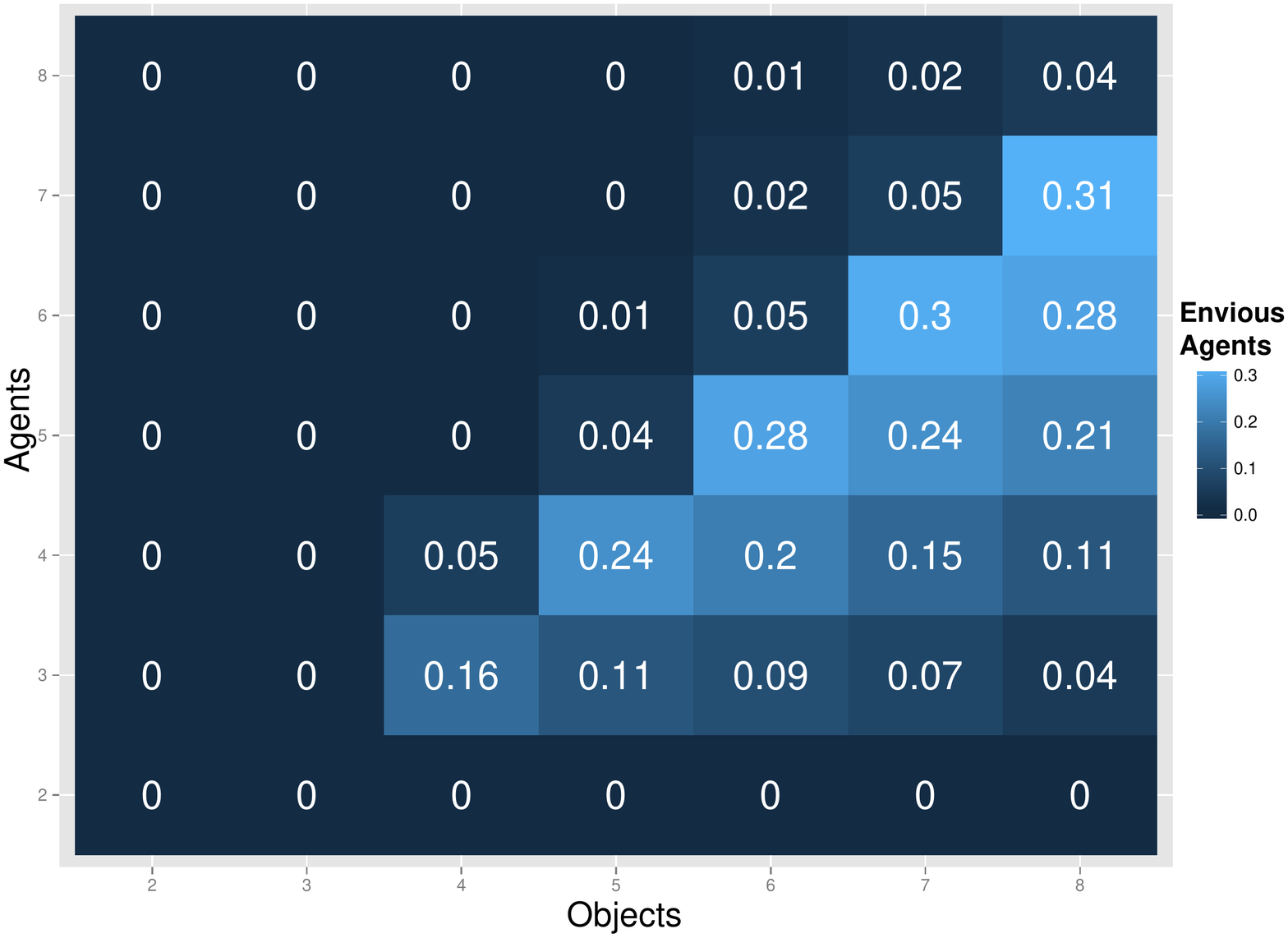}
		\caption{Fraction of envious agents, $\alpha = 0.5$.}
		\label{}
	\end{subfigure}~
	\begin{subfigure}[t]{0.48\textwidth}
		\centering
		\includegraphics[width=\textwidth]{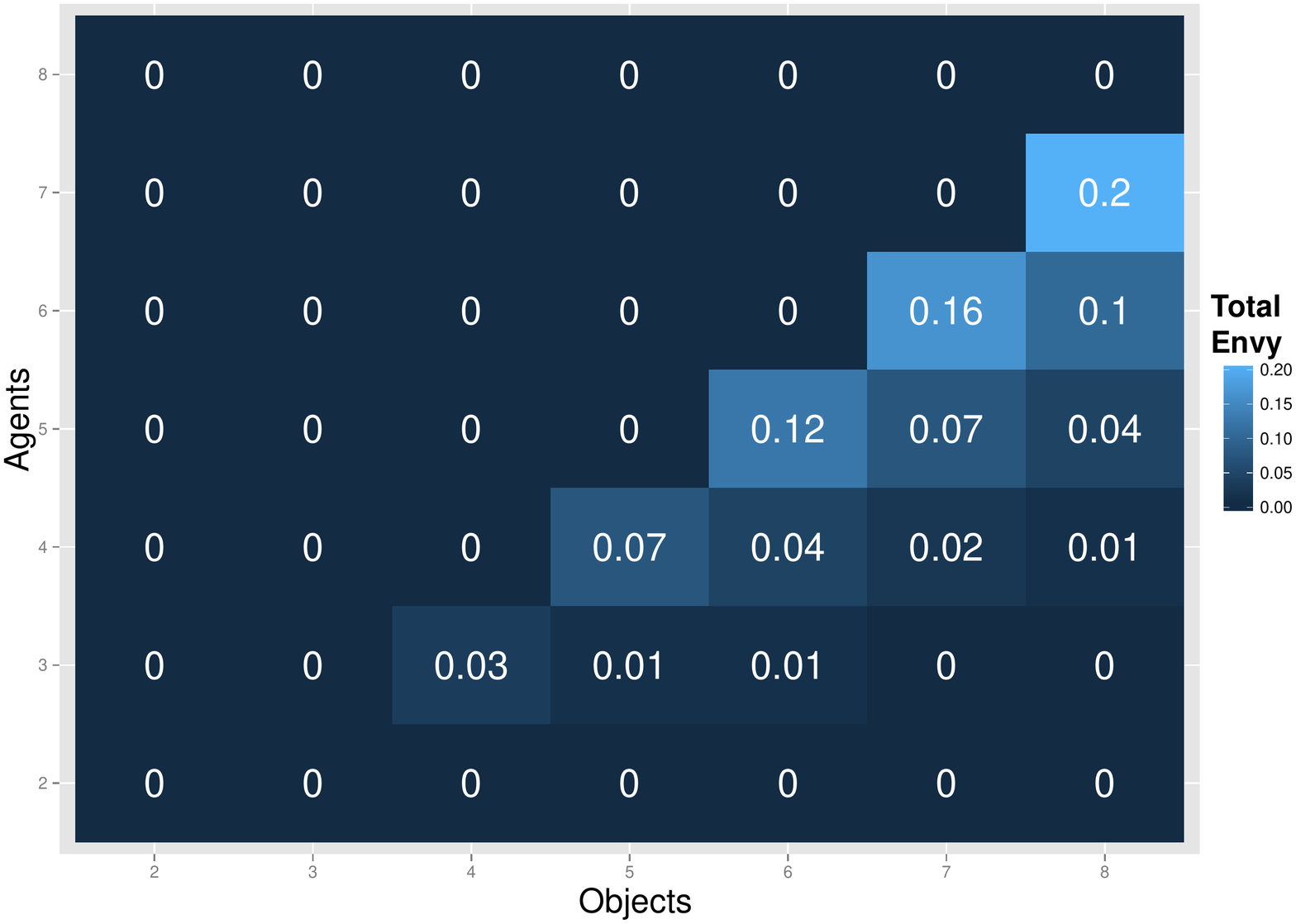}
		\caption{Total envy, $\alpha = 0.5$.}
		\label{}
	\end{subfigure}
	\begin{subfigure}[t]{0.48\textwidth}
		\centering
		\includegraphics[width=\textwidth]{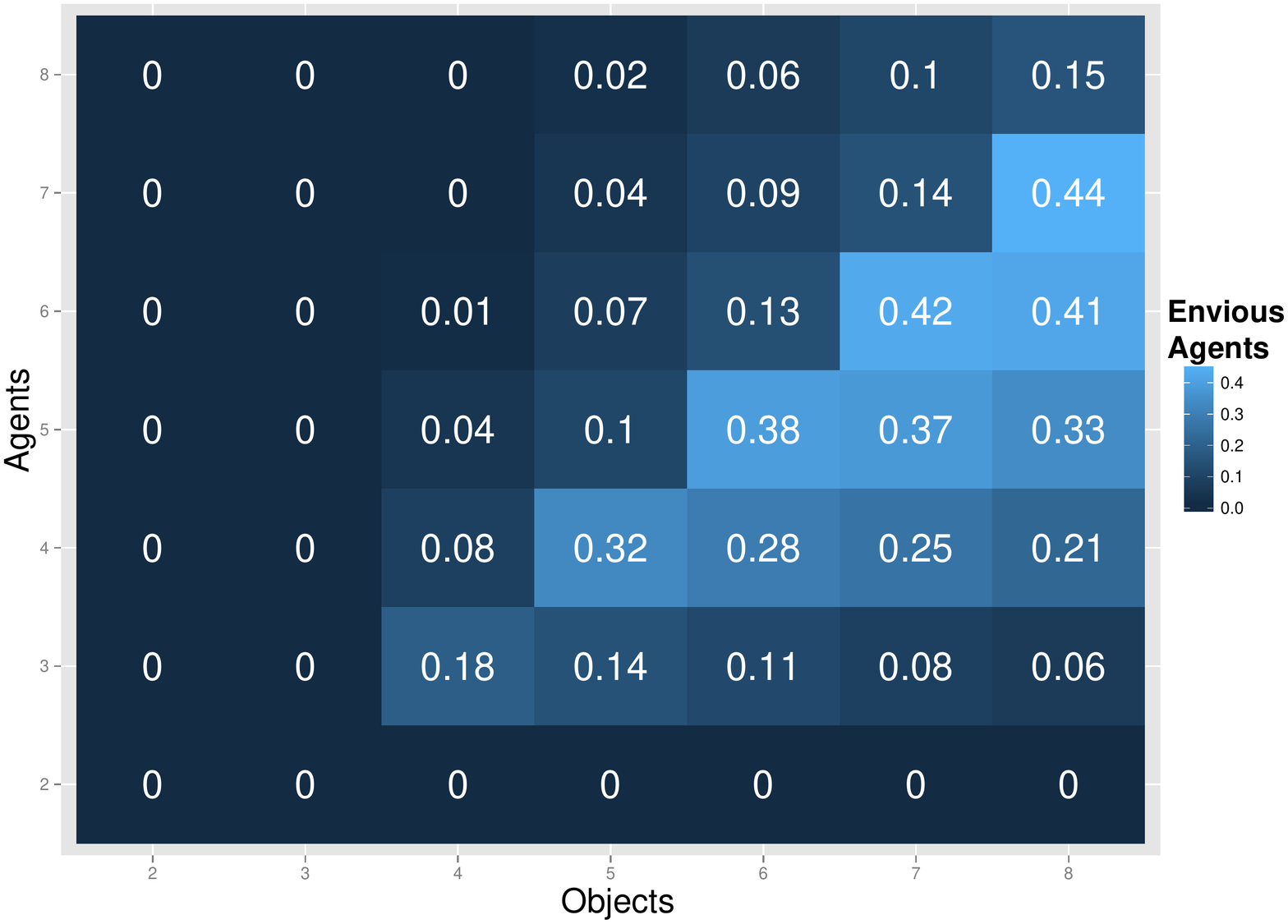}
		\caption{Fraction of envious agents, $\alpha = 1$.}
		\label{}
	\end{subfigure}~
	\begin{subfigure}[t]{0.48\textwidth}
		\centering
		\includegraphics[width=\textwidth]{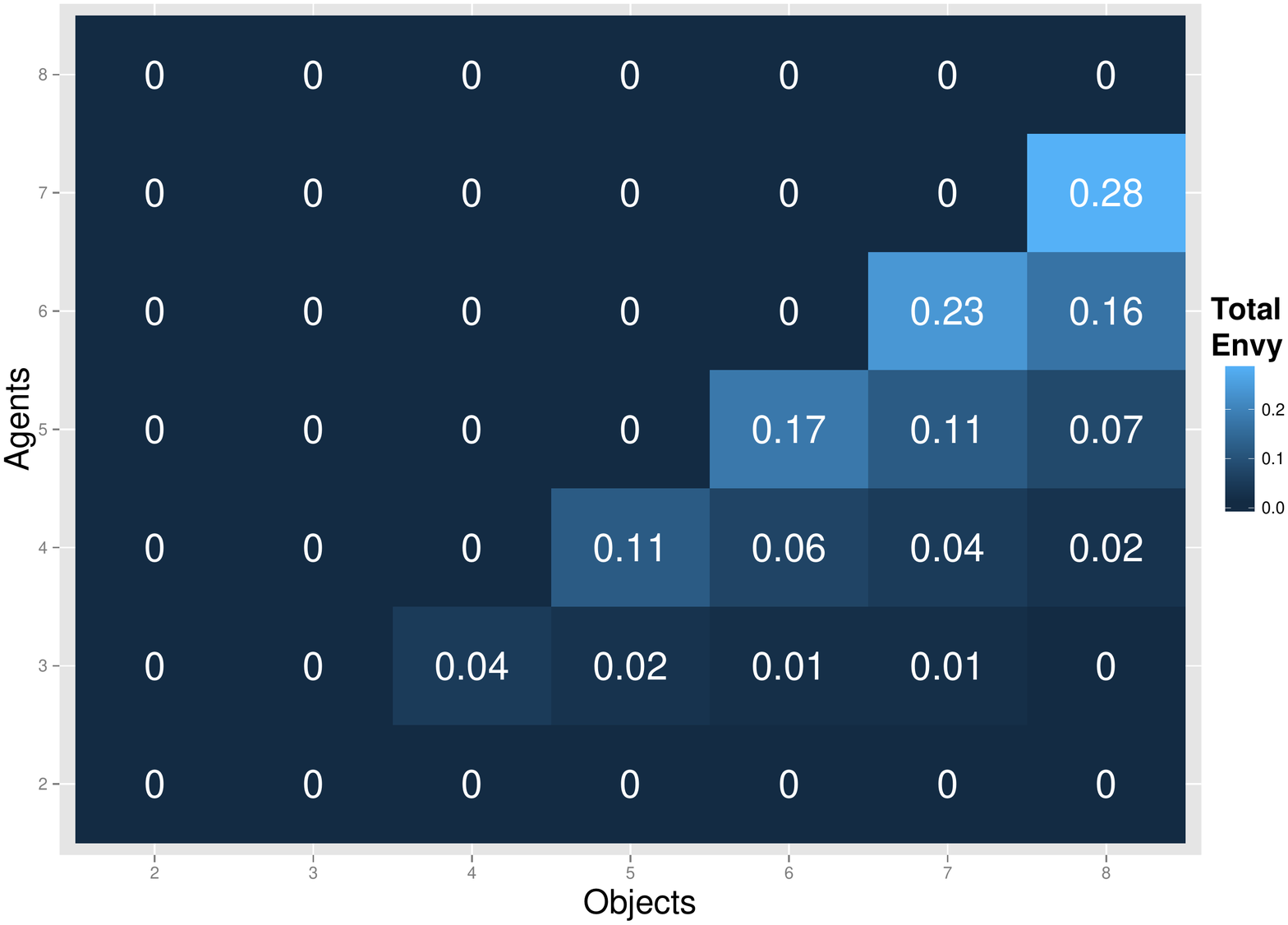}
		\caption{Total envy, $\alpha = 1$.}
		\label{}
	\end{subfigure}
	\begin{subfigure}[t]{0.48\textwidth}
		\centering
		\includegraphics[width=\textwidth]{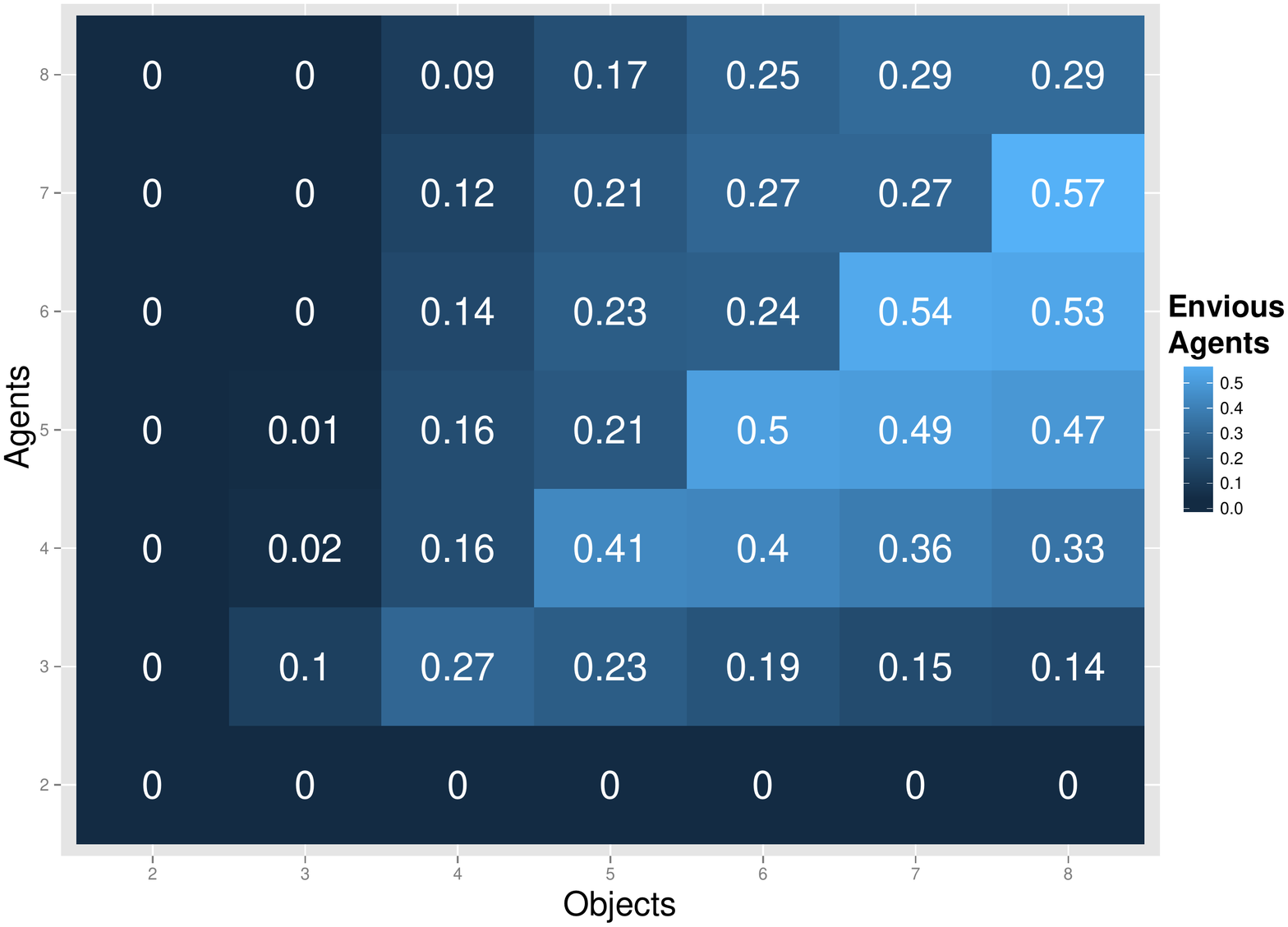}
		\caption{Fraction of envious agents, $\alpha = 2$.}
		\label{}
	\end{subfigure}~
	\begin{subfigure}[t]{0.48\textwidth}
		\centering
		\includegraphics[width=\textwidth]{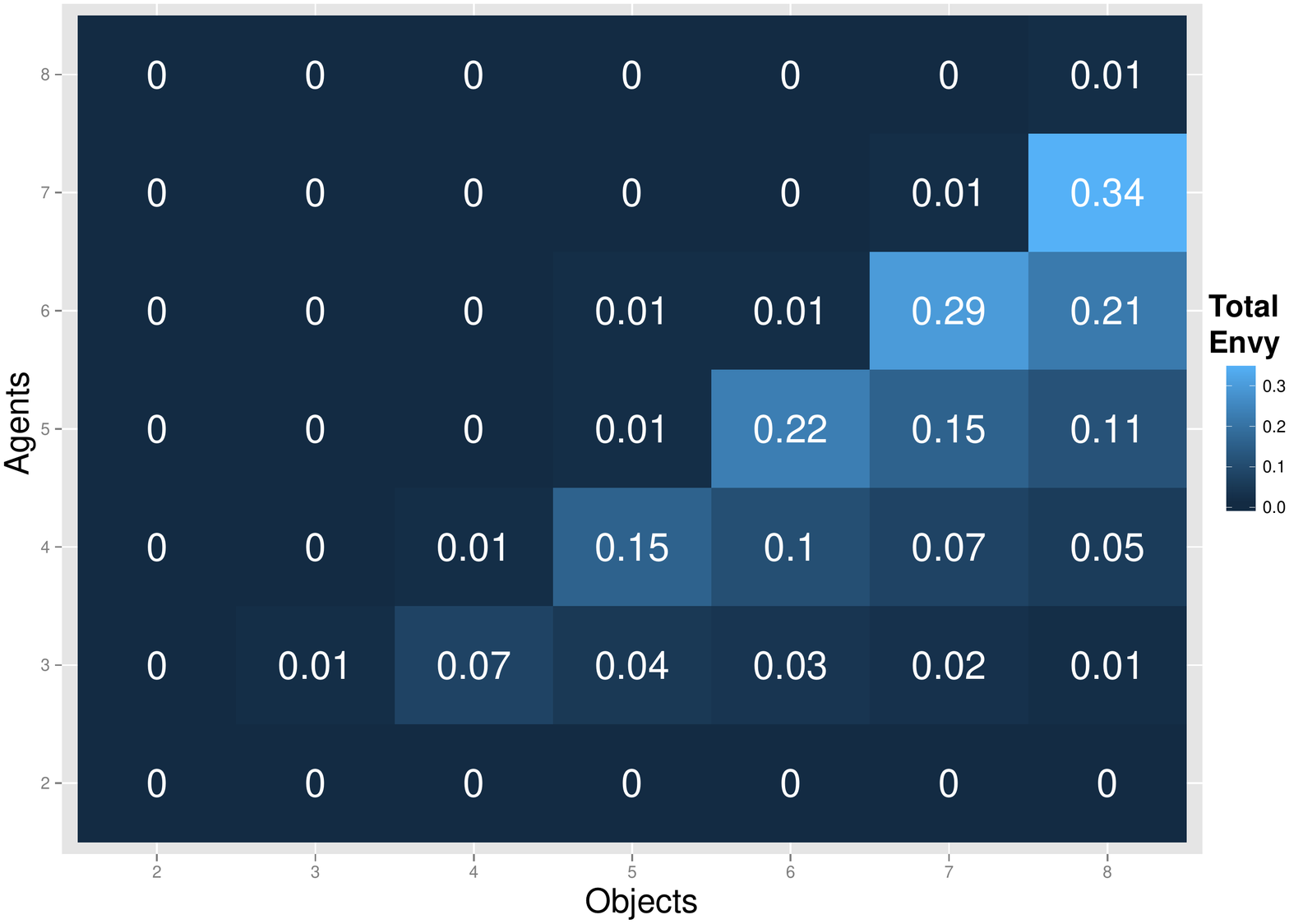}
		\caption{Total envy, $\alpha = 2$.}
		\label{}
	\end{subfigure}
	\caption{The fraction of envious agents and total envy perceived by agents under risk aversion. The total envy is shown up to two decimal points.} 
	\label{fig:envyRiskAverse}
\end{figure*}

\textbf{Envy}:
In Figure \ref{fig:envyRiskAverse} we observe that when $n \geq m$, the fraction of envious agents and total envy grows as $\frac{n}{m} \to 1$. Increasing the risk intensity ($|\alpha|$), the fraction of envious agents increases; however, the total envy among the agents remains considerably low.
For $n < m$, the fraction of envious agents and total envy grows as risk intensity increases. An interesting observation is that envy is maximized when $m = n + 1$, and it decreases as $\frac{m}{n}$ grows. This is mostly due to the choice of using randomized quasi-dictatorial mechanism for implementing RSD where the first dictator receives $m + n -1$ objects and all other agents only receive a single object.
Lastly, we noticed that in all instance where RSD creates envy among the agents, around 25\% of agents bear more than 50\% of envy. That is, few agents feel extremely envious while all other agents are either envyfree or only feel a minimal amount of envy.

\begin{figure*}
	\begin{subfigure}[t]{0.49\textwidth}
		\centering
		\includegraphics[width=\textwidth]{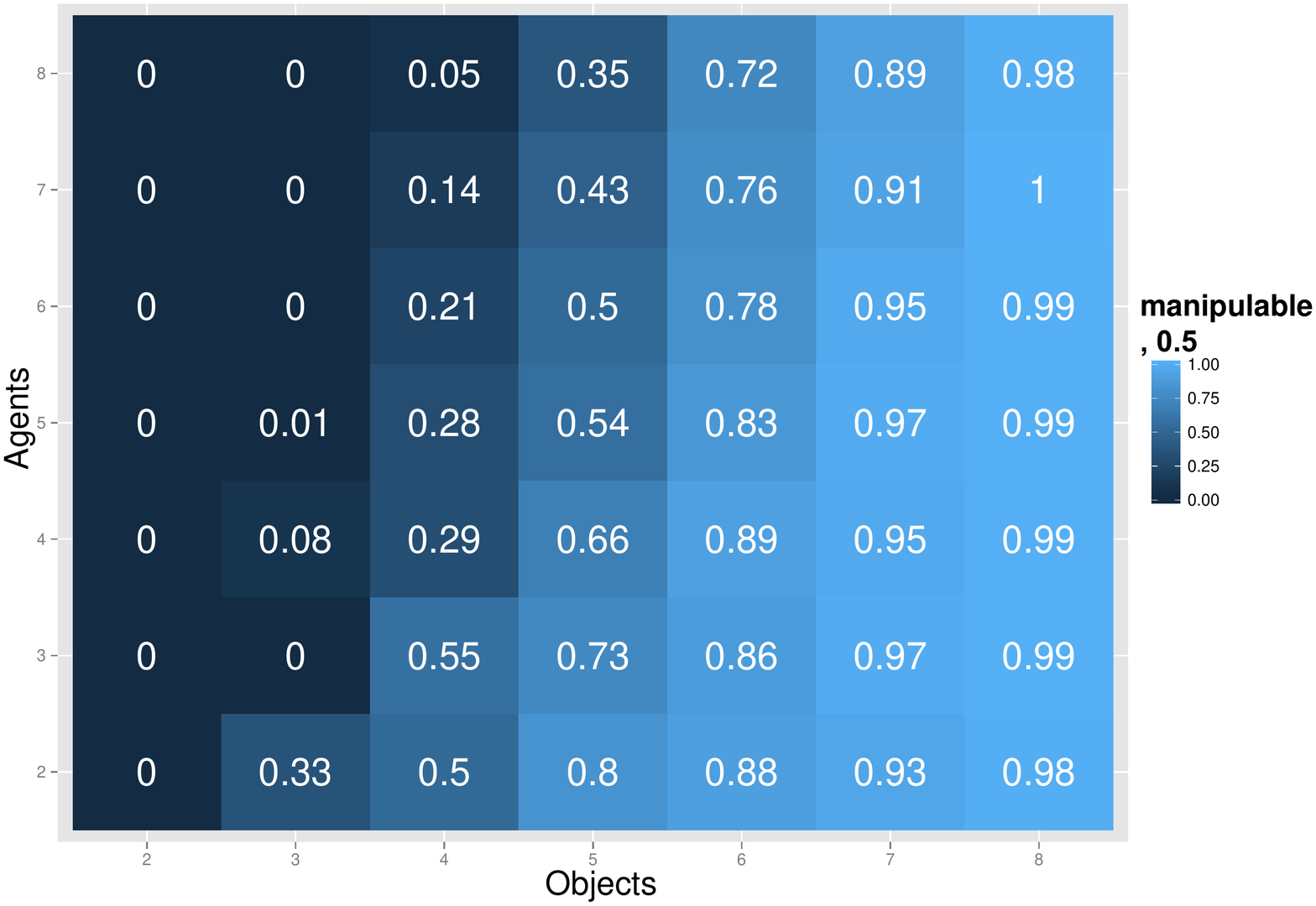}
		\caption{Manipulation, $\alpha = 0.5$.}
		\label{}
	\end{subfigure}~
	\begin{subfigure}[t]{0.49\textwidth}
		\centering
		\includegraphics[width=\textwidth]{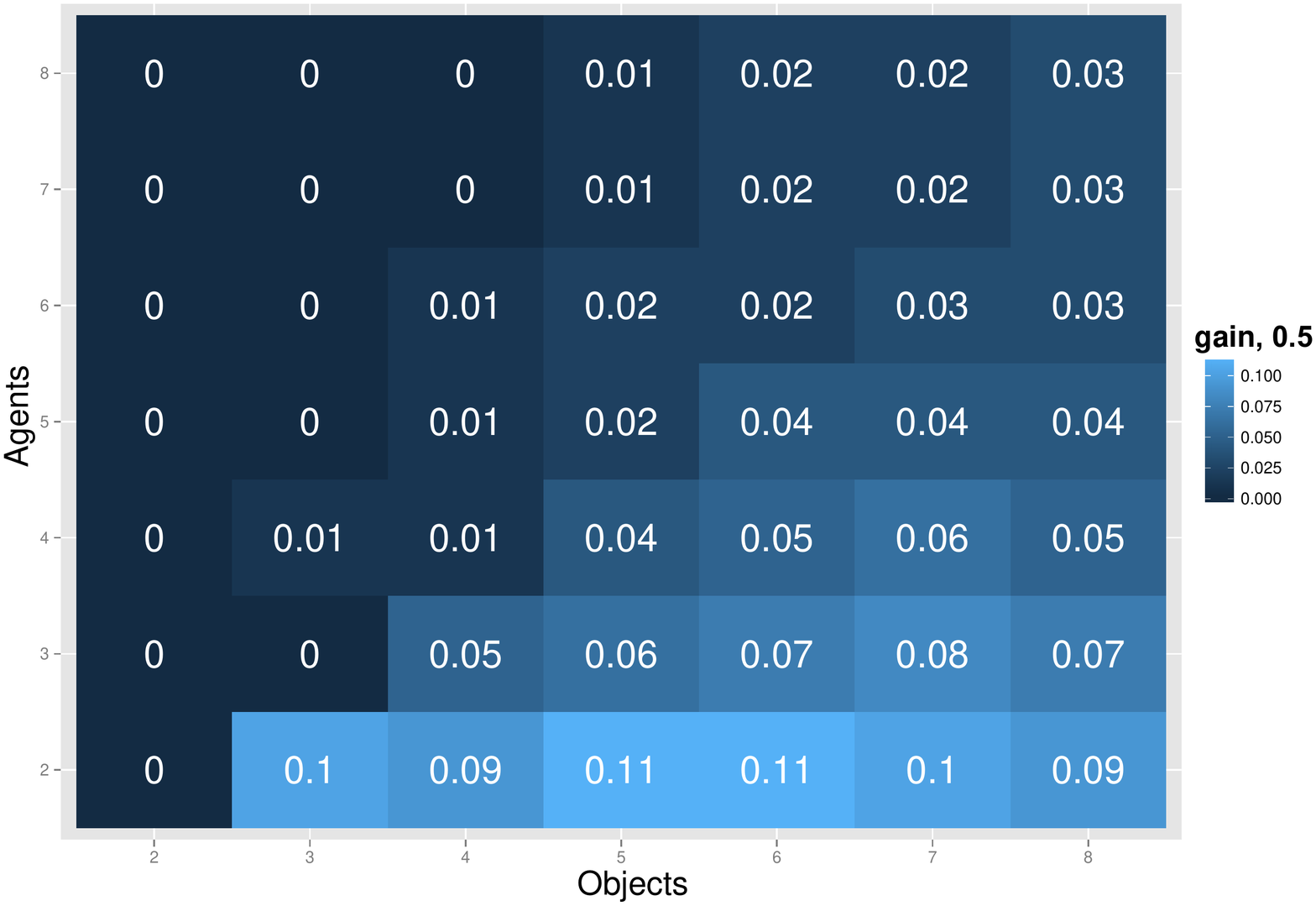}
		\caption{Gain, $\alpha = 0.5$.}
		\label{}
	\end{subfigure}
	\begin{subfigure}[t]{0.49\textwidth}
		\centering
		\includegraphics[width=\textwidth]{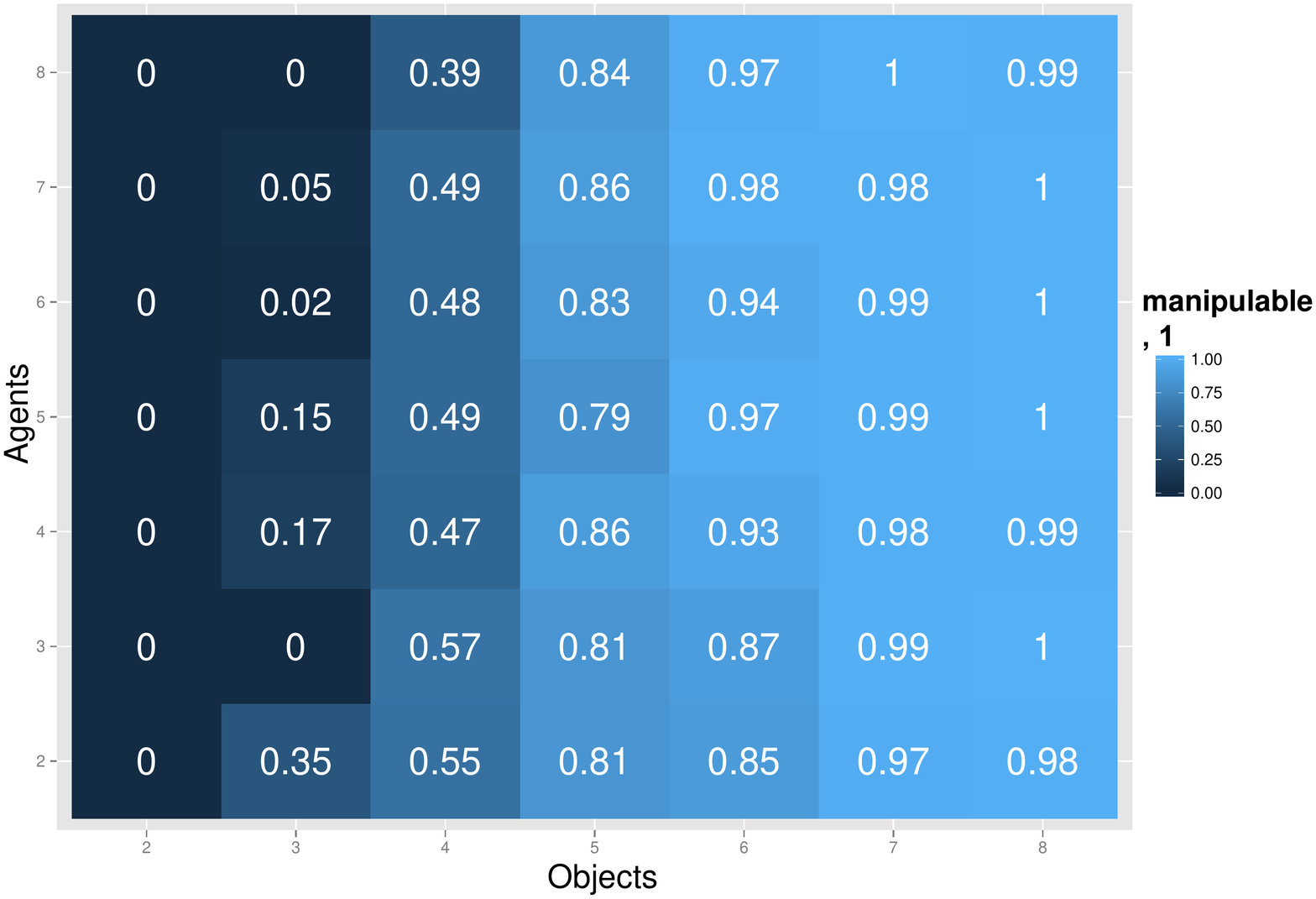}
		\caption{Manipulation, $\alpha = 1$.}
		\label{}
	\end{subfigure}~
	\begin{subfigure}[t]{0.49\textwidth}
		\centering
		\includegraphics[width=\textwidth]{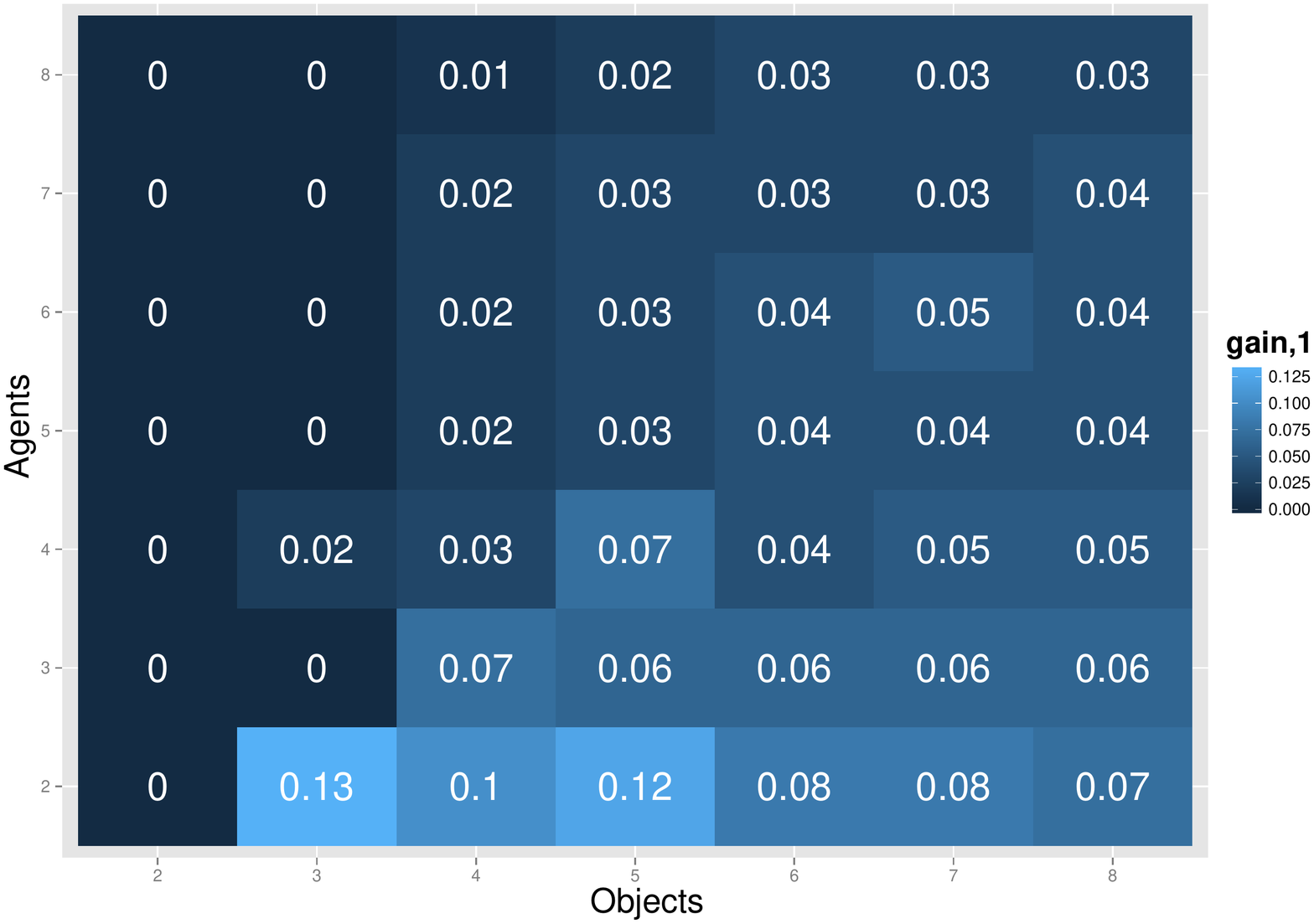}
		\caption{Gain, $\alpha = 1$.}
		\label{}
	\end{subfigure}
	\begin{subfigure}[t]{0.49\textwidth}
		\includegraphics[width=\textwidth]{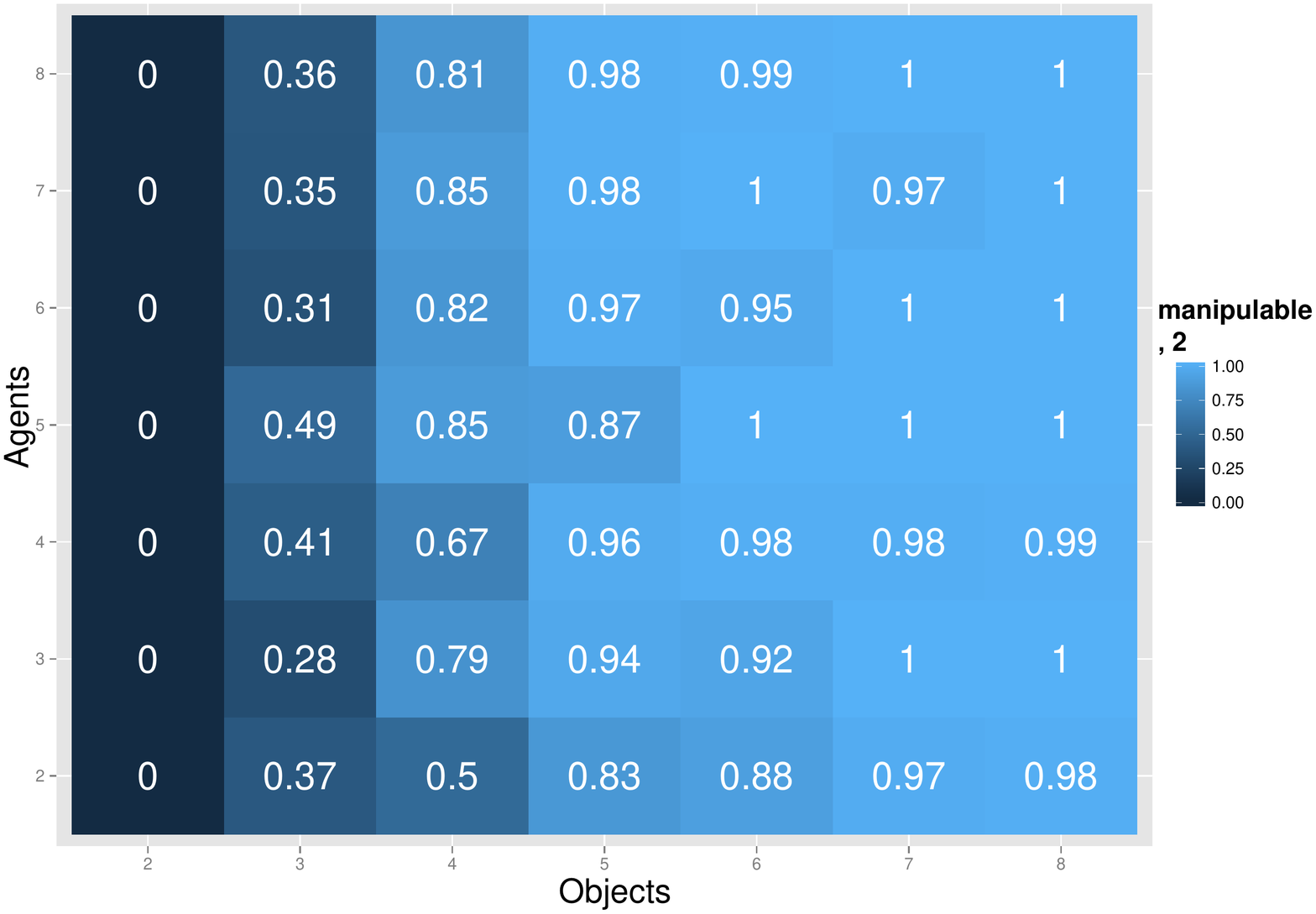}
		\caption{Manipulation, $\alpha = 2$.}
		\label{}
	\end{subfigure}~
	\begin{subfigure}[t]{0.49\textwidth}
		\includegraphics[width=\textwidth]{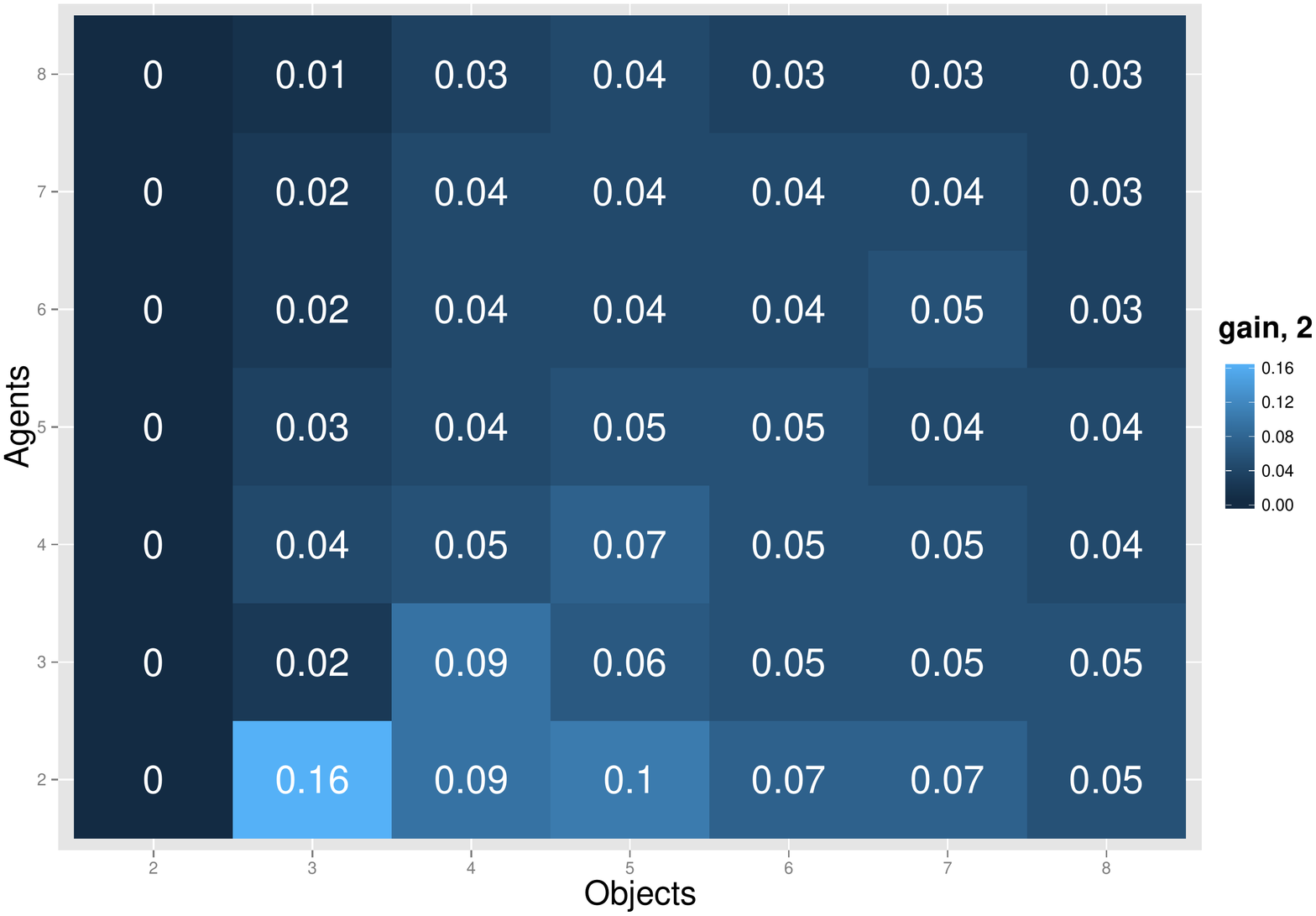}
		\caption{Gain, $\alpha = 2$.}
		\label{}
	\end{subfigure}
	\caption{The fraction of manipulable instances and manipulation gain of PS under risk aversion.} 
	\label{fig:manipRiskAverse}
\end{figure*}

\textbf{Manipulability}: 
Figures \ref{fig:manipRiskAverse} illustrates the manipulability of the PS assignments when agents have risk averse preferences. 
The fraction of manipulable profiles rapidly goes to 1 as $\frac{m}{n}$ grows. Similarly, as agents become more risk averse ($\alpha$ increases) the fraction of manipulable profiles goes to 1 and the manipulation gain increases.

\section{Other Ranking Distribution Models} \label{sec:dist}

In this section, we use variations of two statistical models that are commonly used to capture realistic preference distributions in a population of players. 
Considerable work in computational social choice and machine learning has exploited these statistical models to capture the distribution of ranking preferences in a population of agents \cite{lu2011learning,lu2011robust,Aziz:2015:EUP:2832249.2832402}. 
We will focus on Mallows Models and Polya-Eggenberger Urn Models (Urn)  \cite{mallows1957non,marden1996analyzing,berg1985paradox}.

In Mallows models the population is distributed around a reference ranking proportional to the Kendall-Tau (KT) distance \cite{kendall1938new,kendall1948rank}. Henceforth, preferences closer to the reference ranking are more likely to appear in the population. 
In other words, agents' preferences deviate from the reference ranking with decreasing probability as rankings move away from the reference. 
Mallows models are parametrized by a reference ranking and a dispersion parameter. Formally, given a reference ranking ($\hat{\succ}$) and a dispersion parameter ($\phi$), we have
\begin{gather}
 	P(\succ) = 	\frac{1}{Z} \phi^{KT(\succ, \hat{\succ})}, \quad	\forall \succ \in \mathcal{P}
\end{gather}

where $Z = 1 \cdot (1+\phi) \cdot (1+\phi+\phi^{2}) \ldots (1+\ldots+ \phi^{m-1})$.
When $\phi = 1$ the Mallows model is equivalent to the uniform distribution, and when $\phi = 0$ the distribution mass is entirely on the reference ranking.
It is also possible for an agent population to have multiple references. In these cases, Mallows Mixture models are parametrized by a set of ranking reference with their corresponding dispersion parameters. 


In the Urn distribution model, with every random selection of a preference order the probability of this preference order being selected in subsequent samples increases. Intuitively, we can think of a collection of $m!$ preference orderings and every time an ordering is sampled uniformly from this collection, it will be replaced by two copies of the same preference ordering. 

We use the PrefLib Toolkit \cite{MaWa13a} to generate Mallows and Urn distribution models. In our experiments, we used Mallows model with one reference ranking as well as Mallows mixture models with five reference rankings. Every data point in the figures is averaged over 1,000 samples.


In general, the same patterns under the uniform preference distribution hold for various numbers of agents and objects and when varying the risk parameter and utility functions. 

\begin{figure*}
	\centering
	\begin{subfigure}[t]{0.49\textwidth}
		\centering
		\includegraphics[width=\textwidth]{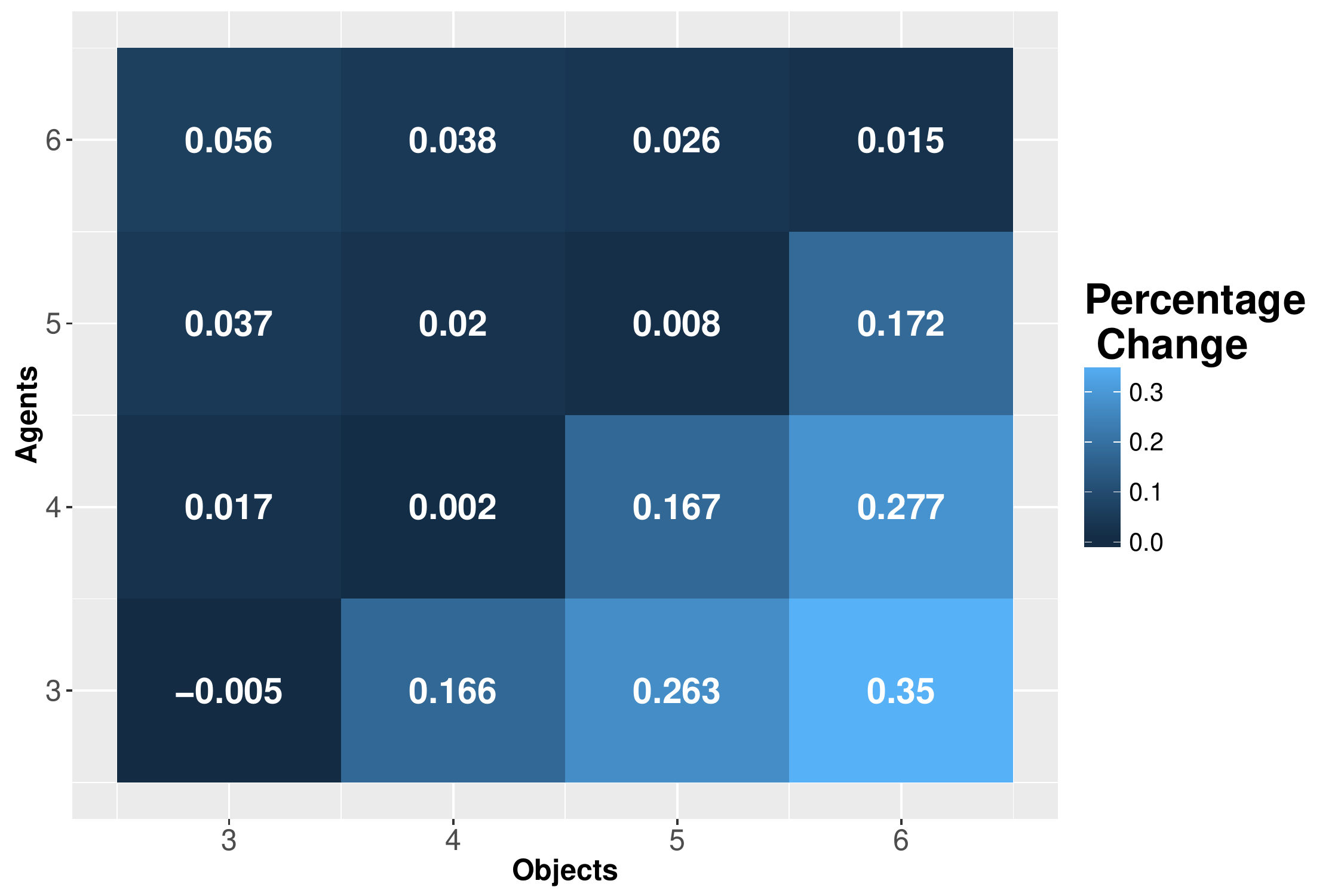}
		\caption{Risk seeking, $\alpha = -1$.}
		\label{swt5-1}
	\end{subfigure}~
	\begin{subfigure}[t]{0.49\textwidth}
		\centering
		\includegraphics[width=\textwidth]{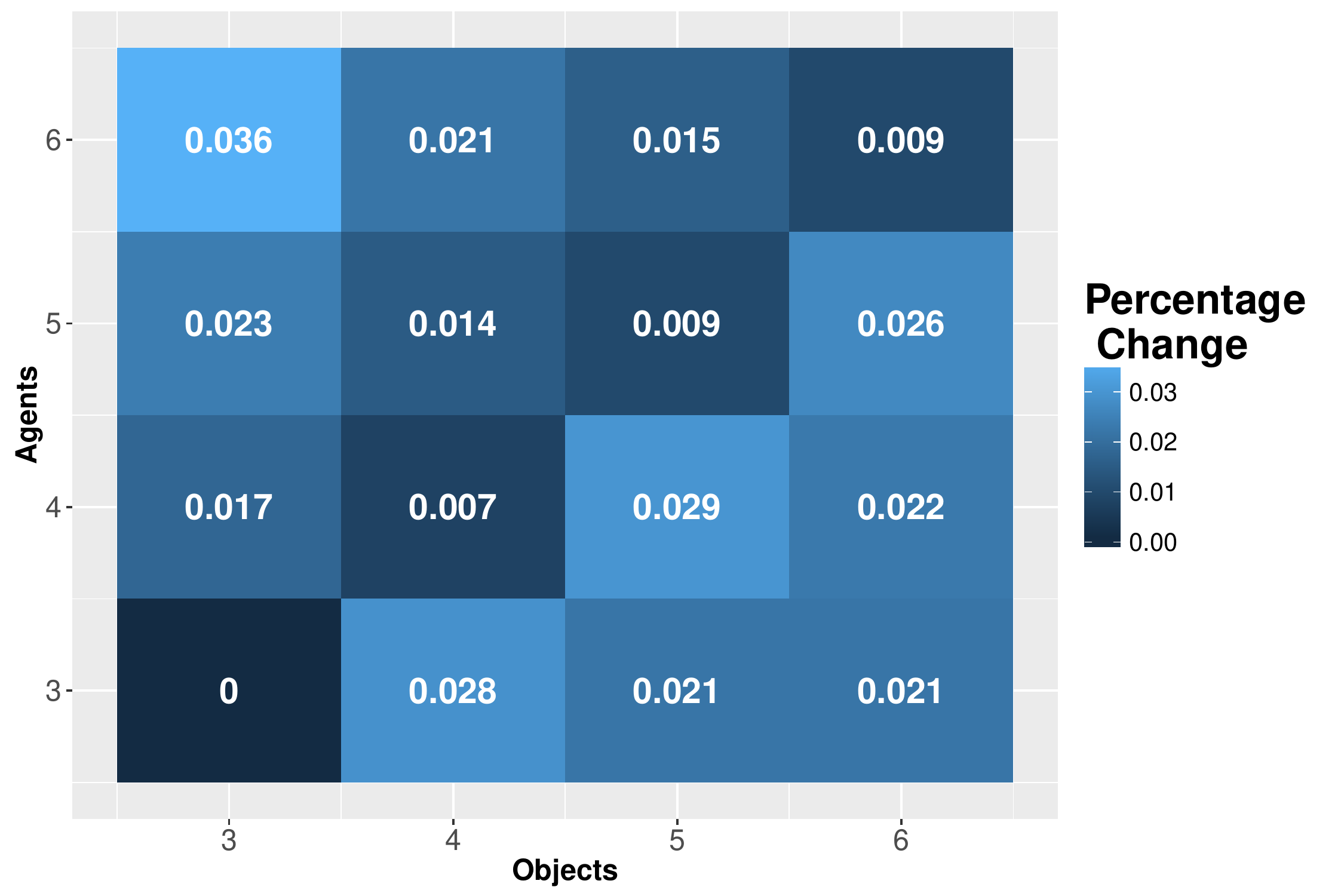}
		\caption{Risk averse, $\alpha = 1$.}
		\label{swt51}
	\end{subfigure}
	\caption{The percentage change in social welfare under the pure Mallows distribution with single reference.} 
	\label{fig:MallowsChange1}
\end{figure*}

\begin{figure*}
	\centering
	\begin{subfigure}[t]{0.49\textwidth}
		\centering
		\includegraphics[width=\textwidth]{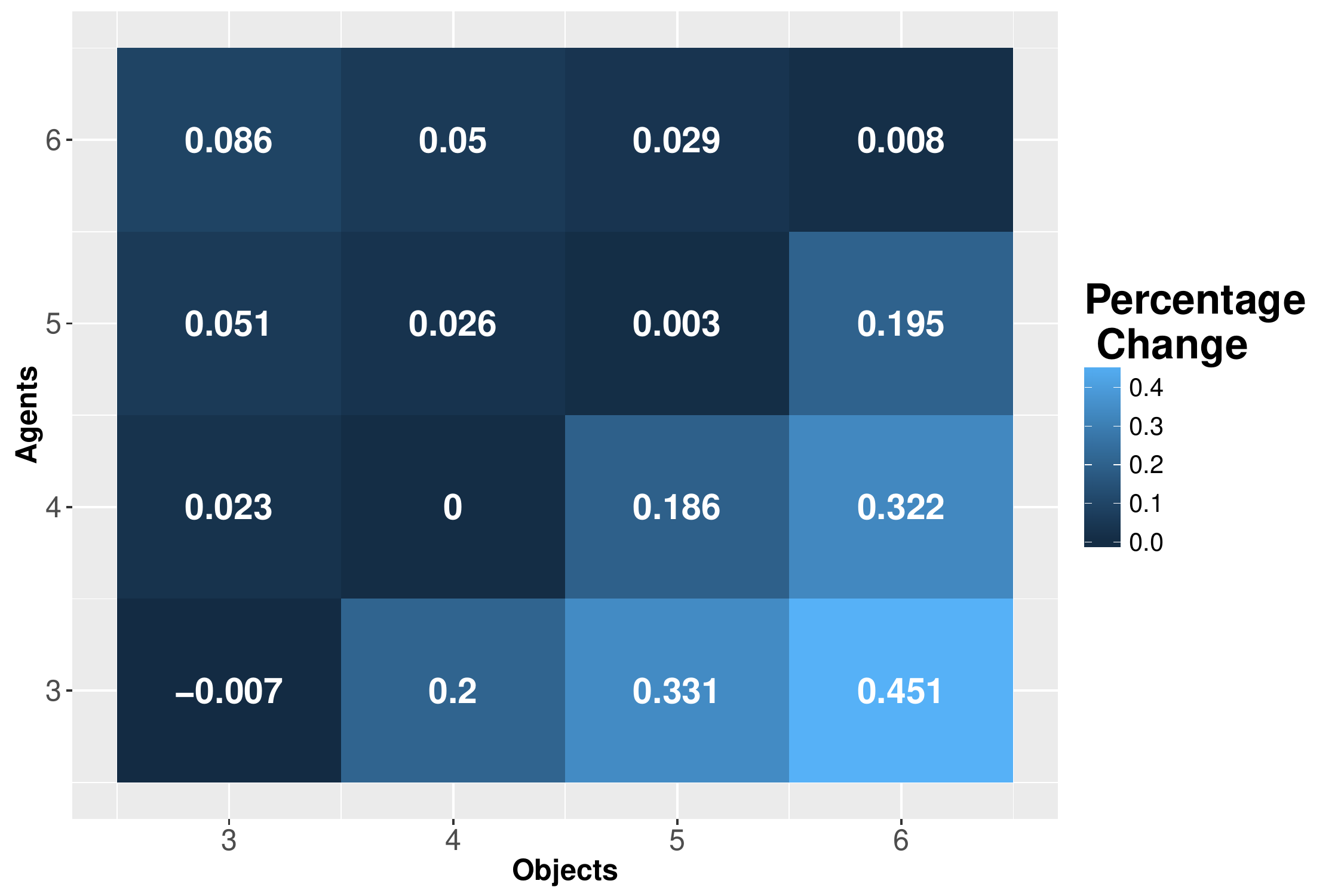}
		\caption{Risk seeking, $\alpha = -1$.}
		\label{swt4-1}
	\end{subfigure}~
	\begin{subfigure}[t]{0.49\textwidth}
		\centering
		\includegraphics[width=\textwidth]{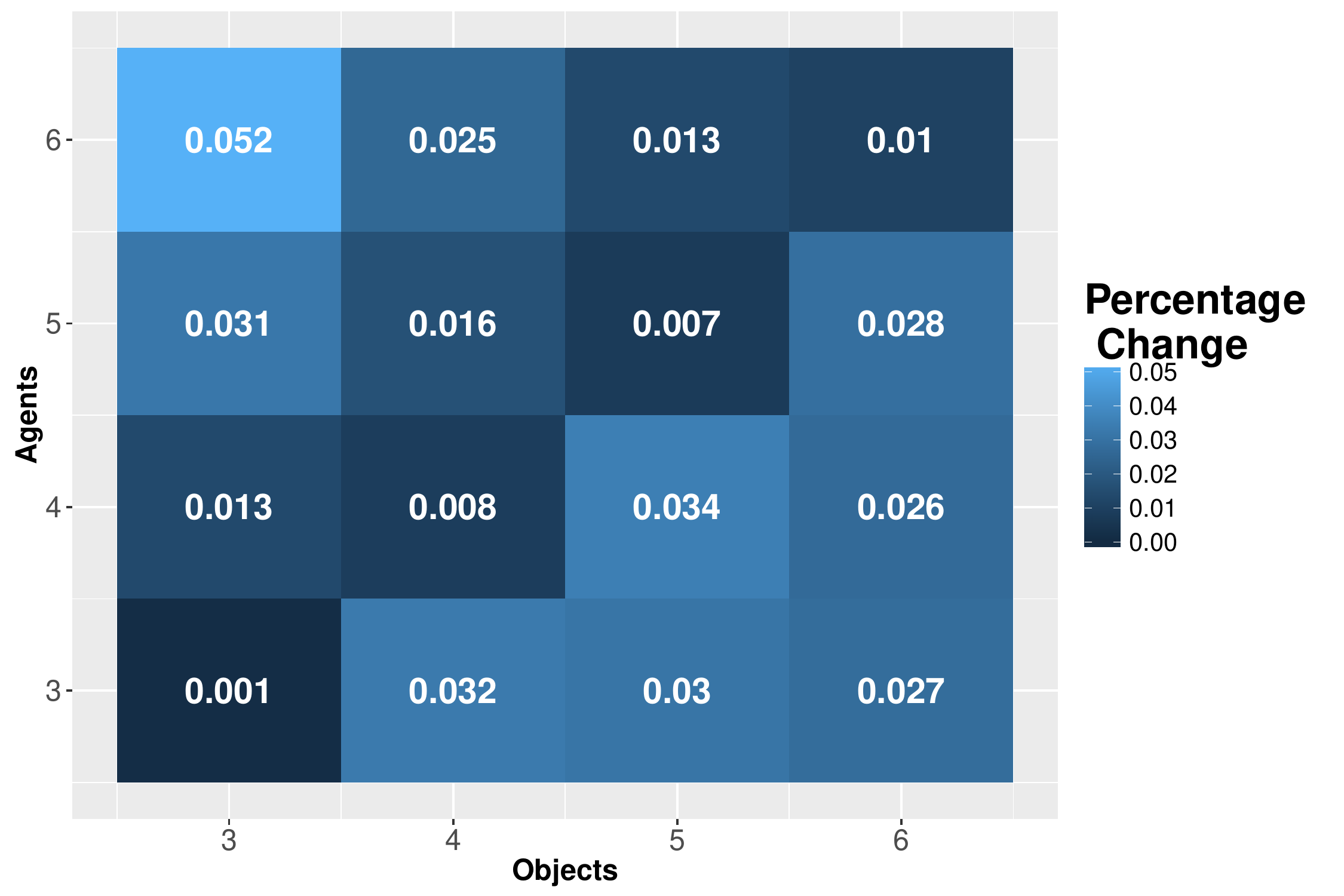}
		\caption{Risk averse, $\alpha = 1$.}
		\label{swt41}
	\end{subfigure}
	\begin{subfigure}[t]{0.49\textwidth}
		\centering
		\includegraphics[width=\textwidth]{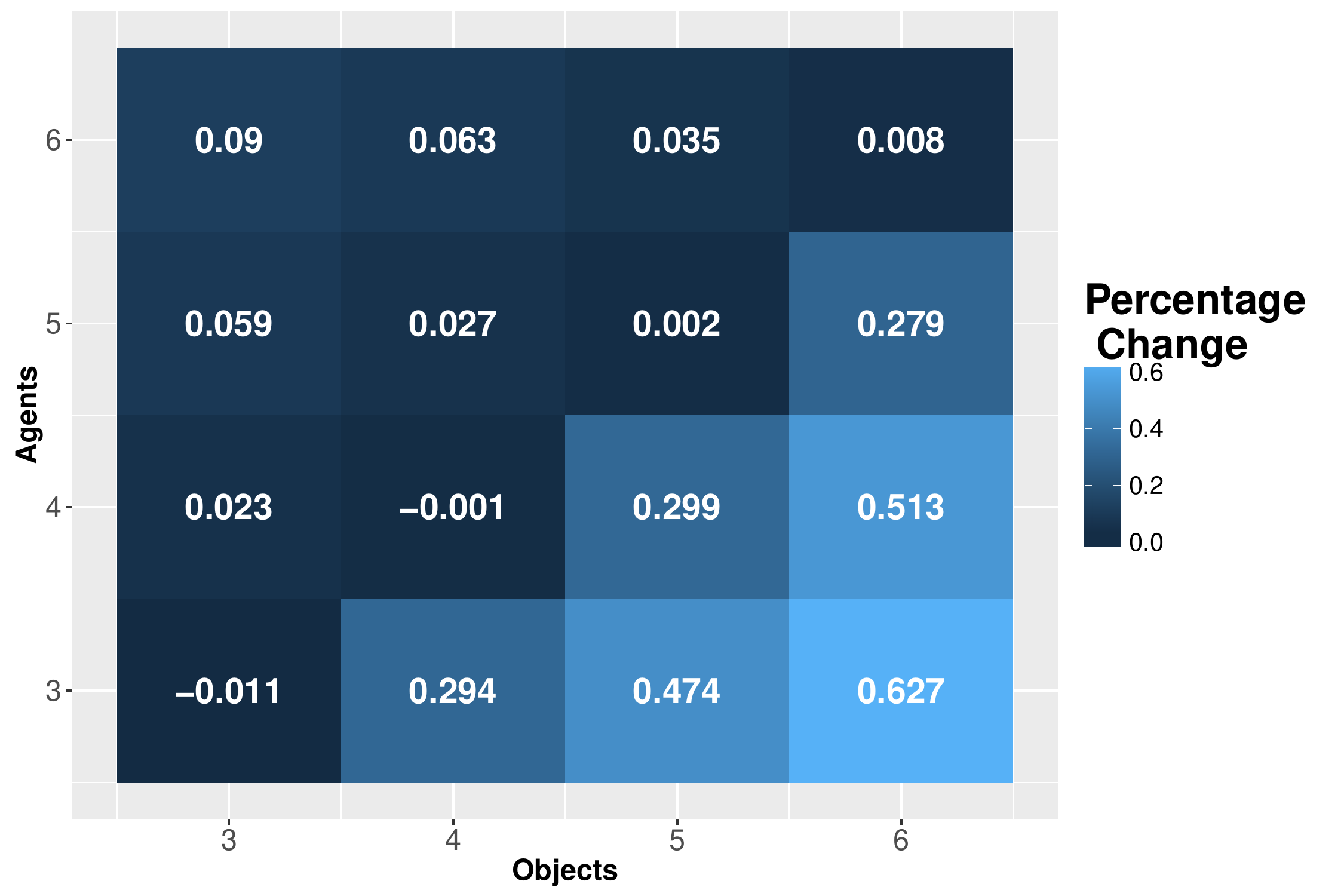}
		\caption{Risk seeking, $\alpha = -2$.}
		\label{swt4-2}
	\end{subfigure}~
	\begin{subfigure}[t]{0.49\textwidth}
		\centering
		\includegraphics[width=\textwidth]{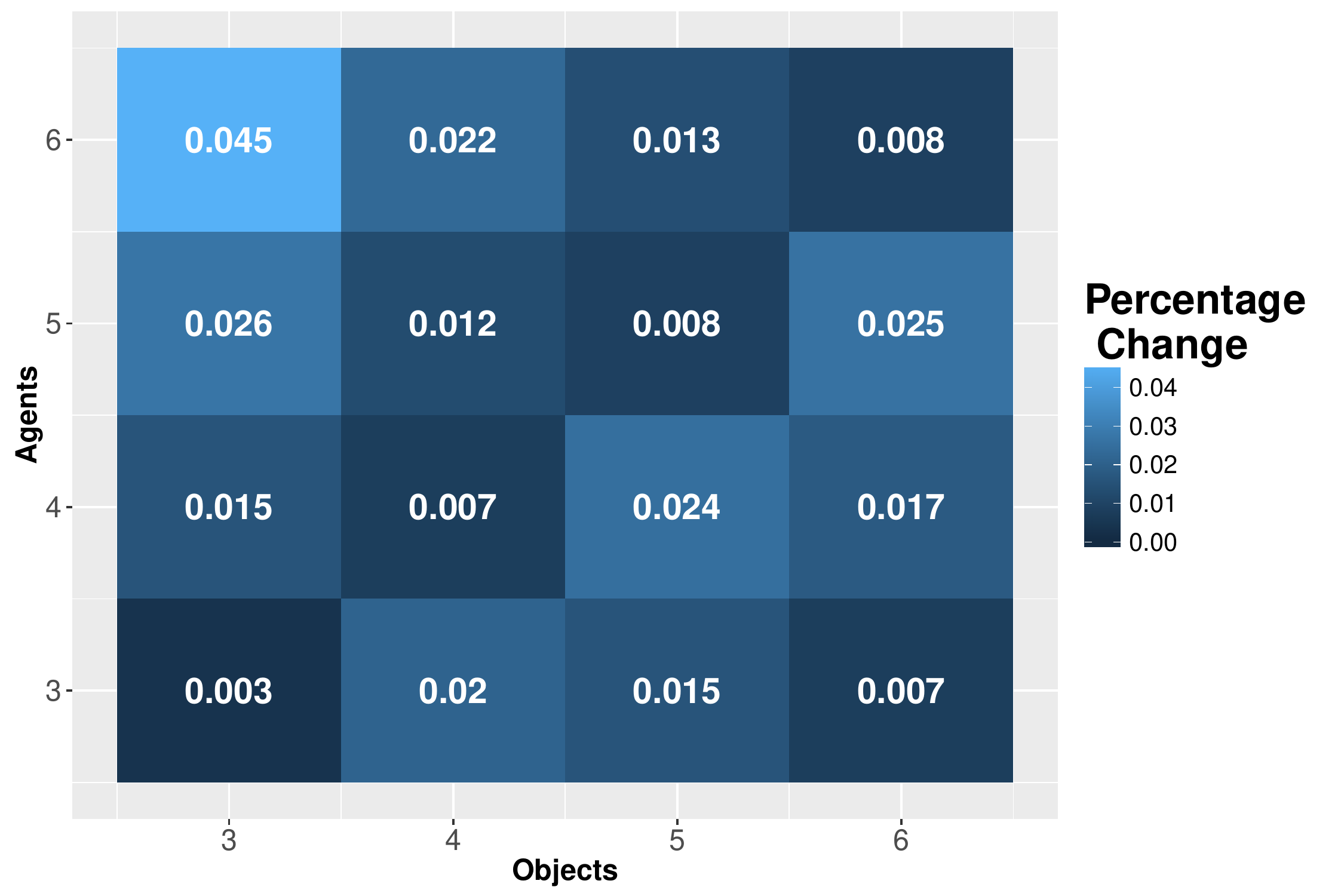}
		\caption{Risk averse, $\alpha = -2$.}
		\label{swt42}
	\end{subfigure}
	\caption{
		The percentage change in social welfare under the Mallows Mixture distribution with five references.} 
	\label{fig:MallowsChange5}
\end{figure*}

\textbf{Social welfare}:
Figures \ref{fig:MallowsChange1} and \ref{fig:MallowsChange5} show the results of our simulation for social welfare when agents' preferences are drawn from Mallows models. These results are consistent with pure Mallows models with single reference rankings (Figure \ref{fig:MallowsChange1}) as well as Mallows mixture models with five references (Figure \ref{fig:MallowsChange5}).
The percentage change in social welfare is infinitesimal when $n \geq m$ for both risk averse and risk seeking populations. Under risk aversion, this percentage change in social welfare remains small when $n < m$. Similar to the uniform populations, the negative values show that in some cases, particularly when $n=m$, RSD assignments outperform those under PS assignments.


\begin{figure*}
	\begin{subfigure}[t]{0.49\textwidth}
		\centering
		\includegraphics[width=\textwidth]{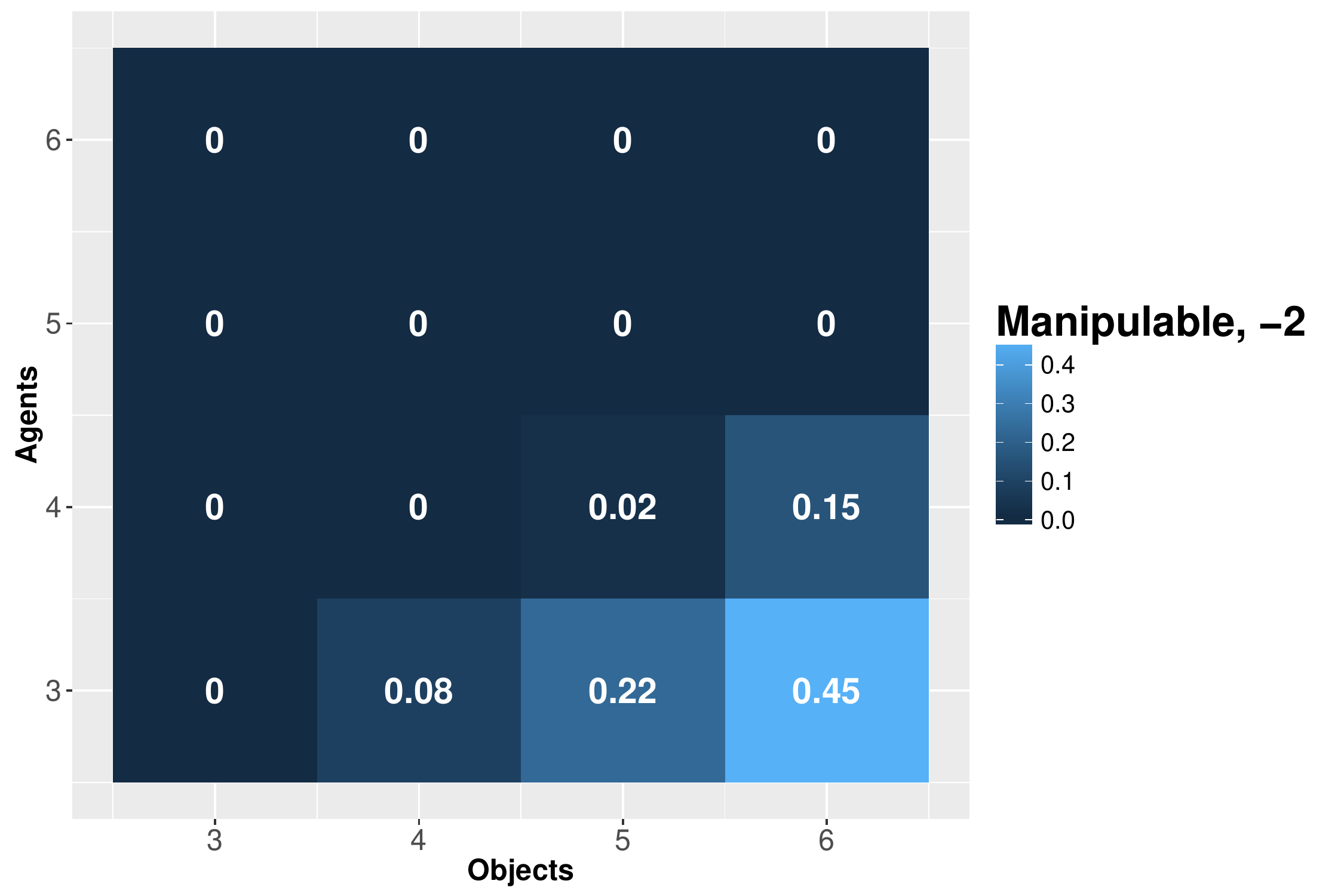}
		\caption{Manipulation, Mallows Mixture, $\alpha = -2$.}
		\label{}
	\end{subfigure}~
	\begin{subfigure}[t]{0.49\textwidth}
		\centering
		\includegraphics[width=\textwidth]{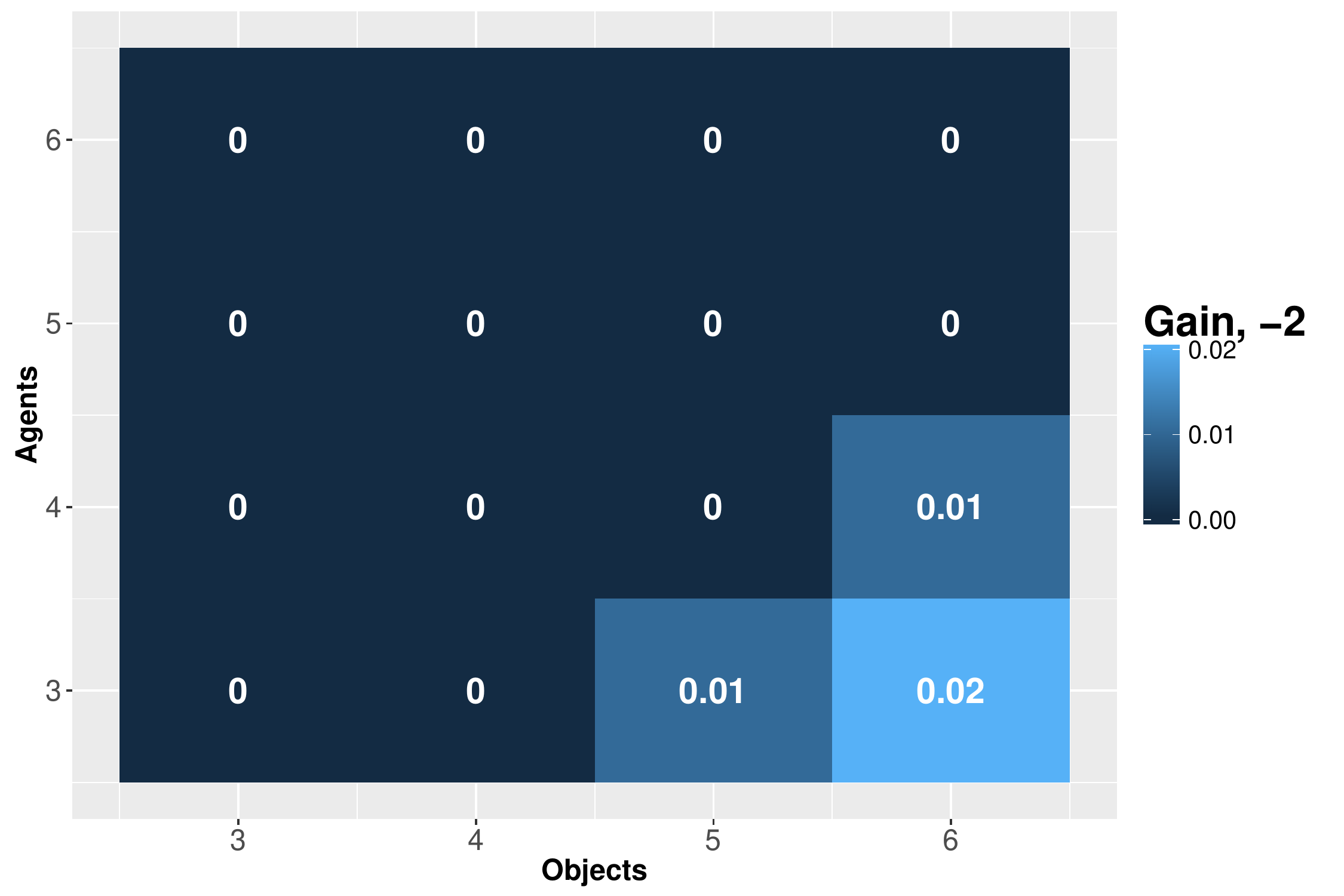}
		\caption{Gain, Mallows Mixture, $\alpha = -2$.}
		\label{}
	\end{subfigure}
	\begin{subfigure}[t]{0.49\textwidth}
		\centering
		\includegraphics[width=\textwidth]{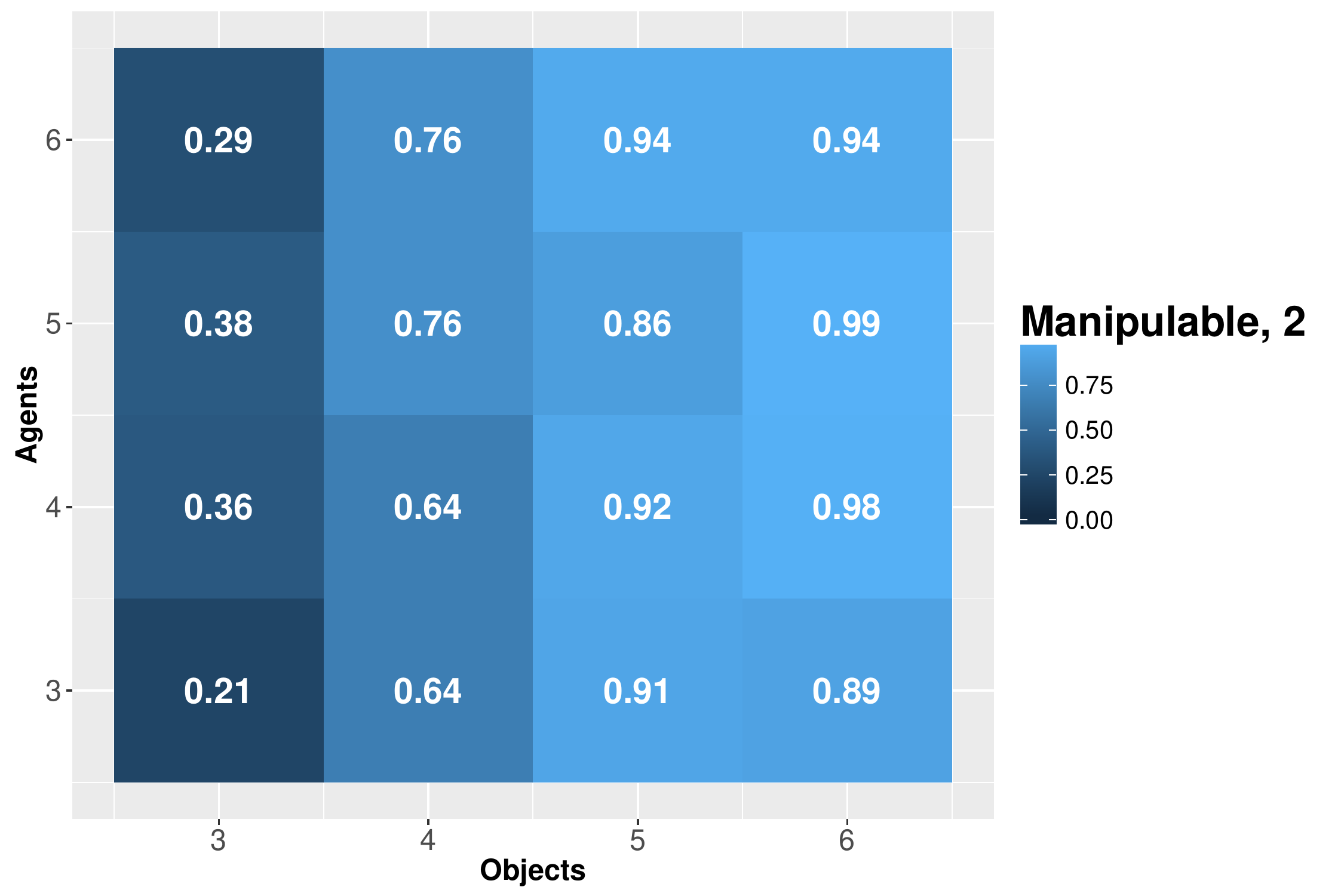}
		\caption{Manipulation, Mallows Mixture, $\alpha = 2$.}
		\label{}
	\end{subfigure}~
	\begin{subfigure}[t]{0.49\textwidth}
		\centering
		\includegraphics[width=\textwidth]{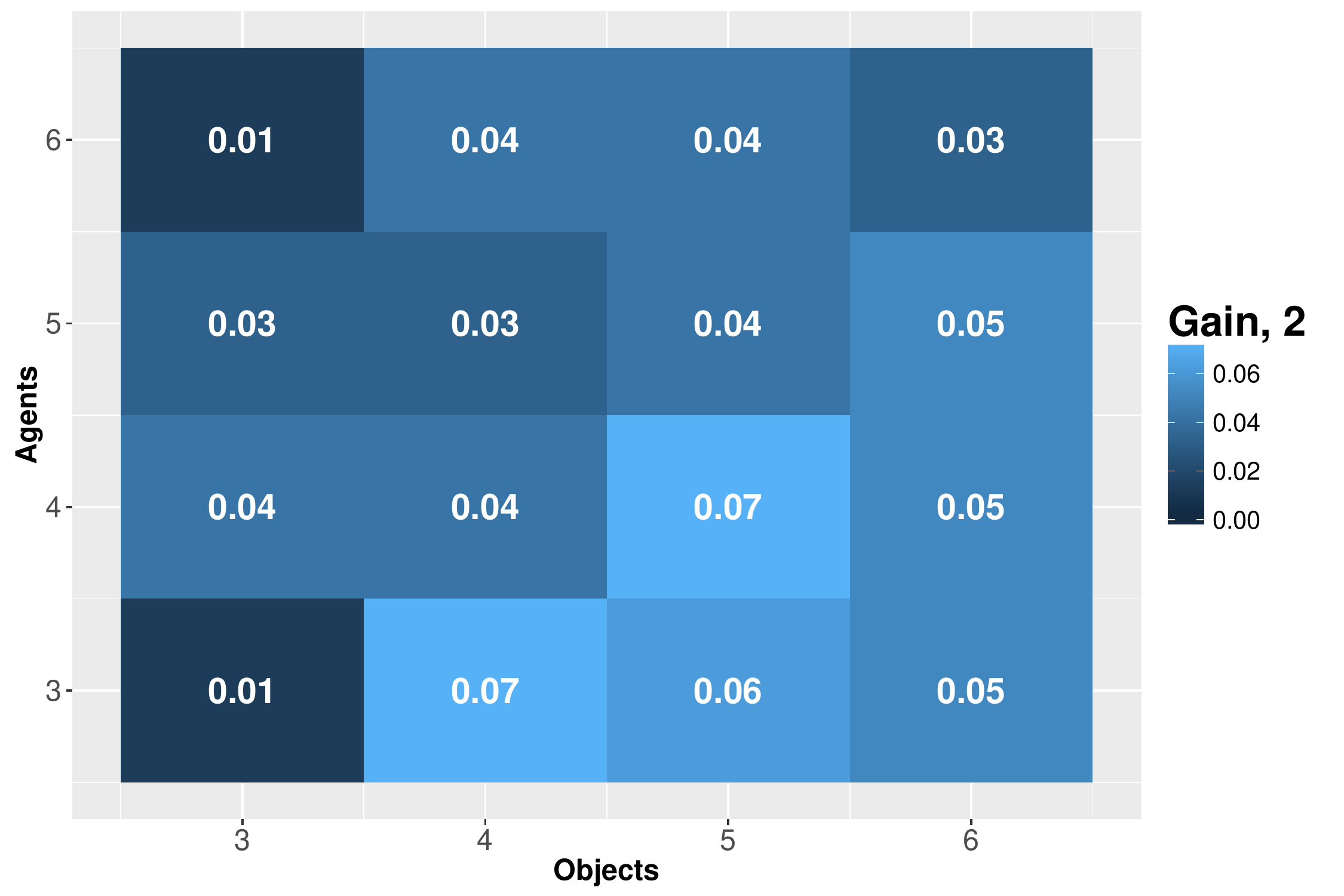}
		\caption{Gain, Mallows Mixture, $\alpha = 2$.}
		\label{}
	\end{subfigure}
	\caption{The fraction of manipulable instances and manipulation gain of PS under the Mallows mixture model with five references.} 
	\label{fig:manipMallow5}
\end{figure*}

\textbf{Manipulability}:
The PS assignments remain very susceptible to manipulation even under more natural assumptions on how the preferences are distributed. Figures \ref{fig:manipMallow5} illustrates the manipulability of the PS assignments and the average gain from manipulation when agents are drawn from Mallows mixture models under risk averse and risk seeking attitudes.  

\begin{figure*}
	\begin{subfigure}[t]{0.49\textwidth}
		\centering
		\includegraphics[width=\textwidth]{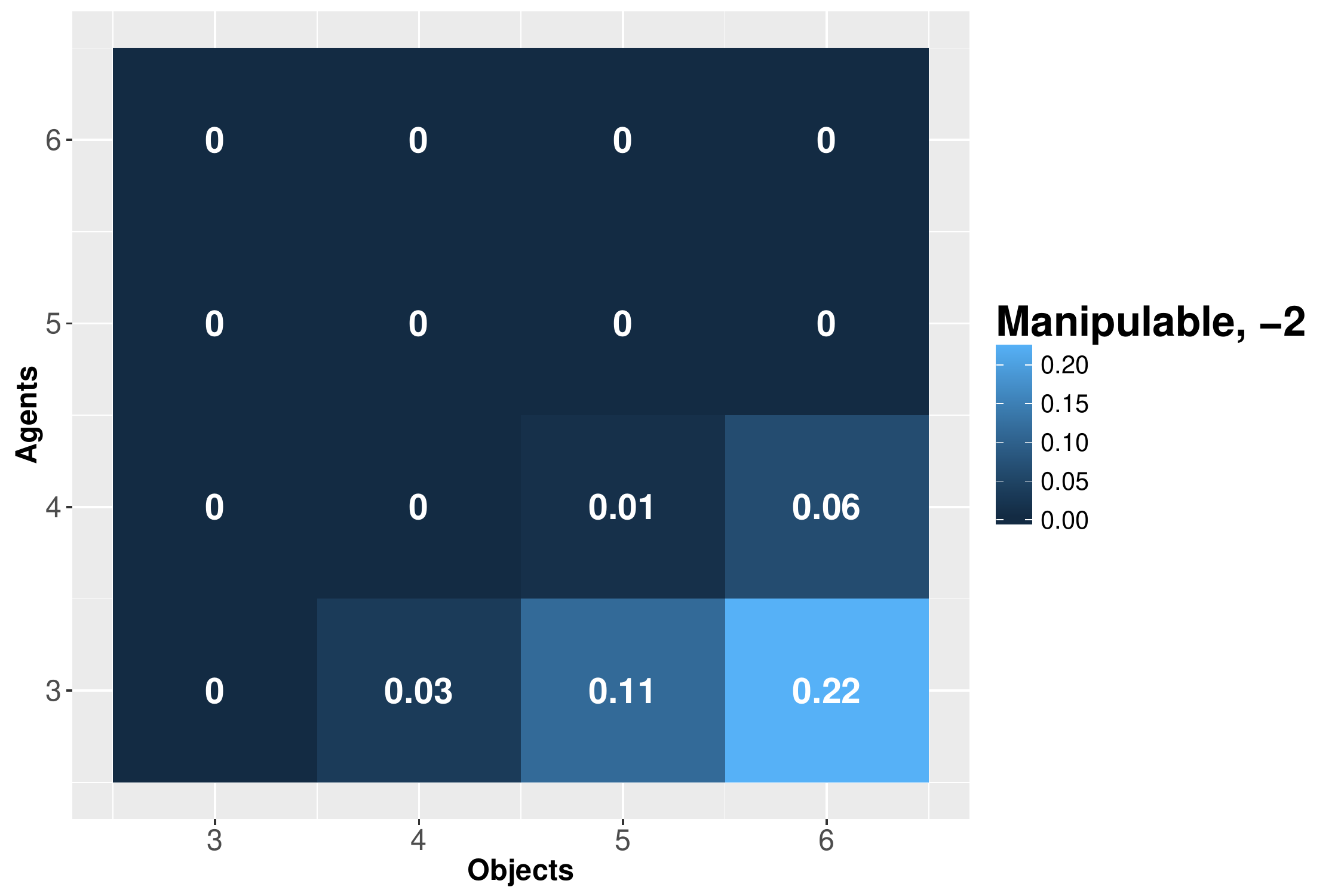}
		\caption{Manipulation, Mallows, $\alpha = -2$.}
		\label{}
	\end{subfigure}~
	\begin{subfigure}[t]{0.49\textwidth}
		\centering
		\includegraphics[width=\textwidth]{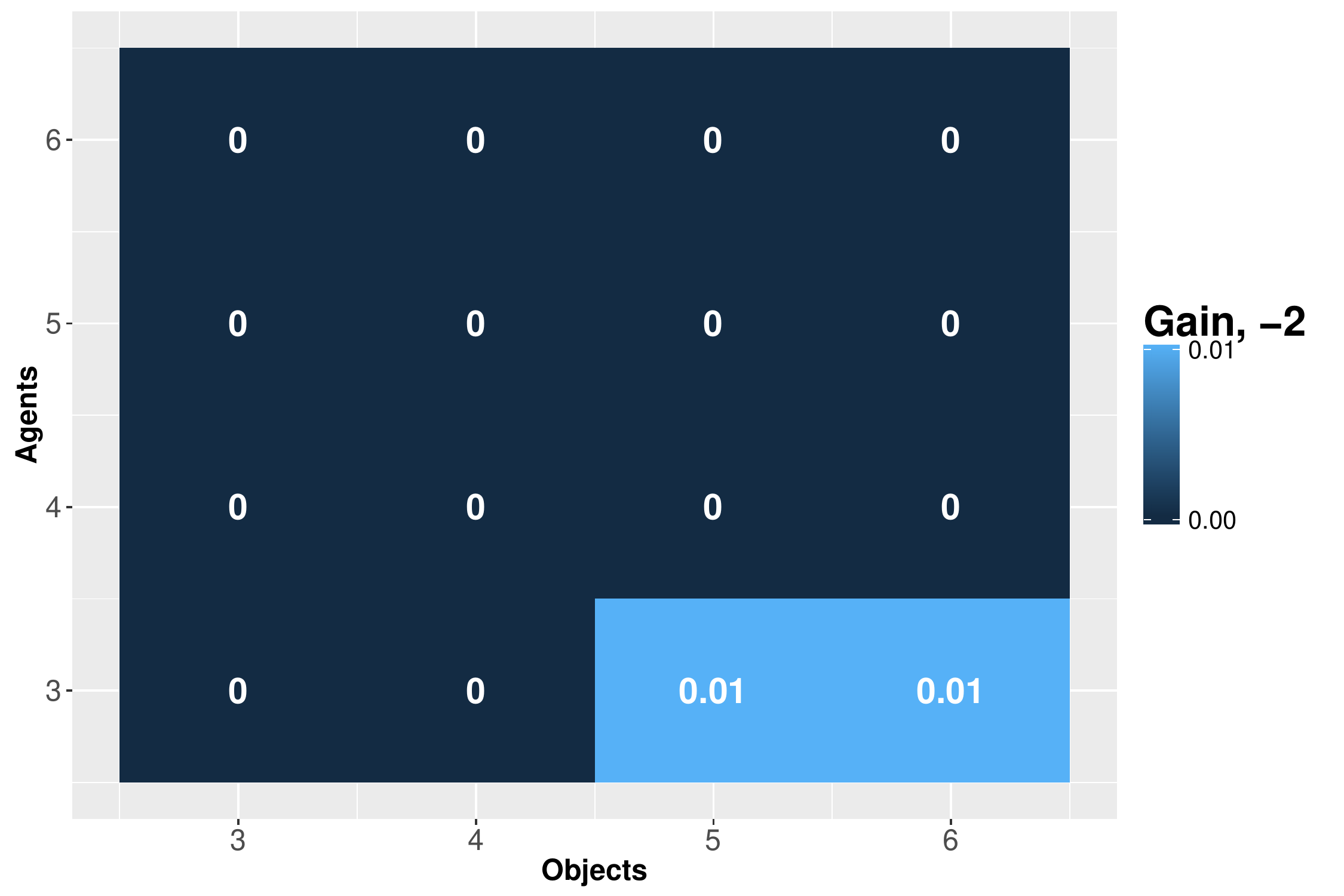}
		\caption{Gain, Mallows, $\alpha = -2$.}
		\label{}
	\end{subfigure}
	\begin{subfigure}[t]{0.49\textwidth}
		\centering
		\includegraphics[width=\textwidth]{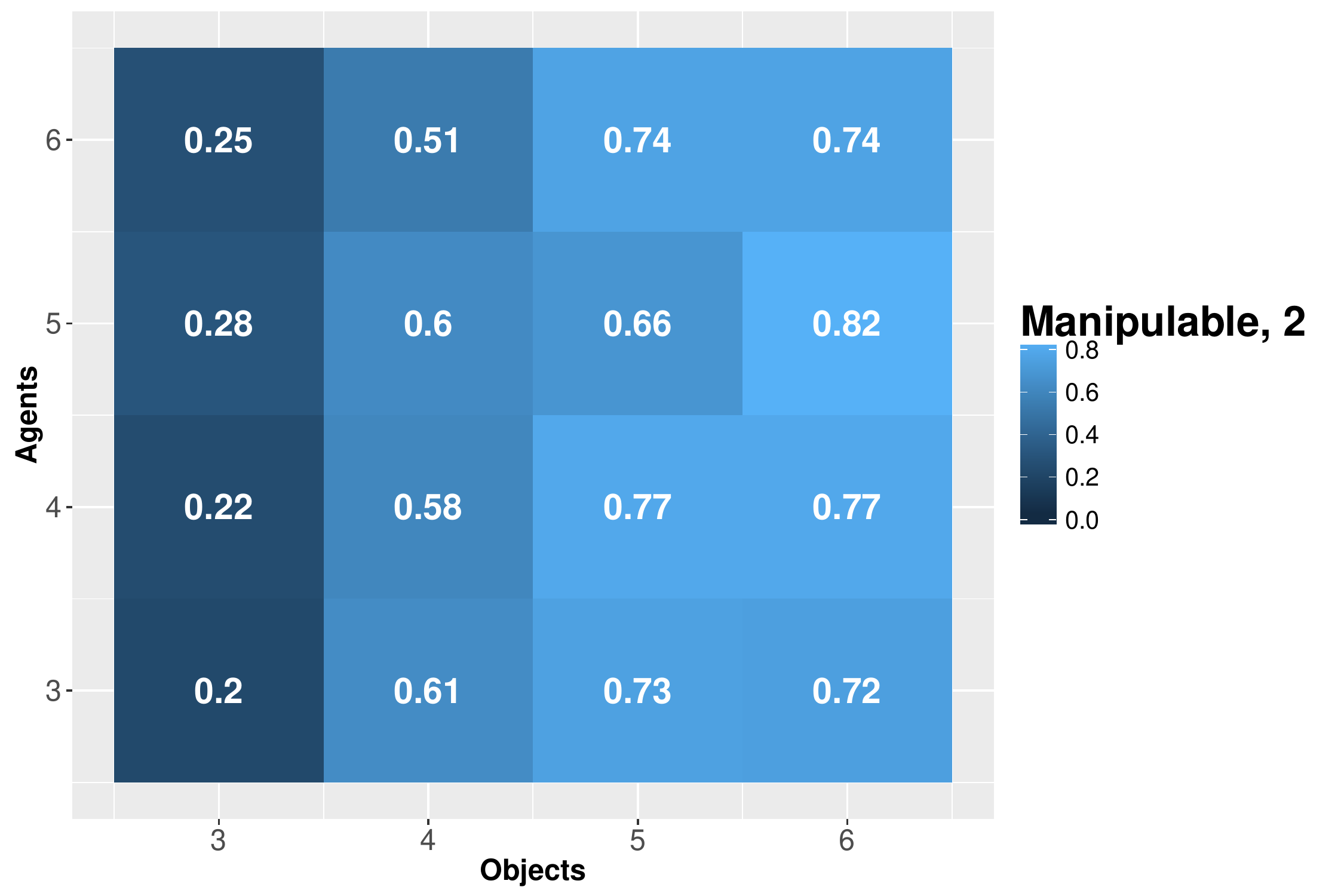}
		\caption{Manipulation, Mallows, $\alpha = 2$.}
		\label{}
	\end{subfigure}~
	\begin{subfigure}[t]{0.49\textwidth}
		\centering
		\includegraphics[width=\textwidth]{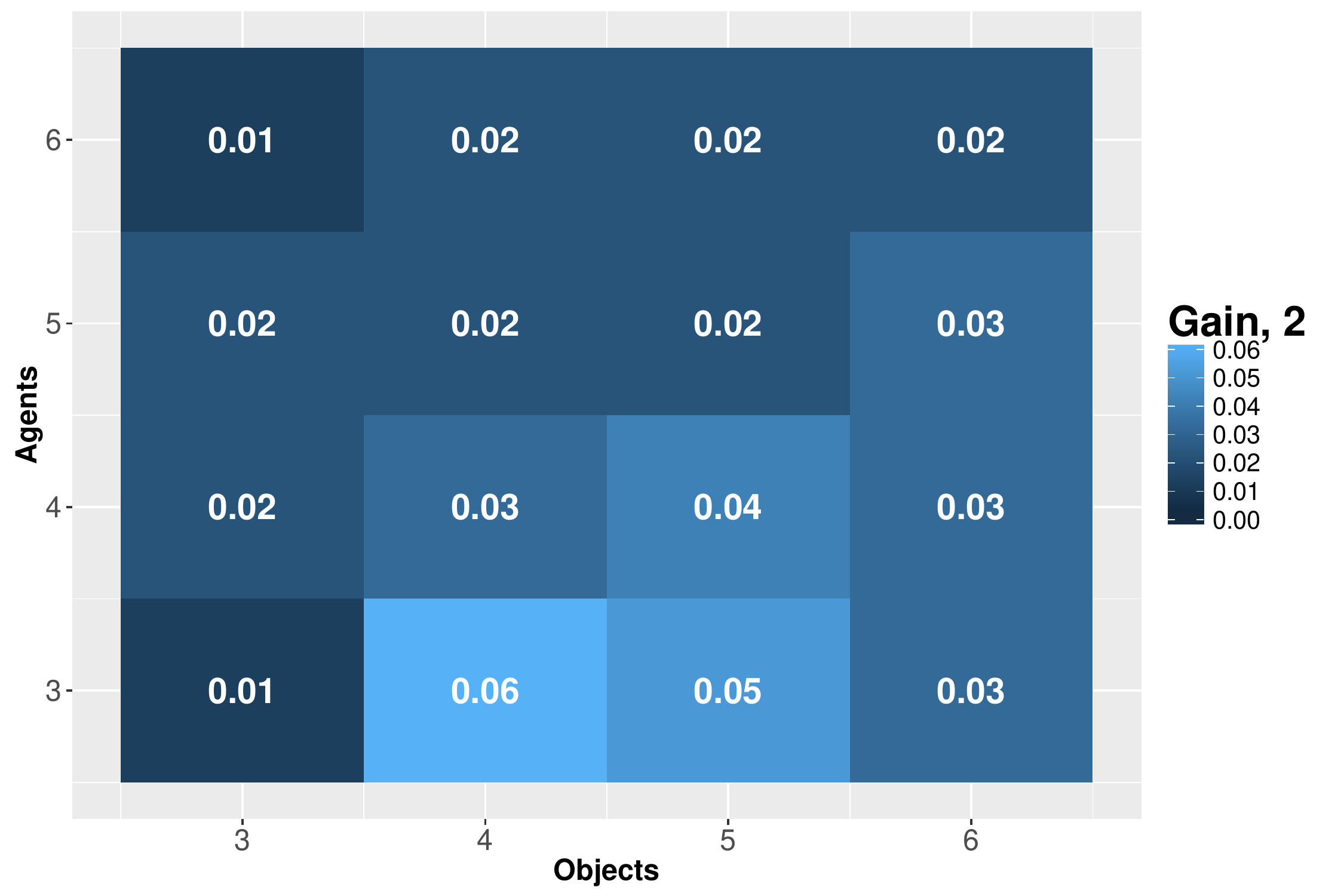}
		\caption{Gain, Mallows, $\alpha = 2$.}
		\label{}
	\end{subfigure}
	\caption{The fraction of manipulable instances and manipulation gain of PS under the pure Mallows model with one reference ranking.} 
	\label{fig:manipMallow1}
\end{figure*}

\begin{figure*}
	\begin{subfigure}[t]{0.49\textwidth}
		\centering
		\includegraphics[width=\textwidth]{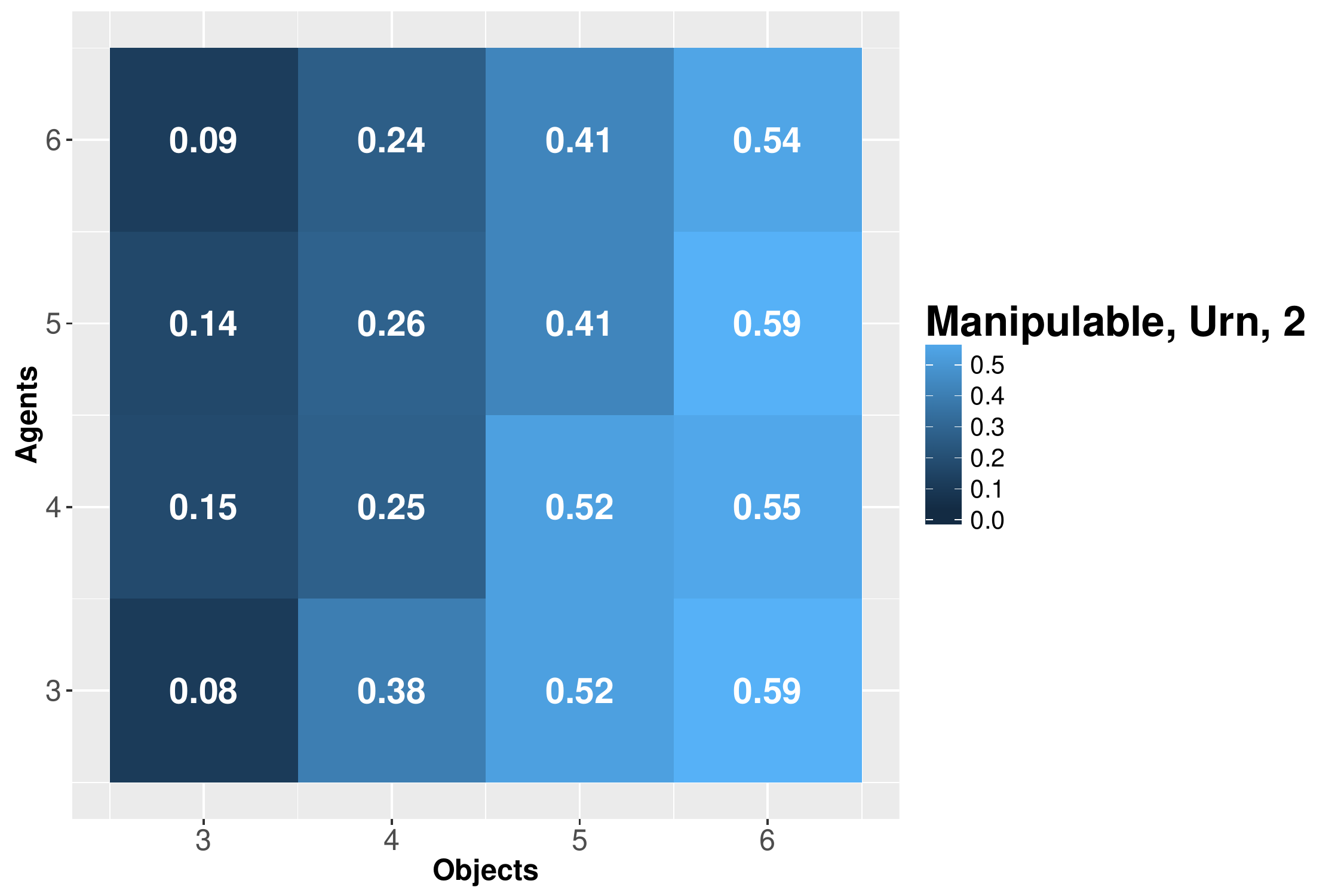}
		\caption{Manipulation, Urn, $\alpha = 2$.}
		\label{}
	\end{subfigure}~
	\begin{subfigure}[t]{0.49\textwidth}
		\centering
		\includegraphics[width=\textwidth]{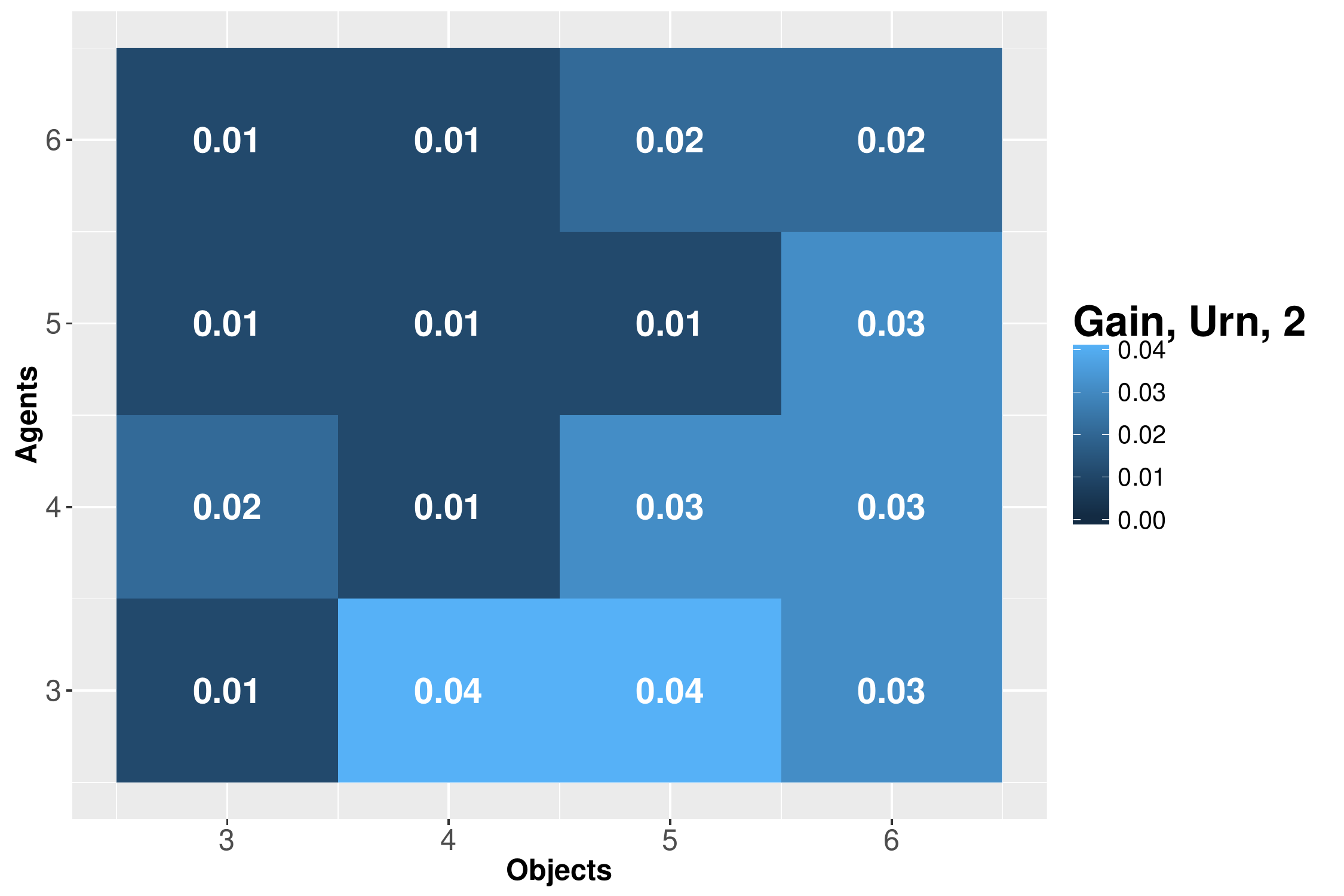}
		\caption{Gain, Urn, $\alpha = 2$.}
		\label{}
	\end{subfigure}
	\caption{The fraction of manipulable instances and manipulation gain of PS under the Polya-Urn model.} 
	\label{fig:manipUrn}
\end{figure*}

The fraction of manipulable profiles and manipulation gain goes to 0 when agents are risk seeking. Under risk aversion, the fraction of manipulable profiles and manipulation gain rapidly increases as $\frac{m}{n}$ grows (Figure \ref{fig:manipMallow5}). These results hold under pure Mallows distribution (Figure \ref{fig:manipMallow1}) as well as under the Polya-Urn model (Figure \ref{fig:manipUrn}) but with slightly slower growth. This is consistent with the fact that, in less diverse populations, agents' preference are more similar and conflicting, and thus manipulation is less likely (even though still significantly considerable) as opposed to more diverse and uniform set of profiles.  

\section{Related Literature}

Assignment problems with ordinal preferences have attracted interest from many researchers.  
Svensson showed that serial dictatorship is the only deterministic mechanism that is strategyproof, nonbossy, and neutral~\cite{svensson1999strategy}. 
Random Serial Dictatorship (RSD) (uniform randomization over all serial dictatorship assignments) satisfies strategyproofness, proportionality, and ex post efficiency~\cite{abdulkadirouglu1998random}. Bogomolnaia and Moulin noted the inefficiency of RSD from the ex ante perspective, and characterized the matching mechanisms based on first-order stochastic dominance~\cite{bogomolnaia2001new}. They proposed the probabilistic serial mechanism as an efficient and envyfree mechanism with regards to ordinal preferences. While PS is not strategyproof, it satisfies weak strategyproofness for problems with equal number of agents and objects. However, PS is strictly manipulable (not weakly strategyproof) when there are more objects than agents \cite{kojima2009random}.
Kojima and Manea, showed that in large assignment problems with sufficiently many copies of each object, truth-telling is a weakly dominant strategy in PS~\cite{kojima2010incentives}. In fact PS and RSD mechanisms become equivalent \cite{che2010asymptotic}, that is, the inefficiency of RSD and manipulability of PS vanishes when the number of copies of each object approaches infinity.

The practical implications of deploying RSD and PS have been the center of attention in many one-sided matching problems~\cite{abdulkadiroglu2009strategy,mennle2014hybrid}. 
In the school choice setting with multi-capacity alternatives, Pathak observed that many students obtained a more desirable random assignment through PS in public schools of New York City~\cite{pathak2006lotteries}; however, the efficiency difference was quite small. These equivalence results and their extensions to all random mechanisms~\cite{liu2013ordinal}, do not hold when the quantities of each object is limited to one.

Other interesting aspects of PS and RSD such as computational complexity and best-response strategies have also been explored \cite{ekici2012equilibrium,aziz2015manipulating,Aziz:2015:EUP:2832249.2832402}.
In this vein, Aziz et al. proved the existence of pure Nash equilibria, but showed that computing an equilibrium is NP-hard \cite{Aziz:2015:EUP:2832249.2832402}.
Nevertheless, Mennle et al. \cite{Mennle:2015:PLM:2832249.2832261} showed that agents can easily find near-optimal strategies by simple local and greedy search.
In the absence of truthful incentives, the outcome of PS is no longer guaranteed to be efficient or envyfree with respect to agents' true underlying preferences, and this inefficiency may result in outcomes that are worse than RSD, especially in `small' markets \cite{ekici2012equilibrium}.
The utilitarian and egalitarian welfare guarantees of RSD have been studied under ordinal and linear utility assumptions \cite{bhalgat2011social,aziz2015egalitarianism}. 
For arbitrary utilities, RSD provides the best approximation ratio for utilitarian social welfare when $m=n$ among all mechanisms that rely only on ordinal preferences \cite{filos2014social}.

\section{Discussion}

We studied the space of general preferences and provided empirical results on the incomparability of RSD and PS. It is worth mentioning that at preference profiles where PS and RSD induce identical assignments, RSD is \emph{sd}-efficient, \emph{sd}-envyfree, and \emph{sd}-strategyproof. However, PS is still highly manipulable. 
We further strengthen this argument by providing an observation in Example \ref{ex:PSnotRSD}:

\begin{example}\label{ex:PSnotRSD}
	Consider the following preference profile $\succ = ((bca), (cab), (bca))$. Table~\ref{tab:equal} shows the prescribed random assignment.
	In this example, with PS as the matching mechanism, agent 1 can misreport her preference as $\succ'_{1}=(cba)$, and manipulate her assignment to $1/4(b), 1/2(c), 1/4(a)$. 
	\begin{table} 
		\centering
		\caption{A random assignment for a preference profile wherein PS and RSD both prescribe an identical matching, i.e. $PS(\succ) = RSD(\succ)$.}
		\begin{tabular}{ c c c c}
			\toprule
			& $a$ & $b$ & $c$ \\ \hline\hline
			$A_1$ & $1/3$ & $1/2$ & $1/6$ \\
			$A_2$ & $1/3$ & $0$ & $2/3$ \\
			$A_3$ & $1/3$ & $1/2$ & $1/6$ \\
			\bottomrule
		\end{tabular}
		\label{tab:equal}
	\end{table}
	It is easy to see that agent 1's misreport improves her expected outcome for all utility models where $\frac{2}{6}u_{1}(c) > \frac{1}{4}u_{1}(b) + \frac{1}{12}u_{1}(a)$ (for example utilities $10, 9, 0$ for $b,c,a$ respectively.).
\end{example}

\begin{figure}
	\centering
	\includegraphics[width=.8\linewidth]{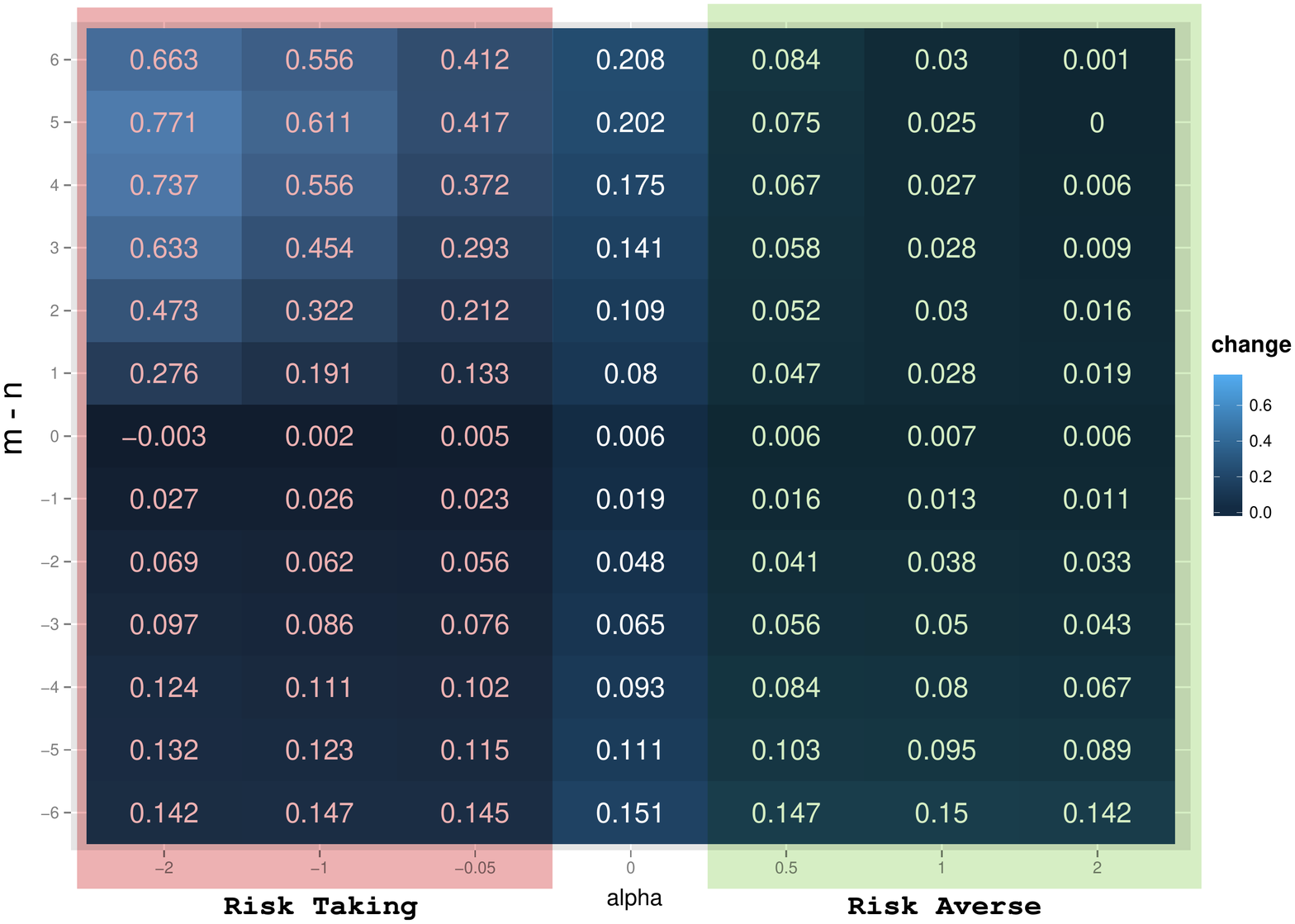}
	\caption{The percentage change in social welfare between RSD and PS for $\alpha\in (-2,-1,-0.5, 0, 0.5, 1, 2)$ and different combinations of $m-n$. Positive $\alpha$ indicates risk averse and negative $\alpha$ risk taking profiles. Linear utility is indicated by $\alpha = 0$. As agents become more risk averse the social welfare gap between RSD and PS closes.}
	\label{fig:changeAll}
\end{figure}

We investigated various utility models according to different risk attitudes. Our findings hold under various assumptions on the population of agents and preference profile distributions.
Our main results are:

\begin{itemize}
	\item In terms of efficiency, the fraction of preference profiles $\succ\in \mathcal{P}^{n}$ for which PS stochastically (or lexicographically) dominates RSD converges to zero as $\frac{n}{m} \to 1$. When instantiating the preferences with actual utility functions, PS allocations are only slightly better than RSD allocations in terms of social welfare when varying $n$ and $m$, particularly under risk averse utilities. In fact, in some cases RSD allocations are superior in terms of social welfare (see Figure \ref{fig:changeAll}).
	
	\item PS is almost 99\% manipulable when $n \leq m$ and the fraction of \emph{sd}- and \emph{ld}- manipulable profiles rapidly goes to 1 as $\frac{m}{n}$ grows. When instantiating the preferences with utility functions, the manipulability of PS increases as agents become more risk averse. Moreover, an agent's utility gain from manipulation also grows when the risk intensity increases.
	
	\item For risk seeking utilities, when $n \geq m$ the fraction of envious agents under all profiles vanishes and RSD becomes envyfree. For risk averse utilities, the fraction of envious agents increases as agents become more risk averse. However, the total amount of envy just slightly grows, and surprisingly, only few agents feel extremely envious while all other agents are either envyfree or only feel a minimal amount of envy.
	
\end{itemize}

An interesting future direction is to study egalitarian social welfare of the matching mechanisms in single and multi unit assignment problems as well as in the full preference domain. Another open direction is to provide a parametric analysis of the matching mechanisms according to the risk aversion factor.

\section{Design Recommendations for Multiagent Systems Practitioners}

Our work in this paper can be used to help guide designers of multiagent systems who need to solve allocation problems. If a designer strongly requires \emph{sd}-efficiency then the theoretical results of PS indicate that it is better than RSD. However, our results show that PS is highly prone to manipulation for various combinations of agents and objects. This manipulation and the possible gain from manipulation become more severe particularly when agents are risk averse, and designers need to take this into consideration.
On the other hand, while RSD does not theoretically guarantee \emph{sd}-efficiency, our results show that it tends to do quite well -- sometimes even outperforming PS in terms of social welfare. RSD also has the added advantage of being \emph{sd}-strategyproof and thus is not prone to the manipulation problems of PS. 

Although computing RSD probabilities (fractional assignments) is \#P-hard \cite{aziz2013computational,saban2015complexity}, RSD is easy to implement in practice.
However, the welfare cost of adopting manipulable mechanisms such as PS raises concern and has real consequences \cite{pathak2011lotteries,budish2012multi}. Even though computing optimal manipulation strategies is computationally hard for the PS mechanism, evidentially individuals can easily figure out how to manipulate such mechanisms using simple greedy heuristics \cite{budish2012multi,Mennle:2015:PLM:2832249.2832261}. Our investigations show that in many instances RSD performs as desirably as PS in terms of social welfare. Conversely, PS assignments are highly susceptible to manipulation especially when agents are risk averse. 

These findings suggest that in multiagent settings where mechanism designers are unsure of sincere reporting of their preferences or when agents are mostly risk averse, the use of RSD is more desirable to ensure truthful reporting while providing reasonable social welfare. However, PS is still a desirable allocation mechanism for its fairness and efficiency properties, particularly in settings where agents are sincere.



\bibliographystyle{spmpsci}      
\bibliography{references}   

\clearpage

\begin{appendices}

\section{Numerical Results} \label{app:numerial}

The following table shows the results of comparing RSD and PS under ordinal preferences for various combinations of agents and objects. Note that in most instances, RSD and PS do not induce the same random allocation.

\begin{longtable}{@{}lllllllll@{}}
	\toprule
	\centering \small
	&            &                & \multicolumn{2}{l}{\textbf{Dominance}} & \textbf{RSD}      & \multicolumn{3}{l}{\textbf{PS manipulability}} \\ \midrule
	\textbf{n} & \textbf{m} & \textbf{Equal} & \textbf{SD}        & \textbf{LD}       & \textbf{weakEnvy} & \textbf{weak}   & \textbf{SD}   & \textbf{LD}  \\ \midrule
	\endhead
	\hline \multicolumn{3}{|r|}{{Continued on the next page}} \\ \hline
	\endfoot
	
	\endlastfoot
	
	2          & 2          & 100\%          & 0\%                & 0\%               & 0\%               & 0\%             & 0\%           & 0\%          \\
	2          & 3          & 27\%           & 18\%               & 29\%              & 23\%              & 31\%            & 31\%          & 31\%         \\
	2          & 4          & 10\%           & 36\%               & 60\%              & 20\%              & 53\%            & 53\%          & 53\%         \\
	2          & 5          & 3\%            & 39\%               & 78\%              & 16\%              & 78\%            & 78\%          & 78\%         \\
	2          & 6          & 1\%            & 45\%               & 90\%              & 13\%              & 87\%            & 87\%          & 87\%         \\
	2          & 7          & 0\%            & 46\%               & 95\%              & 12\%              & 95\%            & 95\%          & 95\%         \\
	2          & 8          & 0\%            & 45\%               & 96\%              & 11\%              & 97\%            & 97\%          & 97\%         \\
	2          & 9          & 0\%            & 47\%               & 96\%              & 11\%              & 100\%           & 100\%         & 100\%        \\
	2          & 10         & 0\%            & 48\%               & 99\%              & 9\%               & 99\%            & 99\%          & 99\%         \\ \hline
	3          & 2          & 100\%          & 0\%                & 0\%               & 0\%               & 0\%             & 0\%           & 0\%          \\
	3          & 3          & 67\%           & 0\%                & 0\%               & 11\%              & 24\%            & 0\%           & 0\%          \\
	3          & 4          & 3\%            & 5\%                & 40\%              & 47\%              & 77\%            & 5\%           & 5\%          \\
	3          & 5          & 0\%            & 4\%                & 75\%              & 46\%              & 96\%            & 26\%          & 27\%         \\
	3          & 6          & 0\%            & 6\%                & 84\%              & 42\%              & 95\%            & 53\%          & 54\%         \\
	3          & 7          & 0\%            & 5\%                & 90\%              & 41\%              & 100\%           & 68\%          & 69\%         \\
	3          & 8          & 0\%            & 5\%                & 93\%              & 39\%              & 100\%           & 80\%          & 83\%         \\
	3          & 9          & 0\%            & 9\%                & 96\%              & 35\%              & 100\%           & 90\%          & 92\%         \\
	3          & 10         & 0\%            & 7\%                & 95\%              & 34\%              & 100\%           & 94\%          & 94\%         \\ \hline
	4          & 2          & 62\%           & 38\%               & 38\%              & 0\%               & 0\%             & 0\%           & 0\%          \\
	4          & 3          & 33\%           & 34\%               & 46\%              & 21\%              & 42\%            & 0\%           & 0\%          \\
	4          & 4          & 21\%           & 3\%                & 8\%               & 27\%              & 72\%            & 0\%           & 0\%          \\
	4          & 5          & 0\%            & 0\%                & 48\%              & 61\%              & 96\%            & 1\%           & 1\%          \\
	4          & 6          & 0\%            & 0\%                & 76\%              & 62\%              & 98\%            & 17\%          & 18\%         \\
	4          & 7          & 0\%            & 0\%                & 84\%              & 62\%              & 100\%           & 33\%          & 35\%         \\
	4          & 8          & 0\%            & 1\%                & 93\%              & 61\%              & 99\%            & 52\%          & 54\%         \\
	4          & 9          & 0\%            & 1\%                & 94\%              & 60\%              & 100\%           & 65\%          & 69\%         \\
	4          & 10         & 0\%            & 2\%                & 95\%              & 56\%              & 100\%           & 79\%          & 85\%         \\ \hline
	5          & 2          & 39\%           & 61\%               & 61\%              & 0\%               & 0\%             & 0\%           & 0\%          \\
	5          & 3          & 8\%            & 34\%               & 83\%              & 27\%              & 66\%            & 0\%           & 0\%          \\
	5          & 4          & 3\%            & 19\%               & 53\%              & 42\%              & 94\%            & 0\%           & 0\%          \\
	5          & 5          & 6\%            & 1\%                & 7\%               & 42\%              & 90\%            & 0\%           & 0\%          \\
	5          & 6          & 0\%            & 0\%                & 58\%              & 69\%              & 100\%           & 0\%           & 0\%          \\
	5          & 7          & 0\%            & 0\%                & 84\%              & 71\%              & 100\%           & 4\%           & 4\%          \\
	5          & 8          & 0\%            & 0\%                & 91\%              & 71\%              & 100\%           & 18\%          & 18\%         \\
	5          & 9          & 0\%            & 0\%                & 94\%              & 71\%              & 100\%           & 32\%          & 36\%         \\
	5          & 10         & 0\%            & 0\%                & 97\%              & 70\%              & 100\%           & 49\%          & 55\%         \\ \hline
	6          & 2          & 21\%           & 79\%               & 79\%              & 0\%               & 0\%             & 0\%           & 0\%          \\
	6          & 3          & 2\%            & 71\%               & 96\%              & 31\%              & 59\%            & 0\%           & 0\%          \\
	6          & 4          & 0\%            & 22\%               & 88\%              & 52\%              & 90\%            & 0\%           & 0\%          \\
	6          & 5          & 0\%            & 9\%                & 46\%              & 59\%              & 98\%            & 0\%           & 0\%          \\
	6          & 6          & 3\%            & 1\%                & 7\%               & 54\%              & 96\%            & 0\%           & 0\%          \\
	6          & 7          & 0\%            & 0\%                & 62\%              & 74\%              & 100\%           & 0\%           & 0\%          \\
	6          & 8          & 0\%            & 0\%                & 89\%              & 74\%              & 100\%           & 1\%           & 1\%          \\
	6          & 9          & 0\%            & 0\%                & 95\%              & 75\%              & 100\%           & 8\%           & 9\%          \\
	6          & 10         & 0\%            & 0\%                & 97\%              & 75\%              & 100\%           & 23\%          & 25\%         \\ \hline
	7          & 2          & 12\%           & 88\%               & 88\%              & 0\%               & 0\%             & 0\%           & 0\%          \\
	7          & 3          & 1\%            & 64\%               & 99\%              & 33\%              & 83\%            & 0\%           & 0\%          \\
	7          & 4          & 0\%            & 26\%               & 97\%              & 57\%              & 99\%            & 0\%           & 0\%          \\
	7          & 5          & 0\%            & 8\%                & 87\%              & 66\%              & 100\%           & 0\%           & 0\%          \\
	7          & 6          & 0\%            & 2\%                & 41\%              & 69\%              & 100\%           & 0\%           & 0\%          \\
	7          & 7          & 1\%            & 1\%                & 6\%               & 61\%              & 99\%            & 0\%           & 0\%          \\
	7          & 8          & 0\%            & 0\%                & 71\%              & 79\%              & 100\%           & 0\%           & 0\%          \\
	7          & 9          & 0\%            & 0\%                & 93\%              & 79\%              & 100\%           & 0\%           & 0\%          \\
	7          & 10         & 0\%            & 0\%                & 96\%              & 78\%              & 100\%           & 5\%           & 6\%          \\ \hline
	8          & 2          & 8\%            & 92\%               & 92\%              & 0\%               & 0\%             & 0\%           & 0\%          \\
	8          & 3          & 0\%            & 63\%               & 100\%             & 34\%              & 76\%            & 0\%           & 0\%          \\
	8          & 4          & 0\%            & 33\%               & 99\%              & 60\%              & 95\%            & 0\%           & 0\%          \\
	8          & 5          & 0\%            & 10\%               & 97\%              & 70\%              & 100\%           & 0\%           & 0\%          \\
	8          & 6          & 0\%            & 4\%                & 83\%              & 74\%              & 100\%           & 0\%           & 0\%          \\
	8          & 7          & 0\%            & 1\%                & 29\%              & 74\%              & 100\%           & 0\%           & 0\%          \\
	8          & 8          & 0\%            & 0\%                & 5\%               & 69\%              & 99\%            & 0\%           & 0\%          \\
	8          & 9          & 0\%            & 0\%                & 70\%              & 81\%              & 100\%           & 0\%           & 0\%          \\
	8          & 10         & 0\%            & 0\%                & 93\%              & 82\%              & 100\%           & 0\%           & 0\%          \\ \hline
	9          & 2          & 3\%            & 97\%               & 97\%              & 0\%               & 0\%             & 0\%           & 0\%          \\
	9          & 3          & 0\%            & 76\%               & 100\%             & 35\%              & 70\%            & 0\%           & 0\%          \\
	9          & 4          & 0\%            & 33\%               & 100\%             & 62\%              & 100\%           & 0\%           & 0\%          \\
	9          & 5          & 0\%            & 19\%               & 99\%              & 72\%              & 100\%           & 0\%           & 0\%          \\
	9          & 6          & 0\%            & 6\%                & 98\%              & 76\%              & 100\%           & 0\%           & 0\%          \\
	9          & 7          & 0\%            & 2\%                & 78\%              & 78\%              & 100\%           & 0\%           & 0\%          \\
	9          & 8          & 0\%            & 0\%                & 26\%              & 78\%              & 100\%           & 0\%           & 0\%          \\
	9          & 9          & 0\%            & 0\%                & 4\%               & 71\%              & 100\%           & 0\%           & 0\%          \\
	9          & 10         & 0\%            & 0\%                & 69\%              & 84\%              & 100\%           & 0\%           & 0\%          \\ \hline
	10         & 2          & 2\%            & 99\%               & 99\%              & 0\%               & 0\%             & 0\%           & 0\%          \\
	10         & 3          & 0\%            & 70\%               & 100\%             & 37\%              & 79\%            & 0\%           & 0\%          \\
	10         & 4          & 0\%            & 46\%               & 100\%             & 63\%              & 98\%            & 0\%           & 0\%          \\
	10         & 5          & 0\%            & 17\%               & 100\%             & 73\%              & 97\%            & 0\%           & 0\%          \\
	10         & 6          & 0\%            & 10\%               & 99\%              & 77\%              & 100\%           & 0\%           & 0\%          \\
	10         & 7          & 0\%            & 2\%                & 95\%              & 79\%              & 100\%           & 0\%           & 0\%          \\
	10         & 8          & 0\%            & 1\%                & 77\%              & 80\%              & 100\%           & 0\%           & 0\%          \\
	10         & 9          & 0\%            & 0\%                & 21\%              & 79\%              & 100\%           & 0\%           & 0\%          \\
	10         & 10         & 0\%            & 0\%                & 4\%               & 73\%              & 100\%           & 0\%           & 0\%          \\ \bottomrule
	\caption{Experimental results over the space of preference profiles. SD (respectively LD) refers to the fraction of profiles where PS Stochastically (Lexicographically) Dominates RSD, and weakEnvy shows the average fraction of agents that are weakly envious under RSD. The last three columns show the fraction of profiles that PS is weakly manipulable, \emph{sd}-manipulable (SD), and \emph{ld}-manipulable (LD).}
	\label{tab:numerialOrdinalResults}
\end{longtable}

%

\end{appendices}

\end{document}